\RequirePackage{snapshot}

\documentclass[11pt,]{lmcs}

\makeatletter
\def\endfront@text{}
\makeatother

	\theoremstyle{plain}
      \newtheorem{lemma}[thm]{Lemma}
      \newtheorem*{thm*}{Theorem}
      \newtheorem*{prop*}{Proposition}
    \theoremstyle{definition}
      \newtheorem{defn}{Definition}

\title[Denotational cost semantics]
{Denotational semantics as a foundation for cost recurrence extraction for
functional languages}

\author{Norman Danner}
\address{Wesleyan University, USA}
\email{ndanner@wesleyan.edu}
\thanks{Norman Danner's research is supported by 
the National Science Foundation under grant number~1618203.  Conflicts of
interest:  none.}

\author{Daniel R. Licata}
\address{Wesleyan University, USA}
\email{dlicata@wesleyan.edu}
\thanks{Daniel R. Licata's research is supported by the National
Science Foundation under grant number~1618203.  Conflicts of interest:  none.}

\usepackage{accents}

\usepackage{bussproofs}
    
\EnableBpAbbreviations
\def\doRule#1{\ifx#1\relax\else\RightLabel{#1}\fi}
\newcommand{\ndAXC}[2][\relax]{\AXC{$\mathstrut$}\doRule{#1}\UIC{#2}}
\newcommand{\ndUIC}[2][\relax]{\doRule{#1}\UIC{#2}}
\newcommand{\ndBIC}[2][\relax]{\doRule{#1}\BIC{#2}}
\newcommand{\ndTIC}[2][\relax]{\doRule{#1}\TIC{#2}}

    \let\proves\vdash

\def\oftype{\mathbin:}
\def\tmoftype#1#2{{#1}\oftype{#2}}
\newcommand{\typepf}[3][{}]{\proves_{#1}\tmoftype{#2}{#3}}

\newcommand{\emptyctx}{\underbar{~}}

\newcommand{\typejudge}[4][{}]{\mbox{${#2}\typepf[{#1}]{#3}{#4}$}}

\newcommand{\typejudgeG}[3][{}]{\typejudge[#1]\Gamma{#2}{#3}}

\newcommand{\typejudgeE}[3][{}]{\typejudge[#1]{\underbar{~}}{#2}{#3}}

    \newcommand{\evalto}{\downarrow}
    \newcommand{\evalin}[3]{\mbox{${#1}\evalto^{#3}{#2}$}}
    \newcommand{\eval}[2]{\mbox{${#1}\evalto {#2}$}}
    \newcommand{\typejudgeV}[2]{\mbox{$\typepf{#1}{#2}$}}

\usepackage{float}
      \makeatletter
  \newcommand{\fs@singlerule}{%
    \fs@plain%
    \def\@fs@mid{\vskip1ex\hrule\vskip1ex}
  }
  \makeatother
  \floatstyle{singlerule}
  \restylefloat{figure}
  \restylefloat{table}
  \newfloat{program}{t}{pgm}
  \floatname{program}{Program}

\usepackage{import}

\usepackage{mathtools}
\usepackage{stmaryrd}
\usepackage{subcaption}
\usepackage{suffix}

\usepackage{tikz-cd}

\usepackage{url}
\usepackage{natbib}

\newcommand*{\doi}[1]{\href{http://dx.doi.org/#1}{doi: #1}}
\hypersetup{colorlinks=true,allcolors=blue}

\usepackage{color}
\usepackage[normalem]{ulem}

\input{macros}

\begin{document}

\begin{abstract}

A standard informal method for analyzing the asymptotic complexity of a
program is to extract a recurrence that describes its cost in terms of the
size of its input, and then to compute a closed-form upper bound on that
recurrence.  We give a formal account of that method for functional programs
in a higher-order language with $\sletkw$-polymorphism. The method consists of
two phases.  In the first phase, a monadic translation is performed to extract
a cost-annotated version of the original program.  In the second phase, the
extracted program is interpreted in a model.  The key feature of this second
phase is that different models describe different notions of size.  This plays
out in several ways.  For example, when analyzing functions that take
arguments of inductive types, different notions of size may be appropriate
depending on the analysis.  When analyzing polymorphic functions, our approach
shows that one can formally describe the notion of size of an argument in
terms of the data that is common to the notions of size for each type instance
of the domain type.  We give several examples of different models that
formally justify various informal cost analyses to show the applicability of
our approach.  

\end{abstract}

\maketitle

\section{Introduction}

The method for analyzing the asymptotic cost of a (functional)
program~$f(x)$ that is typically taught to introductory
undergraduate students is to extract a recurrence~$T_f(n)$ that describes an upper
bound on the cost of~$f(x)$ in terms of the size of~$x$, and then
establish a non-recursive upper bound on~$T_f(n)$ (we will focus on upper
bounds, but much of what we say holds \emph{mutatis mutandis} for lower
bounds, and hence tight bounds).
The goal of this work is to put the process of this informal approach to
cost analysis on firm mathematical footing.
Of course, 
various formalizations of cost analysis have been discussed for
almost as long as there has been a distinct subfield of Programming Languages.
Most of the recent work in this area
is focused on developing formal techniques for cost analysis that enable
the (possibly automated) analysis of as large a swath of programs as possible.
In doing so, the type systems and the logics used grow ever more complex.  
There is work that incorporates size and cost into type information, for
example by employing refinement types or type-and-effect systems.  There is
work that formalizes reasoning about cost in program logics such as separation
logic with time credits.  But as witnessed
by most undergraduate texts on algorithm analysis, complex type systems and
separation logic are not commonly taught.  Instead, a function
of some form (a \emph{recurrence}) 
that computes the cost in terms of the size of the argument
is extracted from the source code.
This is the case for ``simple'' compositional worst-case analyses,
but also more more complex techniques.  For example,
the banker's and physicist's methods of amortized analysis likewise proceed by
extracting a function to describe cost; the notion of cost itself, and the
extraction of a suitably precise cost function, is more complex, but broadly
speaking the structure of the analysis is the same.
That is the space we are
investigating here:  how do we justify that informal process?\footnote{We do
mean the process here---some of the approaches do end up sythesizing
recurrences, but that is almost a side-effect rather than the first step.}
The justification might not itself play a role in applying the
technique informally, any more than we require introductory students to
understand the theory behind a type inference algorithm in order to
informally understand why their programs typecheck.  But certainly that
theory should be settled.
Our approach is through denotational semantics, which,
in addition to justifying
the informal process, also helps to explicate a few
questions, such as why length is an appropriate measure of size for cost
recurrences for polymorphic list functions (a question that is close to, but
not quite the same as, parametricity).

Turning to the technical development, in previous work
\citep{danner-et-al:plpv13,danner-et-al:icfp15} we have developed a recurrence
extraction technique for higher-order functional programs for which the
bounding is provable that is based on work by \citet{danner-royer:ats-lmcs}.
The technique is described as follows:

\begin{enumerate}
\item We define what is essentially a monadic translation into the writer
monad from a call-by-value
\emph{source language} that supports inductive types and structural
recursion (fold) to a call-by-name \emph{recurrence language}; we refer to
programs in the latter language as \emph{syntactic recurrences}.
The recurrence language is axiomatized by a
\emph{size (pre)order} rather than equations.  The syntactic recurrence
extracted from a source-language program~$f(x)$ 
describes both the cost and result of~$f(x)$ in terms of~$x$.

\item We define a \emph{bounding relation} between source language programs
and syntactic recurrences.  The bounding relation is a logical relation that
captures the notion that the syntactic recurrence is in fact a bound on the
operational cost and the result of the source language program.  This notion
extends reasonably to higher-type, where higher-type arguments of a syntactic
recurrence are thought of recurrences that are bounds on the corresponding
arguments of the source language program.  We then prove a \emph{bounding
theorem} that asserts that every typeable program in the source language is
related to the recurrence extracted from it.

\item The syntactic recurrence is interpreted in a model of the recurrence
language.  This is where values are abstracted to some notion of size; e.g.,
the interpretation may be defined so that a value of inductive type~$\delta$
is interpreted by the number of $\delta$-constructors in~$v$.  We call the
interpretation of a syntactic recurrence a \emph{semantic recurrence}, and it
is the semantic recurrences that are intended to match the recurrences that
arise from informal analyses.
\end{enumerate}

In this paper, we extend the above approach in several ways.  First and
foremost, we investigate the models, semantic recurrences, and size abstraction more
thoroughly than in previous work, and show how different models can be
used to formally justify typical informal extract-and-solve cost
analyses.  Second, we add ML-style $\sletkw$-polymorphism, and adapt the
techniques to an environment-based operational semantics, a more
realistic foundation for implementation than the substitution-based
semantics used in previous work.  In recent work, we have extended the
technique for source languages with call-by-name and general
recursion~\citep{kavvos-et-al:popl2020}, and for amortized
analyses~\citep{cutler-et-al:icpf20}; we do not consider these
extensions in the main body of this paper, in order to focus on the above issues in isolation.

Our source language, which we describe in Section~\ref{sec:src-lang}, is a
call-by-value higher-order functional language with inductive datatypes and
structural recursion (fold) and ML-style $\sletkw$-polymorphism.  That is,
$\sletkw$-bound identifiers may be assigned a quantified type, provided that
type is instantiated at quantifier-free types in the body of the
$\sletkw$~expression.  Restricting to polymorphism that is predicative
(quantifiers can be instantiated only with non-quantified types), first-order
(quantifiers range over types, not type constructors), and second-class
(polymorphic functions cannot themselves be the input to other functions) is
sufficient to program a number of example programs, without complicating the
denotational models used to analyze them.  We define an environment-based
operational semantics where each rule is annotated with a cost.  For
simplicity, we only ``charge'' for each unfolding of a recursive call, but the
technique extends easily for any notion of cost that can be defined in terms
of evaluation rules.  We could also replace the rule-based cost annotations
with a ``tick'' construct that the programmer inserts at the code points for
which a charge should be made, though this requires the programmer to justify
the cost model.

The recurrence language in Section~\ref{sec:rec-lang} is a call-by-name
$\lambda$-calculus with explicit predicative polymorphism (via type
abstraction and type application) and an additional type for costs.  Ultimately we
care only about the meaning of a syntactic recurrence, not so much any
particular strategy for evaluating it, and such a focus on mathematical
reasoning makes call-by-name an appropriate formalism.  The choice of
explicit predicative polymorphism instead of $\sletkw$-polymorphism is minor, but
arises from the same concerns:  our main interest is in the models of the
language, and it is simpler to describe models of the former than of the
latter.  To describe the recurrence language
as call-by-name is not quite right, because the verification of the
bounding
theorem that relates source programs to syntactic recurrences
does not require an operational semantics.  Instead it
suffices to axiomatize the recurrence language by a preorder, which we call
the size order.  The size order is defined in
Figure~\ref{fig:synrec-lang-preorder-full}, and a brief glimpse will show the
reader that the axioms primarily consist of directed versions of the standard
call-by-name equations.    This is the minimal set of axioms necessary to
verify the bounding theorem, but as we discuss more fully when we investigate
models of the recurrence language, there is more to the size order than that.
In a nutshell, a model in which the size order axiom for a given type
constructor is non-trivial (i.e., in which the two sides are not actually
equal) is a model that genuinely abstracts that particular type constructor to
a size.

We can think of the cost type in the recurrence language as the ``output'' of
the writer monad, and the recurrence extraction function that we give in
Section~\ref{sec:rec-extraction} as the call-by-value monadic translation of
the source language.  In some sense, then, the recurrence extracted from a
source program is just a cost-annotated version of the program.  However, we
think of the ``program'' part of the syntactic recurrence differently:  it
represents the \emph{size} of the source program.  Thus the syntactic
recurrence simultaneously describes both the cost and size of the original
program, what we refer to as a \emph{complexity}.  It is no surprise that we
must extract both simultaneously, if for no other reason than
compositionality, because if we are to describe cost in terms of size, then
the cost of $f(g(x))$ depends on both the cost and size of $g(x)$.  Thinking
of recurrence extraction as a call-by-value monadic translation gives us
insight into how to think of the size of a function:  it is a mapping from
sizes (of inputs) to complexities (of computing the result on an input of that
size).  This leads us to view size as a form of usage, or
\emph{potential} cost, and it is this last term that we adopt instead of
size.\footnote{We warn the reader that ``potential'' as we use it here
is not related to ``potential'' as it is used in
amortized analysis, though it does seem like potential associated to a data
value gives information about its use cost, so there may be a deeper
connection; we leave this question for future study.}  The bounding relation
$\bounded e E$
that we define in Section~\ref{sec:bounding} is a logical relation between
source programs~$e$ and syntactic recurrences~$E$.  A syntactic recurrence is
really a complexity, and $\bounded e E$ says that the operational cost of~$e$
is bounded by the cost component of~$E$ and that the value of~$e$ is bounded
by the potential component of~$E$.  The Bounding Theorem
(Theorem~\ref{thm:type-preservation}) tells us that every typeable program is
bounded by the recurrence extracted from it.  Its proof is somewhat long and
technical, but follows the usual pattern for verifying the Fundamental Theorem
for a logical relation, and the details are in
Appendix~\ref{app:syn-bounding-thm-proof}.

In the recurrence language, the ``data'' necessary to describe the size has as
much information as the original program; in the semantics we can abstract
away as much or as little of this information as necessary.  After defining
environment models of the recurrence language in Section~\ref{sec:models}
(following \citet{Bruce-et-al:Semantics}), in Section~\ref{sec:examples} we
give several examples to demonstrate that different size abstractions result
in semantic recurrences that formally justify typical extract-and-solve
analyses.  We stress that we are not attempting to analyze the cost of
heretofore unanalyzed programs.  
Our goal is a formal process that mirrors as closely as possible the informal
process we use at the board and on paper.
The main
examples demonstrate analyses where:

\begin{enumerate}
\item The size of a value~$v$ of inductive type~$\delta$ is defined in terms
of the
number of $\delta$-constructors in~$v$.  For example,
a list is measured by its length, a
tree by either its size or its height, etc., enabling typical size-based
recurrences (Section~\ref{sec:constr-counting-model}).
\item The size of a value~$v$ of inductive type~$\delta$ is (more-or-less) the
number of constructors of every inductive type in~$v$.  For example, 
the size of 
a $\snattree$~$t$ is the number of~$\snattree$ constructors in~$t$ (its usual
``size'') along with the maximum number of~$\cnat$ constructors in any node
of~$t$, enabling the analysis of functions with more complex costs, such as
the function that sums the nodes of a $\snattree$
(Section~\ref{sec:cac-model}).
\item A polymorphic function can be analyzed in terms of a notion of size that
is more abstract than that given by its instances
(Section~\ref{sec:merged-model}).  For example, while the
size of a $\snattree$ may be a pair $(k, n)$, where $k$ is the maximum key
value and~$n$ the size (e.g., to
permit analysis of the function that sums all the
nodes), we may want the domain of the 
recurrence extracted from a function of type
$\carr{\stree\alpha}{\rho}$ to be~$\N$,
corresponding to counting only the $\stree{}$ constructors.
\item We make use of the fact that the interpretation of the size order just
has to satisfy certain axioms to derive recurrences for \emph{lower bounds}.
As an example, we parlay this into a formal justification for the informal
argument that $\slmap(f\comp g)$ is more efficient 
than~$(\slmap\,f)\comp(\slmap\,g)$ (Section~\ref{sec:lower-bounds}).
\end{enumerate}

\noindent
These examples end up clarifying the role of the size order, as mentioned
earlier.  It is not just
the rules necessary to drive the proof of the syntactic bounding theorem, but
a non-trivial interpretation of $\szleq_\sigma$ (i.e., one in which 
$e\szleq_\sigma e'$ is valid but not $e = e'$) tells us that we have
a model with a non-trivial size abstraction for~$\sigma$.
This clarification highlights interesting analogies with abstract
interpretation:
(1)~when a
datatype~$\delta = \clfp t F$ 
is interpreted by a non-trivial size abstraction, there is an
abstract interpretation between $\den\delta{}$ (abstract) and
$\den{\sfsubst F\delta}{}$ (concrete); (2)~interpreting the recurrence
extracted from a polymorphic function in terms of a more abstract notion of
size is possible if there is an abstract interpretation between two models.

The remaining sections of the paper discuss recent work in cost analysis and
how our work relates to it, as well as limitations of and future directions
for our approach.

\section{The source language}
\label{sec:src-lang}

The source language that serves as the object language of our recurrence
extraction technique is a higher-order language with inductive
types, structural iteration (fold) over those types, and 
ML-style polymorphism (i.e., predicative polymorphism, with polymorphic
identifiers introduced only in let bindings), with an environment-based
operational semantics that approximates typical implementation.
This generalizes the source language of 
\citet{danner-et-al:icfp15}, which
introduced the technique for a 
monomorphically-typed language with a substitution-based semantics.
We address general recursion in Section~\ref{sec:recursion}.

The grammar and typing rules for expressions are given in
Figure~\ref{fig:src-lang-full}.  Type assignment derives (quantifier-free)
types for expressions given a \emph{type context} that assigns type schemes
(quantified types) to identifiers.  We write $\emptyctx$ for the empty type
context.  Values are not a subset of expressions because, as one would expect
in implementation, a function value consists of a function expression along
with a \emph{value environment}:  a binding of free variables to values.  The
same holds for values of suspension type, and we refer to any pair of an
expression and a value environment for its free variables as a \emph{closure}
(thus we use closure more freely than the usual parlance, in which it is
restricted to functions).  We adopt the notation common in the explicit
substitution literature (e.g., \citet{abadi-et-al:jfp91})
and write~$\cl v\theta$ for a closure with value
environment~$\theta$.   Since the typing for $\smapkw$ and $\smapvkw$
expressions depend on values, this requires a separate notion of typing for
values, which in turn depends on a notion of typing for closures.  These are
defined in Figures~\ref{fig:src-val-typing-full}
and~\ref{fig:src-typing-closures}.  
There is nothing deep in the typing of a closure value~$v\theta$ under
context~$\Gamma$.  Morally the rules
just formalize that~$v$ can be assigned the expected type without regard
to~$\theta$ and that $\theta(x)$ is of type~$\Gamma(x)$.
But since type contexts may assign type schemes,
whereas type assignment only derives types, the formal definition is that
$\theta(x)$ can be assigned any instance of~$\Gamma(x)$.

We will freely assume notation for $n$-ary sums and products and their
corresponding introduction and elimination forms, such as
$\sigma_0\stimes\sigma_1\stimes\sigma_2$,
$\stuple{e_0, e_1, e_2}$, and
$\sproj 1 e$
for
$\sprod{(\sprod{\sigma_0}{\sigma_1})}{\sigma_2}$,
$\spair{\spair{e_0}{e_1}}{e_2}$, and
$\sproj 1 {(\sproj 0 e)}$, respectively.
We write $\fv(e)$ for the free variables of~$e$ and
$\ftv(\tau)$ for the free type variables of~$\tau$.

\begin{figure}
\[
\begin{array}{rrll}
\rho,\sigma\in\Type & ::= & \alpha \mid \sunit \mid \sprod\sigma\sigma \mid \ssum\sigma\sigma \mid \sarr\sigma\sigma \mid \ssusp\sigma \mid \delta & \text{(types)} \\
\delta &::=& \slfp t F & \text{(inductive types)} \\
F &::=& t \mid \alpha \mid \sigma \mid \sprod F F \mid \ssum F F \mid
\sarr\sigma F &\text{(shape functors)} \\
\tau &::=& \sigma \mid \sforall \alpha \tau &\text{(type schemes)} \\
e &::=
  & x \mid \striv \mid \spair e e \mid \sproj i e \mid 
    \sinj i e \mid \scase e x {e_0} x {e_1} \\
 &\mid& \slam x e \mid \sapp e e \mid
    \sdelay e \mid \sforce e \\
 &\mid& \scons \delta e \mid \sdest\delta e \mid \sfold\delta e x e \\
 &\mid& \slet x e e  \\
 &\mid& \smap F y v e  \mid \smapv F y v v
    & \text{(expressions)}
\end{array}
\]
\[
\begin{array}{cc}
\ndAXC{$\typejudge{\sctx,x\oftype\sforall{\vec\alpha}\sigma}{x}{\bindin\sigma{\vec\alpha}{\vec\sigma}}$}
\DisplayProof
&
\ndAXC{$\typejudge\sctx\striv\sunit$}
\DisplayProof
\\[3ex]
  \AXC{$\typejudge\sctx{e_0}{\sigma_0}$}
  \AXC{$\typejudge\sctx{e_1}{\sigma_1}$}
\ndBIC{$\typejudge\sctx{\spair{e_0}{e_1}}{\sprod{\sigma_0}{\sigma_1}}$}
\DisplayProof 
&
  \AXC{$\typejudge\sctx e {\sprod{\sigma_0}{\sigma_1}}$}
\ndUIC{$\typejudge\sctx{\sproj i e}{\sigma_i}$}
\DisplayProof
\\[3ex]
  \AXC{$\typejudge\sctx e {\sigma_i}$}
\ndUIC{$\typejudge\sctx{\sinj i e}{\ssum{\sigma_0}{\sigma_1}}$}
\DisplayProof
&
  \AXC{$\typejudge\sctx e {\ssum{\sigma_0}{\sigma_1}}$}
  \AXC{$\setidx{\typejudge{\sctx,x\oftype\sigma_i}{e_i}{\sigma_i}}{i=0,1}$}
\ndBIC{$\typejudge\sctx{\scase e x {e_0} x {e_1}}{\sigma}$}
\DisplayProof
\\[3ex]
  \AXC{$\typejudge{\sctx,x\oftype\sigma'}{e}{\sigma}$}
\ndUIC{$\typejudge\sctx{\slam x e}{\sarr{\sigma'}{\sigma}}$}
\DisplayProof
&
  \AXC{$\typejudge\sctx e {\sarr{\sigma'}{\sigma}}$}
  \AXC{$\typejudge\sctx {e'} {\sigma'}$}
\ndBIC{$\typejudge\sctx{\sapp e {e'}}{\sigma}$}
\DisplayProof
\\[3ex]
  \AXC{$\typejudge\sctx e\sigma$}
\ndUIC{$\typejudge\sctx{\sdelay e}{\ssusp\sigma}$}
\DisplayProof
&
    \AXC{$\typejudge\sctx e{\ssusp\sigma}$}
\ndUIC{$\typejudge\sctx{\sforce e}{\sigma}$}
\DisplayProof
\\[3ex]
  \AXC{$\typejudge\sctx e {\sfsubst F \delta}$}
\RightLabel{($\delta=\clfp t F$)}
\ndUIC{$\typejudge\sctx{\scons\delta e}{\delta}$}
\DisplayProof
&
  \AXC{$\typejudge\sctx e {\delta}$}
\RightLabel{($\delta=\clfp t F$)}
\ndUIC{$\typejudge\sctx{\sdest\delta e}{\sfsubst F \delta}$}
\DisplayProof
\\[3ex]
\multicolumn{2}{c}{
      \AXC{$\typejudge\sctx {e'}{\delta}$}
      \AXC{$\typejudge{\sctx,x\oftype{\sfsubst{F}{\ssusp\sigma}}}{e}{\sigma}$}
    \RightLabel{($\delta=\slfp t F$)}
    \ndBIC{$\typejudge\sctx{\sfold\delta {e'} x {e}}\sigma$}
\DisplayProof
} 
\\[3ex]
\multicolumn{2}{c}{
  \AXC{$\typejudge\sctx {e'} {\sigma'}$}
  \AXC{$\typejudge{\sctx,x\oftype{\cforall{\vec\alpha}{\sigma'}}}{e}{\sigma}$}
  \AXC{$\vec\alpha$ not free in any~$\Gamma(y)$}
\ndTIC{$\typejudge\sctx{\slet x {e'} {e}}{\sigma}$}
\DisplayProof
}
\\[3ex]
  \AXC{$\typejudge{y\oftype\rho}{v}{\sigma}$}
  \AXC{$\typejudge\sctx{e}{\sfsubst F\rho}$}
\ndBIC{$\typejudge\sctx{\smap F y v e} {\sfsubst F\sigma}$}
\DisplayProof
&
  \AXC{$\typejudge{y\oftype\rho}{v'}{\sigma}$}
  \AXC{$\typejudgeE{v}{\sfsubst F\rho}$}
\ndBIC{$\typejudgeE{\smapv F y {v'} v} {\sfsubst F\sigma}$}
\DisplayProof
\end{array}
\]
\caption{A source language with let polymorphism and inductive
datatypes.  
$\smapkw$ and $\smapvkw$ expressions depend on values, which are
defined in Figure~\ref{fig:src-val-typing-full}.}
\label{fig:src-lang-full}
\end{figure}

\begin{figure}
\[
\begin{array}{rrl}
v &::=& y \mid \striv \mid \spair{v}{v} \mid {\sinj i {v}} \mid 
           \cl*{\slam x e}\theta \mid \cl*{\sdelay e}\theta \mid
           \scons\delta{(v)} \\
\theta &::=& \TmVar\tofinite \Val
\end{array}
\]
\[
\begin{array}{cc}
\ndAXC{$\typejudge{\sctx,{y\oftype\sigma}}{y}{\sigma}$}
\DisplayProof
&
\ndAXC{$\typejudge{\sctx}\striv\sunit$}
\DisplayProof
\\[3ex]
  \AXC{$\setidx{\typejudge{\sctx}{v_i}{\sigma_i}}{i=0,1}$}
\ndUIC{$\typejudge{\sctx}{\spair{v_0}{v_1}}{\sprod{\sigma_0}{\sigma_1}}$}
\DisplayProof
&
  \AXC{$\typejudge{\sctx}{v}{\sigma_i}$}
\ndUIC{$\typejudge{\sctx}{\sinj* i {v}}{\ssum{\sigma_0}{\sigma_1}}$}
\DisplayProof
\\[3ex]
  \AXC{$\closurejudge{\sctx}{\cl*{\slam x e}\theta}{(\sarr{\sigma'}{\sigma})}$}
\ndUIC{$\typejudge{\sctx}{\cl*{\slam x e}\theta}{\sarr{\sigma'}{\sigma}}$}
\DisplayProof
&
  \AXC{$\closurejudge{\sctx}{\cl*{\sdelay e}\theta}{(\ssusp\sigma)}$}
\ndUIC{$\typejudge{\sctx}{\cl*{\sdelay e}\theta}{\ssusp\sigma}$}
\DisplayProof
\\[3ex]
\multicolumn{2}{c}{
  \AXC{$\typejudge{\sctx}{v}{\sfsubst F\delta}$}
\RightLabel{($\delta=\clfp t F$)}
\ndUIC{$\typejudge{\sctx}{\scons\delta{v}}{\delta}$}
\DisplayProof
}
\end{array}
\]
\caption{Grammar and typing rules for values.
For any value environment~$\theta$, it must be that for all~$x$
there is~$\sigma$ such that
$\typejudgeE{\theta(x)}{\sigma}$.
The judgment
$\closurejudge\sctx{\cl e\theta}{\sigma}$ is defined
in Figure~\ref{fig:src-typing-closures}.
}
\label{fig:src-val-typing-full}
\end{figure}

\begin{figure}
\[
\begin{array}{c}
  \AXC{For all~$x\in\dom\theta$:
  \text{ if } $\sctx(x)=\sforall{\vec\alpha}\rho \text { then } \forall\vec\sigma,
  \typejudgeE{\theta(x)}{\substin\rho{\vec\sigma}{\vec\alpha}}$}
\ndUIC{$\theta$ is a $\sctx$-environment}
\DisplayProof
\\[3ex]
  \AXC{$\typejudge{\sctx,\sctx'}{e}{\sigma}$}
  \AXC{$\theta$ is a $\sctx'$-environment}
\ndBIC{$\closurejudge\sctx{\cl e\theta}{\sigma}$}
\DisplayProof
\end{array}
\]
\caption{Typing for closures.}
\label{fig:src-typing-closures}
\end{figure}

\begin{figure}
\begin{gather*}
\begin{aligned}
\sbool &= \ssum\sunit\sunit \\
\sfalse &= \sinj[\sbool]{0}{\striv} \\
\strue &= \sinj[\sbool]{1}{\striv}
\end{aligned}
\qquad
\begin{aligned}
\snat &= \slfp t {\ssum\sunit t} \\
\snzero &= \scons\snat(\sinj 0 \striv) \\
\snsucc\,x &= \scons\snat(\sinj 1 x) \\
\snn n&= \snsucc(\dots\snsucc\snzero)~\text{($n$ $\snsucc$'s)}
\end{aligned}
\\[\baselineskip]
\begin{aligned}
\slist\sigma &= \slfp t {\ssum\sunit{\sprod\sigma t}} \\
\slnil_\sigma &= \scons{\slist\sigma}(\sinj 0 \striv) \\
\slcons_\sigma(x, xs) &= \scons{\slist\sigma}(\sinj 1 ({\spair x {xs}}))
\end{aligned}
\qquad
\begin{aligned}
\stree\sigma &= \slfp t {\ssum\sunit{\sprod\sigma {\sprod t t}}} \\
\stemp_\sigma &= \scons{\stree\sigma}(\sinj 0 \striv) \\
\stnode_\sigma(x, t_0, t_1) 
  &= \scons{\stree\sigma}(\sinj 1 {\stuple{x, t_0, t_1}})
\end{aligned}
\end{gather*}
\caption{Some standard types in the source language.}
\label{fig:src-lang-examples}
\end{figure}

Inductive types are defined by shape functors, ranged over by the
metavariable~$F$; a generic inductive type has the form
$\slfp t F$.  
If $F$ is a shape functor and $\sigma$ a type, then
$F[\sigma]$ is the result of substituting $\sigma$ for free occurrences
of~$t$ in~$F$ (the $\mu$ operator binds~$t$, of course).
Formally a shape functor is just a type, and so when
certain concepts are defined by induction on type, they are automatically
defined for shape functors as well.
In the syntax for shape functors, $t$ is a fixed type variable, and
hence simultaneous nested definitions are not allowed.  That is, types
such as $\slfp t {\csum{\cunit}{\slfp* s {\csum\cunit{\cprod s t}}}}$
are forbidden.  However, an inductive type can be used inside of other
types via the constant functor $(\sigma)$ production of $F$, e.g.\ coding
the type $\slist{(\slist \alpha)}$ as $\slfp t {\csum \cunit {\cprod {(
  \slfp t {\csum \cunit {\cprod \alpha t}})} t}}$.  This restriction is
just to simplify the presentation of the languages and models, and
lifting it does not require fundamental changes.
Figure~\ref{fig:src-lang-examples} gives a number of types and values
that we will use in examples.  We warn the reader that because most
models of the recurrence language have non-standard interpretations of
inductive types, the types $\sigma$ and $\slfp t \sigma$ may be treated
very differently even when $t$ is not free in~$\sigma$.  Thus it can
actually make a real difference whether we define $\sbool$ to be
$\ssum\sunit\sunit$ or $\clfp t {\ssum\sunit\sunit}$ (if every
type that would be defined by an ML 
$\mathtt{datatype}$ declaration were implemented as a possibly degenerate inductive type).

For every inductive type~$\delta=\slfp t F$ there is an associated
constructor~$\sconskw_\delta$, destructor~$\sdestkw_\delta$, and
iterator~$\sfoldkw_\delta$.  Thought of informally as term constants, the
first two have the typical types~${\carr{F[\delta]}{\delta}}$
and~${\carr\delta{F[\delta]}}$, but the type of $\sfoldkw_\delta$ is somewhat
non-standard: ${\carr{\carr\delta{(\carr{F[\ssusp\sigma]}{\sigma})}}{\sigma}}$.
We use suspension types of the form $\ssusp\sigma$ primarily to delay
computation of recursive calls in evaluating $\sfoldkw$ expressions.  This
is not necessary for any theoretical concerns, but rather practical:
without something like this, implementations of standard programs
would have unexpected costs.  We will return to this when we
discuss the operational semantics.  We also observe that in this informal
treatment, the types are not polymorphic; in our setting, polymorphism and
inductive types are orthogonal concerns.

The $\smapkw_F$ and $\smapvkw_F$ constructors are used to define the
operational semantics of $\sfoldkw_{\slfp t F}$.  The latter 
witnesses functoriality of shape functors.
Informally speaking, evaluation of~$\smapv F y {v'} v$ 
traverses a value~$v$ of
type~$F[\delta]$, applying a function~$y\mapsto v'$ 
of type~$\carr\delta\sigma$ to each
inductive subvalue of type~$\delta$ to obtain a value of
type~$F[\sigma]$.
$\smapkw_F$ is a technical tool for defining this action when
$F$ is an arrow shape, in which case the value of type~$F[\delta]$ is
really a delayed computation and hence is represented by an arbitrary
expression.
Because the definition of
$\smapkw_F$ and $\smapvkw_F$ depend on values, we must define them
(and their typing) simultaneously with terms.  
Furthermore, evaluation of $\smapvkw_{\carr\rho F}$
results in a function closure value that contains a $\smapkw_F$ expression,
and the function closure itself is, as usual, an ordinary
$\lambda$-expression.  This is also the reason that $\smapkw$ and~$\smapvkw$
are part of the language, rather than just part of the metalanguage used to
define the operational semantics.  They are not intended to be used in program
definitions though, so we make the following definition:

\begin{defn}[Core language]
\label{defn:core-lang}
The \emph{core language} consists of the terms of the source language that are
typeable not using $\smapkw$ or $\smapvkw$.
\end{defn}

\begin{figure}
\[
\begin{array}{cc}
\multicolumn{2}{c}{
  \ndAXC{$\evalin{\cl x\theta}{\theta(x)} 0$}
  \DisplayProof
}
\\
\multicolumn{2}{c}{
  \ndAXC{$\evalin{\cl\striv\theta}\striv 0$}
  \DisplayProof
}
\\[3ex]
  \AXC{$\evalin{\cl{e_0}\theta}{v_0}{n_0}$}
  \AXC{$\evalin{\cl{e_1}\theta}{v_1}{n_1}$}
\ndBIC{$\evalin{\cl{\spair{e_0}{e_1}}\theta}{\spair{v_0}{v_1}}{n_0+n_1}$}
\DisplayProof
&
  \AXC{$\evalin {\cl e\theta} {\spair{v_0}{v_1}} n$}
\ndUIC{$\evalin{\cl*{\sproj i e}\theta}{v_i}{n}$}
\DisplayProof
\\[3ex]
  \AXC{$\evalin {\cl e\theta} v n$}
\ndUIC{$\evalin{\cl*{\sinj i e}\theta}{\sinj i v} n$}
\DisplayProof
&
  \AXC{$\evalin {\cl e\theta} {\sinj i {v_i}} {n}$}
  \AXC{$\evalin {\cl {e_i} {\bindin\theta  x{v_i}}} {v} {n_i}$}
\ndBIC{$\evalin {\cl*{\scase e x {e_0} x {e_1}}\theta} {v} {n+n_i}$}
\DisplayProof
\\[3ex]
\ndAXC{$\evalin{\cl*{\slam x e}\theta}{\cl*{\slam x e}\theta} 0$}
\DisplayProof
&
  \AXC{$\evalin {\cl{e_0}\theta} {\cl*{\slam x {e_0'}}{\theta'}} {n_0}$}
  \AXC{$\evalin {\cl{e_1}\theta} {v_1} {n_1}$}
  \AXC{$\evalin {\cl {e_0'} {\bindin{\theta'} {x} {v_1}}} v {n}$}
\ndTIC{$\evalin {\cl*{\sapp{e_0}{e_1}}\theta} v {n_0+n_1+n}$}
\DisplayProof
\\[3ex]
\ndAXC{$\evalin{\cl*{\sdelay e}\theta}{\cl*{\sdelay e}\theta}{0}$}
\DisplayProof
&
    \AXC{$\evalin {\cl e\theta} {\cl*{\sdelay e'}{\theta'}} n$}
    \AXC{$\evalin {\cl{e'}{\theta'}} v {n'}$}
\ndBIC{$\evalin{\cl*{\sforce e}\theta} v {n+n'}$}
\DisplayProof
\\[3ex]
  \AXC{$\evalin {\cl e \theta} v n$}
\ndUIC{$\evalin{\cl*{\scons\delta e}\theta}{\scons\delta v} n$}
\DisplayProof
&
  \AXC{$\evalin {\cl e \theta} {\scons\delta v} n$}
\ndUIC{$\evalin {\cl*{\sdest\delta e}\theta} v n$}
\DisplayProof
\\[3ex]
\multicolumn{2}{c}{
  \AXC{$\evalin {\cl{e'}\theta} {\scons\delta{v'}} {n'}$}
  \AXC{$\eval
    {\smapv F y {\cl*{\sdelay*{\sfold\delta y x e}}{\theta}} {v'}} 
    {v''}$}
  \AXC{$\evalin {\cl e {\bindin\theta x {v''}}} {v} n$}
\ndTIC{$\evalin {\cl*{\sfold\delta{e'} x e}{\theta}} v {n' + n + 1}$}
\DisplayProof
}
\\[3ex]
\multicolumn{2}{c}{
  \AXC{$\evalin {\cl{e'}\theta} {v'} {n'}$}
  \AXC{$\evalin {\cl e {\bindin\theta x{v'}}} v n$}
\ndBIC{$\evalin {\cl*{\slet x {e'} e}\theta} v {n'+n}$}
\DisplayProof
}
\end{array}
\]
\caption{The operational cost semantics for the source language.  
We only define evaluation for closures~$\cl e\theta$ such that
$\typejudgeE{\cl e\theta}{\sigma}$ for some~$\sigma$, and hence
just write~$\cl e\theta$.
The semantics
for $\sfoldkw$ depends on the semantics for $\smapkw$, which is given
in Figure~\ref{fig:src-map-semantics}.}
\label{fig:src-lang-semantics}
\end{figure}

\begin{figure}
\[
\begin{array}{cc}
\multicolumn{2}{c}{
  \AXC{$\evalin{\cl e\theta}{v''}{n}$}
  \AXC{$\eval{\smapv F y {v'} {v''}}{v}$}
\ndBIC{$\evalin{\cl*{\smap F y {v'} e}\theta}{v}{n}$}
\DisplayProof
}
\\
\ndAXC{$\eval {\smapv t y {v'} {v}} {\esubst {v'} {\substfor {v} y}}$}
\DisplayProof
&
\ndAXC{$\eval {\smapv \sigma y {v'} {v}} {v}$}
\DisplayProof
\\[3ex]
  \AXC{$\setidx{\eval {\smapv {F_i} y {v'} {v_i'}} {v_i}}{i=0,1}$}
\ndUIC{$\eval {\smapv {\sprod{F_0}{F_1}} y {v'} {\spair {v_0'}{v_1'}}} {\spair{v_0}{v_1}}$}
\DisplayProof
&
  \AXC{$\eval {\smapv {F_i} y {v'} {v}} {v_i}$}
\ndUIC{$\eval {\smapv {\ssum{F_0}{F_1}} y {v'} {\sinj i {v}}} {\sinj i {v_i}}$}
\DisplayProof
\\[3ex]
\multicolumn{2}{c}{
\ndAXC{$\eval {\smapv {\sarr\rho F} y {v'} {\cl*{\slam x e}\theta}}
              {\cl*{\slam x {\smap F y {v'} e}}\theta}$}
\DisplayProof
}
\end{array}
\]
\caption{The operational semantics for the source language $\smapkw$ and
$\smapvkw$ constructors.  Substitution of values is defined
in Figure~\ref{fig:src-val-subst}.}
\label{fig:src-map-semantics}
\end{figure}

\begin{figure}
\[
\begin{aligned}[t]
\substin y v y &= v \\
\substin\striv v y &= \striv \\
\substin {\spair{v_0}{v_1}} v y &= \spair{\substin{v_0} v y}{\substin{v_1} v y} \\
\substin{(\sinj i {v'})} v y &= \sinj* i {\substin{v'} v y}
\end{aligned}
\qquad
\begin{aligned}[t]
\substin{(\cl*{\slam x e}\theta)} v y &= \cl*{\slam x e}{(\bindin\theta y v)} \\
\substin{(\cl*{\sdelay e}\theta)} v y &= \cl*{\sdelay e}{(\bindin\theta y v)} \\
\substin{(\scons\delta{v'})} v y &= \scons\delta{(\substin{v'} v y)}
\end{aligned}
\]
\caption{Substitution of values for identifiers in values,
$\substin{v'}{v}{y}$.}
\label{fig:src-val-subst}
\end{figure}

The operational cost semantics for the language is defined in
Figure~\ref{fig:src-lang-semantics} and its dependencies, 
which define a relation
$\evalin {\cl e \theta} v n$, where $e$ is a (well-typed) expression,
$\theta$ a value environment, $v$ a value, and $n$ a non-negative integer.
As with closure values,
we write a closure with expression~$e$ and value environment~$\theta$
as~$\cl e\theta$, and opt for this notation for
compactness (a more typically presentation might
be $\theta\proves\evalin e v n$).
The intended meaning is that under value environment~$\theta$, the term~$e$
evaluates to the value~$v$ with cost~$n$.
A value environment that needs to be spelled out will be written
$\lenv
 \bindto{x_0}{v_0},\dots,\bindto{x_{n-1}}{v_{n-1}}
 \renv$ 
or more commonly~$\lenv\bindto{\vec x}{\vec v}\renv$.
We write $\theta\{y \mapsto v\}$ for extending a value environment
$\theta$ by binding the (possibly fresh) variable $y$ to $v$.  
Value environments are part of the language, so
when we write $\cl e\theta$, the bindings are
not immediately applied.
However, we use a substitution notation~$\substin{}{v}{y}$ for
defining the semantics of $\smapvkw_t$ because this
is defined in the metalanguage as a metaoperation.
Using explicit environments,
rather than a metalanguage notation for substitutions, adds a certain
amount of syntactic complexity.  The payoff is a semantics that more closely
reflects typical implementation.

Our approach to charging some amount of cost for each step of the evaluation,
where that amount may depend on the main term former, is standard.  
Recurrence extraction is parametric in these choices.
We observe
that our environment-based semantics permits us to charge even for looking up
the value of an identifier, something that is difficult to codify in a
substitution-based semantics.
Our particular choice to charge one unit of cost for each unfolding of 
a~$\sfoldkw$, and no cost for any other form, is admittedly ad-hoc, but
gives expected costs with a minimum of bookkeeping fuss, especially when it
comes to the semantic interpretations of the recurrences.
Another common alternative is to define a tick operation
$\stickkw : \sarr\alpha\alpha$ which charges a unit of cost, as done by
\citet{danielsson:popl08} and others.  This requires the user to annotate the
code at the points for which cost should be charged, which increases the load
on the programmer, but allows her to be specific about exactly what to count
(e.g., only comparisons).  It is straightforward to adapt our approach to that
setting.

\begin{figure}
\begin{minipage}{\textwidth}
\begin{verbatim}
fun member(t : bst, x : int) : bool =
    case t of
         E => false
       | N(y, t0, t1) => if x = y then true
                         else if x < y then member(t0, x)
                         else member(t1, x)
\end{verbatim}
\subcaption{Binary search tree membership function in ML.}
\label{fig:susp-fold-ml}
\end{minipage}
\\
\begin{minipage}{\textwidth}
\begin{verbatim}
fun member(t : bst, x : int) : bool =
    case t of
         E => false
       | N(y, t0, t1) => if x = y then true
                         else if x < y then member(t0, x)
                         else member(t1, x)
\end{verbatim}
\subcaption{Binary search tree membership without suspensions.}
\label{fig:susp-fold-no-susp}
\end{minipage}
\\
\begin{minipage}{\textwidth}
\begin{verbatim}
fun member(t, x) = 
    treefold (fn (y, r0, r1) => if x = y then true
                                else if x < y then force r0
                                else force r1) t
\end{verbatim}
\subcaption{Binary search tree membership with suspensions.}
\label{fig:susp-fold-susp}
\end{minipage}
\caption{Using suspension types to control evaluation of recursive calls
in fold-like constructs.}
\label{fig:susp-fold}
\end{figure}

The reason for suspending the
recursive call in the semantics of~$\sfoldkw$
is to ensure that typical recursively-defined functions
that do not always evaluate all recursive calls still
have the expected cost.  For example, consider membership testing in a
binary search tree.  Typical ML code for such a function might look
something like the code in
Figure~\ref{fig:susp-fold-ml}.
This function is linear in the height of~$t$, because
the lazy evaluation of conditionals ensures that
at most one recursive call is evaluated.  If we were to implement |member|
with a
|fold| operator for trees that does not suspend the recursive call
(so the step function would have 
type~$\sarr{\sprod{\sprod\sint\sbool}\sbool}{\sbool}$), as
in Figure~\ref{fig:susp-fold-no-susp},
then the recursive calls~|r0| and~|r1| are evaluated at each step, leading
to a cost that is linear in the size of~$t$, rather than the height.  Our
solution is to ensure that the recursive calls are delayed, and only
evaluated when the corresponding branch of the conditional is
evaluated, so the step function has type
$\sarr{\sprod{\sprod\sint{\ssusp\sbool}}{\ssusp\sbool}}{\sbool}$, in which
case the code looks something like that of
Figure~\ref{fig:susp-fold-susp}.  This issue does not come up when
recursive definitions are allowed only at function type, as is typical in
call-by-value languages.
However, in order to simplify the construction of models, here we restrict 
recursive definitions to the use of $\sfoldkw_\delta$ (i.e. to structural
recursion only, rather than general recursion), which must be permitted
to have any result type.

Given the complexity of the language, it behooves us to verify type
preservation.  For this cost
is irrelevant, so we write $\eval{\cl e\theta}{v}$ to mean
$\evalin{\cl e\theta}{v}{n}$ for some~$n$.  Remember that we 
our notion of closure includes expressions of any type, so this theorem does
not just state type preservation for functions.


\begin{thm}[Type preservation]
\label{thm:type-preservation}
If $\typejudgeE {\cl e\theta}{\sigma}$ and $\eval {\cl e\theta}{v}$,
then $\typejudgeV v \sigma$.
\end{thm}
\begin{proof}
See Appendix~\ref{app:src-lang-type-preservation}.
\end{proof}

\section{The recurrence language}
\label{sec:rec-lang}

The recurrence language is defined in Figure~\ref{fig:rec-lang-full}.  It is a
standard system of predicative polymorphism with explicit type abstraction and
application.  Most of the time we will elide type annotations from variable
bindings, mentioning them only when demanded for clarity.
The types and terms corresponding to those of
Figure~\ref{fig:src-lang-examples} are given in
Figure~\ref{fig:cpy-lang-examples}.
This is the language into which we will extract syntactic
recurrences from the source language.  A syntactic recurrence is more-or-less
a cost-annotated version of a source language program.  As we are interested
in the value (denotational semantics) of the recurrences and not in
operational considerations, we think of the recurrence language in a more
call-by-name way (although, as we will see, the main mode of reasoning is
with respect to an ordering, rather than equality).  

\begin{figure}
\[
\begin{array}{rrll}
\rho,\sigma & ::= & 
  \alpha \mid \C \mid \cunit 
  \mid \cprod\sigma\sigma \mid \csum\sigma\sigma \mid \carr\sigma\sigma 
  \mid \delta & \text{(types)} \\
\delta &::=& \clfp t F & \text{(inductive types)} \\
F &::=& t \mid \sigma 
  \mid \cprod F F \mid \csum F F \mid \carr \sigma F &\text{(shape functors)} \\
\tau &::= & \sigma \mid \cforall\alpha\tau & \text{(type schemes)} \\
e &::=& x \mid 0 \mid 1 \mid e + e \mid \ctriv \mid \cpair e e \mid \cproj i e
\\
     &    &\mid \cinj i e \mid \ccaset e x {\sigma_0} {e_0} x {\sigma_1} {e_1} \\
  &\mid& \clam* x \sigma e \mid \capp e e \mid \ctylam \alpha e \mid \ctyapp e \sigma \\
  &\mid& \ccons \delta {} \mid \cdest\delta {} \mid 
        \cfoldkw_\delta
      & \text{(expressions)}
\end{array}
\]
\[
\begin{array}{cc}
\multicolumn{2}{c}{
\ndAXC{$\typejudge\cctx 0 \C$}
\DisplayProof
\quad
\ndAXC{$\typejudge\cctx 1 \C$}
\DisplayProof
\quad
  \AXC{$\typejudge\cctx{e_0}{\C}$}
  \AXC{$\typejudge\cctx{e_1}{\C}$}
\ndBIC{$\typejudge\cctx{e_0+e_1}{\C}$}
\DisplayProof
}
\\
\ndAXC{$\typejudge{\cctx,x\oftype\tau}{x}{\tau}$}
\DisplayProof
&
\ndAXC{$\typejudge\cctx\ctriv\cunit$}
\DisplayProof
\\[3ex]
  \AXC{$\typejudge\cctx{e_0}{\sigma_0}$}
  \AXC{$\typejudge\cctx{e_1}{\sigma_1}$}
\ndBIC{$\typejudge\cctx{\cpair{e_0}{e_1}}{\cprod{\sigma_0}{\sigma_1}}$}
\DisplayProof 
&
  \AXC{$\typejudge\cctx e {\cprod{\sigma_0}{\sigma_1}}$}
\ndUIC{$\typejudge\cctx{\cproj i e}{\sigma_i}$}
\DisplayProof
\\[3ex]
  \AXC{$\typejudge\cctx e {\sigma_i}$}
\ndUIC{$\typejudge\cctx{\cinj i e}{\csum{\sigma_0}{\sigma_1}}$}
\DisplayProof
&
  \AXC{$\typejudge\cctx e {\csum{\sigma_0}{\sigma_1}}$}
  \AXC{$\setidx{\typejudge{\cctx,x\oftype\sigma_i}{e_i}{\sigma_i}}{i=0,1}$}
\ndBIC{$\typejudge\cctx{\ccaset* e x {\sigma_i} {e_i}}{\sigma}$}
\DisplayProof
\\[3ex]
  \AXC{$\typejudge{\cctx,x\oftype\sigma}{e}{\sigma'}$}
\ndUIC{$\typejudge\cctx{\clam* x \sigma e}{\carr\sigma{\sigma'}}$}
\DisplayProof
&
  \AXC{$\typejudge\cctx e {\carr{\sigma'}{\sigma}}$}
  \AXC{$\typejudge\cctx {e'} {\sigma'}$}
\ndBIC{$\typejudge\cctx{\capp e {e'}}{\sigma}$}
\DisplayProof
\\[3ex]
  \AXC{$\typejudge\cctx e \tau$}
\RightLabel{$\alpha\notin\ftv(\cctx)$}
\ndUIC{$\typejudge\cctx{\ctylam \alpha e}{\cforall \alpha \tau}$}
\DisplayProof
&
  \AXC{$\typejudge\cctx e {\cforall \alpha \tau}$}
\RightLabel{($\sigma$ quantifier-free)}
\ndUIC{$\typejudge\cctx{\ctyapp e\sigma }{\substin\tau\sigma \alpha}$}
\DisplayProof
\\[3ex]
  \AXC{$\typejudge\cctx e {\sfsubst F \delta}$}
\RightLabel{($\delta = \clfp t F$)}
\ndUIC{$\typejudge\cctx{\ccons{\delta}e}
                       {{\delta}}$}
\DisplayProof
&
  \AXC{$\typejudge\cctx e \delta$}
\RightLabel{($\delta = \clfp t F$)}
\ndUIC{$\typejudge\cctx{\cdest{\delta} e}
                       {{\sfsubst{F}{\delta}}}$}
\DisplayProof
\\[3ex]
\multicolumn{2}{c}{
  \AXC{$\typejudge\cctx{e'}{\delta}$}
  \AXC{$\typejudge{\cctx,x\oftype\sfsubst F \sigma}
                  {e}
                  {\sigma}$}
\RightLabel{($\delta = \clfp t F$)}
\ndBIC{$\typejudge\cctx{\cfold{\delta}{e'} {(x : \sfsubst F \sigma)} e }
                       {\sigma}$}
\DisplayProof
}
\end{array}
\]
\caption{The recurrence language grammar and typing.}
\label{fig:rec-lang-full}
\end{figure}

\begin{figure}
\begin{gather*}
\begin{aligned}[t]
\cbool &= \csum\cunit\cunit \\
\cfalse &= \cinj[\cbool]{0}{\ctriv} \\
\ctrue &= \cinj[\cbool]{1}{\ctriv}
\end{aligned}
\qquad
\begin{aligned}[t]
\cnat &= \clfp t {\csum\cunit t} \\
\cnzero &= \ccons\cnat(\cinj 0 \ctriv) \\
\cnsucc\,x &= \ccons\cnat(\cinj 1 x)
\end{aligned}
\\[\baselineskip]
\begin{aligned}[t]
\clist\sigma &= \clfp t {\csum\cunit{\cprod\sigma t}} \\
\clnil_\sigma &= \ctylam{\vec\alpha}
                        {\ccons{\clist\sigma}
                               {\vec\alpha}(\cinj 0 \ctriv)} \\
\clcons_\sigma(x, xs) 
  &= \ctylam{\vec\alpha}
            {\ccons{\clist\sigma}{\vec\alpha}(\cinj 1 ({\cpair x {xs}}))}
\end{aligned}
\qquad
\begin{aligned}[t]
\ctree\sigma &= \clfp t {\csum\cunit{\cprod\sigma {\cprod t t}}} \\
\ctemp_\sigma 
  &= \ctylam{\vec\alpha}{\ccons{\ctree\sigma}{\vec\alpha}(\cinj 0 \ctriv)} \\
\ctnode_\sigma(x, t_0, t_1) 
  &= \ctylam{\vec\alpha}
            {\ccons{\ctree\sigma}{\vec\alpha}(\cinj 1 {\ctuple{x, t_0, t_1}})}
\end{aligned}
\end{gather*}
\caption{Some standard types in the recurrence language corresponding
to those in Figure~\ref{fig:src-lang-examples}.}
\label{fig:cpy-lang-examples}
\end{figure}

\begin{figure}
\[
\begin{array}{ccc}
\multicolumn{3}{c}{
\ccongctx ::= \cconghole 
          \mid \ccongctx + e \mid e + \ccongctx
          \mid \cproj i\ccongctx 
          \mid \ccaset* \ccongctx x {\sigma_i} {e_i}
          \mid \capp\ccongctx e 
          \mid \cdest\delta\ccongctx 
          \mid \cfold\delta \ccongctx x e
}
\\[1ex]
\RightLabel{\infrulelbl{refl}}
\ndAXC{$\prejudge \cctx e e \tau$}
\DisplayProof
&&
  \AXC{$\prejudge\cctx{e_0 }{ e_1}{\tau}$}
  \AXC{$\prejudge\cctx{e_1 }{ e_2}{\tau}$}
\RightLabel{\infrulelbl{trans}}
\ndBIC{$\prejudge\cctx{e_0 }{ e_2}{\tau}$}
\DisplayProof
\\[3ex]
\multicolumn{3}{c}{
  \AXC{$\typejudge{\cctx,x\oftype\tau'}{\ccongctx[x]}{\tau}$}
  \AXC{$\prejudge\cctx{e_0}{ e_1}{\tau}$}
\RightLabel{\infrulelbl{mon}}
\ndBIC{$\prejudge\cctx{\ccongctx[e_0] }{\ccongctx[e_1]}{\tau}$}
\DisplayProof
}
\\[1ex]
\RightLabel{\infrulelbl{idl}}
\ndAXC{$0 + e =_\C e$}
\DisplayProof
&
\RightLabel{\infrulelbl{idr}}
\ndAXC{$e + 0 =_\C e$}
\DisplayProof
&
\RightLabel{\infrulelbl{assoc}}
\ndAXC{$e_0 + (e_1 + e_2) =_\C (e_0 + e_1) + e_2$}
\DisplayProof
\\[1ex]
\multicolumn{3}{c}{
    \RightLabel{\infrulelbl{beta-times}}
    \ndAXC{$\prejudge\cctx{ e_i}{\cproj* i {e_0,e_1} }{\sigma}$}
    \DisplayProof
} 
\\[1ex]
\multicolumn{3}{c}{
    \RightLabel{\infrulelbl{beta-plus}}
    \ndAXC{$\prejudge\cctx{ e_i}{\ccaset* {\cinj i e} x {\sigma_i} {e_i}}{\sigma}$}
    \DisplayProof
} 
\\[1ex]
\multicolumn{3}{c}{
    \RightLabel{\infrulelbl{beta-to}}
    \ndAXC{$\prejudge\cctx{ \substin e {e'} x}{\capp* {\clam* x {\sigma'} e} {e'} }{\sigma}$}
    \DisplayProof
} 
\\[1ex]
\multicolumn{3}{c}{
\RightLabel{\infrulelbl{beta-delta}}
\ndAXC{$\prejudge\cctx{ e}
                      {\cdest\delta{(\ccons\delta  e)} }
                      {{\sfsubst{\substin F {\vec\sigma}{\vec\alpha}}{\delta}}}$}
    \DisplayProof
} 
\\[1ex]
\multicolumn{3}{c}{
\RightLabel{\infrulelbl{beta-fold}}
\ndAXC{$\prejudge\cctx
                 {
                   \substin e
                            {(\cmap{F}
                                   {\delta}
                                   {y}
                                   {\cfold\delta y x {e}}
                                   {e'})}
                            {x}
                 }
                 {
                   \cfold\delta {(\ccons\delta{e'})} x e
                 }
                 {\sigma}$}
    \DisplayProof
} 
\\[3ex]
\multicolumn{3}{c}{
  \AXC{$\typejudge\cctx{\ctylam\alpha e}{\cforall\alpha\tau}$}
\RightLabel{\infrulelbl{beta-all}}
\ndUIC{$\prejudge\cctx{e}{\ctyapp*{\ctylam\alpha e}{\sigma}}{\substin\tau\sigma\alpha}$}
\DisplayProof
} 
\end{array}
\]
\caption{The size order relation that defines the semantics of the recurrence
language.  The macro $\cmap F\rho y {e'} e$ is defined in
Figure~\ref{fig:rec-lang-map-macro}.}
\label{fig:synrec-lang-preorder-full}
\end{figure}

\begin{figure}
\begin{align*}
\cmap t \rho r {e'} e &= \subst{e'}{e}{x} \\
\cmap {\tau_0} \rho y {e'} e &= e \\
\cmap*{\csum{F_0}{F_1}} \rho y {e'} e
  &= \ccaset* e x {F_i[\rho]} {\cinj* i {\cmap {F_i} \rho y {e'} x}} \\
\cmap*{\cprod{F_0}{F_1}} \rho y {e'} e
  &= \cpair{\cmap{F_0} \rho y {e'} {\cproj 0 e}}
           {\cmap{F_1} \rho y {e'} {\cproj 1 e}} \\
\cmap*{\carr{\sigma_0} F}\rho  y {e'} e
  &= \clam{(x\oftype\sigma_0)}{\cmap F \rho y {e'} {\capp e x}} \\
\end{align*}
\caption{The macro $F[\rho; y.e', e]$.}
\label{fig:rec-lang-map-macro}
\end{figure}

\subsection{The cost type}
The recurrence language has a \emph{cost type}~$\C$.  As we discuss in 
Section~\ref{sec:rec-extraction}, we can think of recurrence extraction as a
monadic translation into the writer monad, where the ``writing'' action is to
increment the cost component.  Thus it suffices to ensure
that $\C$ is a monoid, though in our examples of models, it is usually
interpreted as a set with more structure (e.g., the natural numbers adjoined
with an ``infinite'' element).

\subsection{The ``missing'' pieces from the source language}
There are no suspension types, nor term constructors corresponding to $\sletkw$,
$\smapkw$, or $\smapvkw$.  We are primarily interested in the denotations of
expressions in the recurrence language, not in carefully accounting for the
cost of evaluating them.  Of course, it is convenient to have the standard
syntactic sugar $\cletmulti{x_0=e_0,\dots,x_{n-1}=e_{n-1}}{e}$ for
$\substin{e}{e_0,\dots,e_{n-1}}{x_0,\dots,x_{n-1}}$.  Because of the way in
which the size order is axiomatized, this must be defined as a substitution,
not as a $\beta$-expansion.  Likewise, we still need a construct that
witnesses functoriality of shape functors, but it suffices to do so with a
metalanguage macro $\cmap F \rho x {e'} e$ that is defined in
Figure~\ref{fig:rec-lang-map-macro}.

\subsection{Datatype constructor, destructor, and fold}

The constructor, destructor, and fold terms are similar to those in the source
language, though here we use the more typical type for $\cfoldkw_\delta$.
Since type abstraction and application are explicit in the recurrence
language, it may feel a bit awkward that $\beta$-reduction for types seems to
change these constants; for example,
$\ctyapp{(\ctylam\alpha{\dotsb\cfold{\clist\alpha}{e'}x{e}\dotsb})}{\sigma}$
would convert to~${\dotsb\cfold{\clist\sigma}{e'}x{e}\dotsb}$.  The right way
to think of this is that there is really a single constant $\cfoldkw$ that
maps inductive types to the corresponding operator---in effect, we write
$\cfoldkw_\delta$ for $\cfoldkw\,\delta$, and so the substitution of a type
for a type variable does not change the constant, but rather the argument
to~$\cfoldkw$.

The choice as to whether to implement datatype-related constructs
as term formers or as
constants and whether they should be polymorphic or not
is mostly a matter of convenience.  The choice here meshes better
with the definitions of environment models we use in
Section~\ref{sec:models}, but using constants does little other than force us
to insert some semantic functions into some definitions.  However, one place
where this is not quite the whole story is for~$\cfoldkw_\delta$, which one
might be tempted to implement as a term constant of type
$\cforall{\vec\alpha\beta}{\carr{\carr{(\carr{\sfsubst F \beta}{\beta})}
                                      {\delta}} {\beta}}$,
where $\vec\alpha = \ftv(\delta)$.
The typing we have chosen is equivalent with respect to any
standard operational or denotational semantics.  However, our denotational
semantics will be non-standard, and the choice turns out to matter in the
model construction of Section~\ref{sec:merged-model}.  There, we show how to
identify type abstraction with size abstraction in a precise sense.  Were we
to use the polymorphic type for~$\cfoldkw_\delta$, then even when $\delta$ is
monomorphic, $\cfoldkw_\delta$ would still have polymorphic type (for the
result), and that would force us to perform undesirable size abstraction on
values of the monomorphic type.

\subsection{The size order}
The semantics of the recurrence language is described in terms of size
orderings~$\szleq_\tau$ that is defined in
Figure~\ref{fig:synrec-lang-preorder-full} for each type~$\tau$ (although the
rules only define a preorder, we will continue to refer to it as an order).
The syntactic recurrence extracted from a program of type~$\sigma$ has the
type $\cprod\C{\ptrans\sigma}$.  The intended interpretation of
$\ptrans\sigma$ is the set of sizes of source language values of
type~$\sigma$.  We expect to be able to compare sizes, and that is the role of
$\szleq_\sigma$.  Although~$\szleq_\C$ is more appropriately thought of as an
ordering on costs, general comments about~$\szleq$ apply equally
to~$\szleq_\C$, so we describe it as a size ordering as well to reduce
verbosity.\footnote{\citet{avanzini-dal-lago:icfp17} perform cost analysis by
representing execution cost directly and then measuring the size of the
result, so thinking of~$\C$ as a set of sizes (of costs?) may not be
unreasonable.}  The relation~$\szleq_\C$ just requires that $\C$ be a monoid
(i.e., have an associative operation with an identity).  In particular, there
are no axioms governing~$1$ needed to prove the bounding theorem; it is not
even necessary to require that $0\not=1$ or even $0\szleq_\C 1$.

Let us gain some intuition behind the axioms for~$\szleq_\sigma$.
On the one hand, they are just directed versions of
the standard call-by-name equational calculus that one might expect.
In the proof of the Syntactic Bounding Theorem
(Theorem~\ref{thm:syn-bounding}), that is
the role they play.  
But there is more going on here than that.
The intended interpretation of the axioms is that the 
introduction-elimination round-trips that they describe 
provide a possibly less precise
description of size than what is started with.
It may help to analogize with abstract interpretation here:  an
introductory form serves as an abstraction, whereas an elimination form
serves as a concretization.  In practice, the interpretation of products, sums, and arrows do not
perform any abstraction, and so in the models we present 
in Section~\ref{sec:models}, the corresponding axioms are witnessed by
equality (e.g., for~\infruleref{beta-times},
$\den{e_i}{} = \den{\cproj i {\cpair {e_0}{e_1}}}{}$).  
That is not the case for datatypes and type quantification, so
let us examine this in more detail.

Looking forward to definitions from Section~\ref{sec:rec-extraction}, 
if $\sigma$ is a source
language type, then $\ptrans\sigma$ is the \emph{potential type} corresponding
to~$\sigma$ and is intended to be interpreted as a set of sizes for $\sigma$
values.  
It happens that $\ptrans{\slist\sigma} = \clist{\ptrans\sigma}$, so a
$\slist\sigma$ value~$vs = [v_0,\dots,v_{n-1}]$ is extracted as a list of
$\ptrans\sigma$ values, each of which represents the size of one of
the $v_i$s.
Hence a great deal of information is preserved about the original
source-language program.  But
frequently the interpretation (the denotational semantics of the recurrence
language) abstracts
away many of those details.  For example, we might interpret $\clist{\ptrans\sigma}$ as $\N$
(the natural numbers), $\clnil$ as~$0$, and $\clcons$ as successor (with
respect to its second argument), thereby yielding a semantics in which each
list is interpreted as its length (we define two such ``constructor-counting''
models in Sections~\ref{sec:constr-counting-model}
and~\ref{sec:cac-model}).  Bearing in mind that
$\clist{\ptrans\sigma} = \clfp t F$ with 
$F = \csum\cunit{\cprod{\ptrans\sigma}{t}}$, the
interpretation of $F[\clist{\ptrans\sigma}]$ must be
$\den\cunit + (\den{\ptrans\sigma}\cross\N)$.  Let us assume 
that $+$ and $\cross$ are given their usual interpretations (though
as we will see in Section~\ref{sec:examples}, that is often not sufficient).
For brevity we will write~$\cconskw$ and~$\cdestkw$ for
$\cconskw_{\clist{\ptrans\sigma}}$ and~$\cdestkw_{\clist{\ptrans\sigma}}$.
Thus
$\cconskw(y, n) = 1+n$ represents the size of a
source-language list that is built using~$\sconskw_{\slist\sigma}$ when applied
to a head of size~$y$ (which is irrelevant) and a tail of size~$n$.
The question is, what should the value of $\cdestkw(1+n)$
be?  
It ought to somehow describe all
possible pairs that are mapped to~$1+n$ by~$\cconskw$.
Ignoring the possibility that it is an element of~$\den\cunit{}$ (which
seems obviously wrong), and assuming the $\N$-component ought to be~$n$,
no one pair $(y', n)\in\den{\ptrans\sigma}\cross\N$
seems to do the job.  However, if we assume the existence of an
maximum element~$\infty$ of~$\den{\ptrans\sigma}{}$, then
$(\infty, n)$ is an upper bound on all pairs~$(y', n)$ such that
$\cconskw(y', n) = 1+n$, and so it seem reasonable
to set~$\cdestkw(1+n) = (\infty, n)$.  But in this case,
the round trip is not an identity because
$\cdestkw(\cconskw(y, n)) = (\infty, n) \geq (y, n)$, and so
\infruleref{beta-delta} and~\infruleref{beta-fold} are not
witnessed by equality.

Turning to type quantification, the standard interpretation
of~$\cforall\alpha\sigma$ is $\prod_{U\in\bfU}\den{\sigma}{\env*{\alpha}{U}}$ for a
suitable index set~$\bfU$ (in the setting of predicative
polymorphism, this does not pose any foundational difficulties), and the
interpretation of a polymoprhic program is the $\bfU$-indexed tuple of all of
its instances.  Let
${\slam{xs}{e}}\oftype {\sarr{\slist\alpha}{\rho}}$ be a polymorphic
program in the source language.  The recurrence extracted from it
essentially has the form ${\ctylam\alpha{{\clam{xs}{E}}}} \oftype
\cforall\alpha{\carr{\clist\alpha}{\cprod\C{\ptrans\rho}}}$.  
Let us consider a denotational semantics in which $\den{\clist\sigma}{} =
\N\cross\N$, where $(k, n)$ describes a $\slist\sigma$ value with maximum
component size~$k$ and length~$n$ (this is a variant of the semantics
in Section~\ref{sec:cac-model}).  We are then in the conceptually unfortunate
situation that the analysis of this polymorphic recurrence depends on its
instances, which are defined in terms of not only the
list length, but also the sizes of the list values.  Parametricity tells us
that the list value sizes are irrelevant, but our model fails to convey that.
Instead, we really want to interpret the type of the recurrence as
$\N\to(\den\C\cross\den{\ptrans\rho})$, where the domain corresponds to list
length.  This is a non-standard interpretation of quantified types, and so the
interpretations of quantifier introduction and elimination will also be
non-standard.  In our approach to solving this problem,
those interpretations in turn depend on the existence of a
Galois connection between~$\N$ and~$\N\cross\N$, for example mapping the
length~$n$ (quantified type) to $(\infty, n)$ (an upper bound on
instances),
and we might map $(k, n)$ (instance type)
to $n$.  The round trip for type quantification corresponds to $(k, n)
\mapsto n \mapsto (\infty, n)$, and hence~\infruleref{beta-all} is not
witnessed by equality (we deploy the usual
conjugation with these two functions in order to propogate the inequality to
function types while respecting contravariance).
We describe an instance of this sort of model construction
in Section~\ref{sec:merged-model}, although there we are not able to
eliminate the $\bfU$-indexed product, and so the type of the recurrence
is interpreted by $\prod_{U\in\bfU}\N\to(\den\C{}\cross\den{\ptrans\rho}{})$.

\section{Recurrence extraction}
\label{sec:rec-extraction}

A challenge in defining recurrence extraction is that computing only
evaluation cost is insufficient for enabling compositionality, because the
cost of~$f(g(x))$ depends on the size of~$g(x)$ as well as its cost.  To drive
this home, consider a typical higher-order function such as 
\begin{verbatim}
map = fn (f, xs) => fold (fn (x, r) => f x :: r) []
\end{verbatim}
The cost of $\slmap(f, xs)$ depends on the cost of evaluating $f$ on
the elements of~$xs$, and hence (indirectly) on the sizes of the elements
of~$xs$.
And since $\slmap(f, xs)$ might itself be an argument to another
function (e.g.\ another |map|), we also need to predict the
sizes of the elements of $\slmap(f, xs)$ which depends on the size of
the output of~$f$.  
Thus, to analyze |map|, we should be given a
recurrence for the cost and size of $f(x)$ in terms of the size of $x$,
from which we produce a recurrence that gives the cost and size of
$\slmap(f, xs)$ in terms of the size of $xs$.  We call the size of
the value of an expression that expression's \emph{potential}, because
the size of the value determines what future (potential) uses of that value will
cost (\emph{use cost} would be another reasonable term for
potential).  

Motivated by this discussion, we define
translations~$\ptrans\cdot$ from source language
types to complexity types and
$\ctrans\cdot$ from source
language terms to recurrence language terms so that if $e\oftype\sigma$,
then $\ctrans e\oftype\C\cross\ptrans\sigma$.  In the recurrence language,
we call 
an expression of type $\ptrans{\sigma}$ a \emph{potential}
and an expression of type $\C\cross\ptrans\tau$ a \emph{complexity}.
We abbreviate $\C \times \ptrans \tau$ by $\ctrans \tau$. 
The first component of $\ctrans e$ is intended to be an upper bound on
the cost of
evaluating~$e$, and the second component of~$\ctrans e$ is intended
to be an upper bound on the potential of~$e$.  
The weakness of the size order
axioms only allows us to conclude ``upper bound'' syntactically (hence the
definition of the bounding relations in Figure~\ref{fig:bounding-reln-type}),
though one can define models of the recurrence language in which the
interpretations are exact.
The potential of a type-level~$0$ expression 
is a measure of the size of that value to which it evaluates, because that is
how the value contributes to the cost of future computations.  And as we just
described, the potential of a type-$\sarr\rho\sigma$ function~$f$ is itself a
function from potentials of type~$\rho$ (upper bounds on
sizes of arguments~$x$ of~$f$) to
complexities of type~$\sigma$ (an upper bound on the
cost of evaluating~$f(x)$ and the size of the
result).

Returning to $\slmap :
\sarr{\sprod{(\sarr\rho\sigma)}{\slist\rho}}{\slist\sigma}$,
its potential should describe what future uses of
$\slmap$ will cost, in terms of the potentials of its arguments.  
In this call-by-value setting, the arguments will already have been evaluated,
so their costs do not play into the potential of~$\slmap$ (the recurrence that
is extracted from an application expression will take those costs into
account).
The above discussion suggests that
$\ptrans{\sarr{\sprod{(\sarr\rho\sigma)}{\slist \rho}}{\slist\sigma}}$
ought to be
$\carr{\cprod{(\carr{\ptrans\rho}{\cprod\C\sigma})}{\ptrans{\slist\rho}}}
      {\cprod\C{\ptrans{\slist\sigma}}}$.
For the argument function, we are provided a recurrence that maps
$\rho$-potentials to $\sigma$-complexities. For the argument
list, we are provided a $(\slist\rho)$-potential.  Using these,
the potential of $\slmap$ must give the cost for doing the whole
$\mathtt{map}$ and give a $(\slist\sigma)$-potential for the value.
This illustrates how the potential of a higher-order function is itself a
higher-order function.  

Since $\ptrans\rho$ has as much ``information'' as~$\rho$, syntactic recurrence
extraction does
not abstract values as sizes (e.g., we do not replace
a list by its length).  This permits us to prove a general bounding
theorem independent of the particular abstraction (i.e., semantics)
that a client may wish to
use.  Because of this, the complexity translation has
a succinct description.  For any monoid $(\C,+,0)$, the writer monad~\citep{wadler:popl92:essence} $\C
\times -$ is a monad with
\[
\begin{array}{l}
\creturnkw(E) := (0,E) \\
E_1 \cbindkw E_2 := \cpair{\cproj 0 {E_1} + \cproj 0 {(E_2(\cproj 1 {E_1}))}}
                         {\cproj 1 {(E_2(\cproj 2 {E_1}))}}
\end{array}
\]
The monad laws follow from the monoid laws for $\C$.  Thinking of $\C$
as costs, these say that the cost of $\creturnkw(e)$ is zero, and that
the cost of bind is the sum of the cost of $E_1$ and the cost of $E_2$
on the potential of $E_1$.  The complexity translation is then a
call-by-value monadic translation from the source language into the
writer monad in the recurrence language, where source expressions that
cost a step have the ``effect'' of
incrementing the cost component, using the monad operation
\[
\ckeyw{incr}(E\oftype\C) \oftype \cprod \C \cunit := \cpair E\ctriv.
\]

\begin{figure}
\begin{gather*}
\ctrans\tau = \cprod\C{\ptrans\tau}
\\
\begin{aligned}[t]
\ptrans\alpha &= \alpha \\
\ptrans\sunit &= \cunit \\
\ptrans{\sprod{\sigma_0}{\sigma_1}} &=
  \cprod{\ptrans{\sigma_0}}{\ptrans{\sigma_1}} \\
\ptrans{\ssum{\sigma_0}{\sigma_1}} &=
  \csum{\ptrans{\sigma_0}}{\ptrans{\sigma_1}} \\
\ptrans{\sarr{\sigma_0}{\sigma_1}} &=
  \carr{\ptrans{\sigma_0}}{\ctrans{\sigma_1}}
\end{aligned}
\qquad
\begin{aligned}[t]
\ptrans{\ssusp\sigma} &= \ctrans\sigma \\
\ptrans{\slfp t F} &= \clfp t {\ptrans F} \\
\ptrans{\sforall\alpha\tau} &= \cforall\alpha{\ptrans\tau}
\end{aligned}
\end{gather*}
\caption{The complexity and potential translation of types.  Remember that
although we have a grammar for structure functors~$F$, they are actually just
a subgrammar of the small types, so we do not require a separate translation
function for them.}
\label{fig:ctrans-ptrans-types}
\end{figure}


\begin{figure}
\begin{center}
\begin{tabular}{ll}
$\ptrans\tau$      & Potential translation of types. \\
$\ctrans\tau$       & Recurrence translation of types. \\
$\ctrans e$         & Recurrence extraction of expressions. \\
$\ptrans\sctx(x) = \ptrans{\sctx(x)}$   & Potential translation of contexts. \\
$\ccost E$ & $\cproj 0 E$ (cost component of~$E$) \\
$\cpot E$ & $\cproj 1 E$ (potential component of~$E$) \\
$\costpluscpy c E$  & $\cpair{c + \ccost E}{\cpot E}$ (``adding cost'')
\end{tabular}
\end{center}
\caption{Notation related to recurrence language expressions and recurrence
extraction.}
\label{fig:rec-lang-notation}
\end{figure}

\begin{figure}
\[
\begin{aligned}
\\
\ctrans {\typejudge{\sctx, x \oftype\sforall{\vec\alpha}{\sigma}} x {\tysubst*\sigma{\vec\sigma}{\vec\alpha}}}
  &= \cpair 0 {\ctyapp x {\ptrans{\vec\sigma}}} \\
\ctrans{\typejudge\sctx\striv\sunit} &= \cpair 0 {\ctriv} \\
\ctrans{\typejudge\sctx{\spair{e_0}{e_1}}{\sprod{\sigma_0}{\sigma_1}}} 
  &= \cpair{c_0+c_1}{\cpair{p_0}{p_1}} \\
\ctrans{\typejudge\sctx{\sproj i e}{\sigma_i}} 
  &= \cpair{c}{\cproj i p} \\
\ctrans{\typejudge\sctx{\sinj i e}{\ssum{\sigma_0}{\sigma_1}}} 
  &= \cpair{c}{\sinj i p} \\
\ctrans{\typejudge\sctx{\scase* e {x} {e}}{\sigma}}
  &= \costpluscpy{c}{\ccaset* p x {\ptrans{\sigma_i}} {\ctrans{e_i}}} \\
  &\qquad\text{($\typejudge\sctx e {\ssum{\sigma_0}{\sigma_1}}$)} \\
\ctrans{\typejudge\sctx{\slam x e}{\sarr{\sigma'}{\sigma}}}
  &= \cpair 0 {\clam* x {\ptrans{\sigma'}} {\ctrans e}} \\
\ctrans{\typejudge\sctx{\sapp{e_0}{e_1}}{\sigma}} 
  &= \costpluscpy{(c_0+c_1)} {\capp{p_0}{p_1}} \\
\ctrans{\typejudge\sctx{\sdelaykw(e)}{\ssusp\sigma}} 
  &= \cpair 0 {\ctrans e} \\
\ctrans{\typejudge\sctx{\sforcekw(e)}{\sigma}} 
  &= \costpluscpy{c}{p} \\
\ctrans{\typejudge\sctx
                  {\scons\delta e}
                  {\delta}} 
  &= \cpair{c}{\ccons{\ptrans\delta}p} \\
\ctrans{\typejudge\sctx
                  {\sdest\delta e}
                  {\sfsubst F\delta}} 
  &= \cpair{c}{\cdest{\ptrans\delta}p} \\
\ctrans{\typejudge\sctx{\sfold\delta {e'} x {e}}{\sigma}}
  &= \costpluscpy{c'}
       {\cfold{\ptrans\delta}
                {p'}
                {(x : {\sfsubst {\ptrans F}{\ctrans\sigma}})}
                {\costpluscpy{1}{\ctrans{e}}}} \\
  &\qquad\text{($\delta = \slfp t F$)} \\
\ctrans{\typejudge\sctx{\slet x {e_0} {e_1}}{\sigma_1}}
    &= \costpluscpy{c_0}
                   {\substin {\ctrans{e_1}} {\ctylam{\vec\alpha}{p_0}} {x}} \\
    &\qquad\text{($\typejudge{\sctx,x\oftype\sforall{\vec\alpha}{\sigma_0}}
                             {e_1}
                             {\sigma_1}$)} \\
\end{aligned}
\]
\caption{The recurrence extraction function on source language terms.
On the right-hand sides, $(c, p) = \ctrans e$ and $(c_i, p_i) = \ctrans{e_i}$
(note that $\ctrans e$ is always a pair).
}
\label{fig:ctrans-expr}
\end{figure}

We write out the translation of types in 
Figure~\ref{fig:ctrans-ptrans-types} and the recurrence extraction function
explicitly in Figure~\ref{fig:ctrans-expr}.  There is a certain amount of
notation involved, which we summarize in Figure~\ref{fig:rec-lang-notation}.
Recurrence extraction is
defined only for typeable terms and only for terms
in the core language (Definition~\ref{defn:core-lang}).

For an ordinary function type $\sarr {\sigma_0} {\sigma_1}$, the
translation $\carr {\ptrans{\sigma_0}} {\ctrans{\sigma_1}}$ i.e. $\carr
{\ptrans{\sigma_0}} {\cprod {\C} {\ptrans{\sigma_1}}}$  includes a cost
component in the codomain.  In contrast, a polymorphic function type
$\sforall \alpha \tau$ is translated to $\cforall \alpha {\ptrans{\tau}}$,
which does not include a cost component.  The reason for this
discrepancy is that polymorphic functions in the source language are
introduced by $\slet{x}{e'}{e}$, which evaluates $e'$ to a value before
binding $x$ to a polymorphic version of that value.  Thus, the elements
of a polymorphic function type incur no immediate cost when they are
instantiated (at an occurrence of a variable).

Our first order of business is to verify that recurrences extracted from
terms in the source language are themselves typeable in the recurrence
language.  For a source-language context~$\sctx =
x_0\oftype\tau_0,\dots,x_{n-1}\oftype\tau_{n-1}$, write $\ptrans\sctx$
for $x_0\oftype\ptrans{\tau_0},\dots,
x_{n-1}\oftype\ptrans{\tau_{n-1}}$.  For both the source and recurrence
languages, we do not explicitly notate the free type variables of a
typing derivation.  However, the intended invariant of the translation
is that a source language derivation $\typejudge \sctx e \tau$ with free
type variables $\vec{\alpha}$ is translated to a recurrence language
derivation $\typejudge{\ptrans\sctx}{\ctrans e}{\ctrans\sigma}$ that
also has free type variables $\vec{\alpha}$.

\begin{prop}[Typeability of extracted recurrences]
\label{prop:extracted-recurrences-typeable}
If $\typejudge\sctx e\sigma$ is in the core language, then 
$\typejudge{\ptrans\sctx}{\ctrans e}{\ctrans\sigma}$.
\end{prop}
\begin{proof}
See Appendix~\ref{app:typeability-extracted-recurrences}.
\end{proof}

\section{The bounding relation and the syntactic bounding theorem}
\label{sec:bounding}

We now turn to the bounding relation, which is a logical relation that is the
main technical tool that relates source programs to
recurrences.
In this section we will refer to source and recurrence language programs
extensively, and so we will adopt the convention that $E$, $E'$,
etc.\ are metavariables for recurrence language terms.  The bounding
relation $\bounded[\sigma] {\cl e\theta} E$ is defined in
Figure~\ref{fig:bounding-reln-type}.
and is intended to mean that 
$\ccost E$ is a bound on the evaluation cost of~$\cl e\theta$ and
$\cpot E$ is a bound on the value to which $\cl e\theta$ evaluates.
Bounding of values is defined by an auxiliary relation
$\vbounded[\sigma] v E$.  This latter relation morally should be defined by
induction on~$\sigma$, declaring that a value is bounded by a potential if its
components are bounded by corresponding computations on that potential.  Of
course, function values are defined in terms of arbitrary
expressions, and so $\vboundedby[\sarr\rho\sigma]$ must be defined in terms of
$\boundedby[\sigma]$.  The standard way to do so for a logical relation is to
declare that $\slam x e$ is bounded as a value by~$E$ if whenever 
a value~$v'$ is bounded as a value by~$E'$,
$\substin e {v'} x$ is bounded as an expression by~$\capp E {E'}$, and we
adapt that same idea to our setting here.
A naive
approach to defining~$\vboundedby[\delta]$ for $\delta = \slfp t F$
would have us define $\vbounded[\delta] v E$ in a way that depends on
$\vboundedby[{\sfsubst F \delta}]$, which is not a smaller type.  
If we did not permit arrows in shape functors, we could get around this by
counting $\delta$-constructors in~$v$.  Instead we must take a more
general approach.  In Figure~\ref{fig:bounding-reln-shape} we
define by induction on the structure function~$F$ the
relations~$\boundedby[F,\rho]$ and~$\vboundedby[F, \rho]$ 
that correspond to bounding at type~$\sfsubst F\rho$.
We then define $\vboundedby[\slfp t F]$ in terms 
of~$\vboundedby[F, \slfp t F]$. 

The source language permits evaluation of closures with open
type (in particular, when evaluating a $\sletkw$-binding), so the bounding
relation is phrased in terms of open types.  Value bounding at open type is
defined in terms of all of its instances by closed monomorphic types---we do
not enforce any parametricity properties here.  Because source language type
contexts assign type schemes to identifiers, the standard approach of
extending a logical relation on closed terms to open terms by substituting
related values requires us to also define a notion of value bounding at type
schemes, and we again take this to be in terms of instances at closed
types.

We present the relations as a formal derivation system of an
inductive definition, because the proofs of
Lemmas~\ref{lem:bounding-map} and~\ref{lem:bounding-fold} (technical lemmas
needed for the proof of Theorem~\ref{thm:syn-bounding}, the bounding theorem)
rely on a
well-founded notion of subderivation.  A least relation closed under
these rules (which contain a negative occurrence of the relation being
defined in the $\sarr{}{}$ rule) exists because the type subscript gets
smaller in all bounding premises ($\bounded{}{}$ or $\vbounded{}{}$, not
$\szleq$, which is the previously defined size relation on recurrence
language terms).  The premise types are smaller for an ordering that
considers all substitution instances~$\substin\tau\rho\alpha$ of~$\tau$ with
a closed monomorphic type~$\rho$ to be smaller than the polymorphic type
$\sforall \alpha \tau$ or a type with a free variable $\alpha.\tau$;
this ordering is sufficient because of the restriction to predicative
polymorphism.  Although the derivations are infinitary as a result of
the clauses corresponding to arrow types and shapes, it is
straightforward to assign ordinal ranks to derivations so that the rank
of any derivation is strictly larger than the rank of any of its
immediate subderivations, justifying such a proof by induction on
derivations.

\begin{figure}
\hbox to\textwidth{\underline{Expression bounding at open (monomorphic) types:}\hfill}
\[
  \AXC{$\evalin {\cl e\theta} v n$}
  \AXC{$n\szleq_{\C} \ccost E$}
  \AXC{$\vbounded[\sigma] v {\cpot E}$}
\ndTIC{$\bounded[\sigma]{\cl e\theta}E$}
\DisplayProof
\]
\bigbreak
\hbox to\textwidth{\underline{Value bounding at open (monomorphic) types:}\hfill}
\[
  \AXC{For all closed $\rho$:  
       $\vbounded[\substin\sigma\rho\alpha]{v}{\substin E \ptrho\alpha}$}
\ndUIC{$\vbounded[\sigma]{v}{E}$}
\DisplayProof
\]
\bigbreak
\hbox to\textwidth{\underline{Value bounding at closed (monomorphic) types:}\hfill}
\[
\begin{array}{cc}
\multicolumn{2}{c}{
\AXC{$\vphantom {}$}\UIC{$\vbounded[\sunit]{\striv}{E}$}
\DisplayProof
}
\\[3ex]
  \AXC{$\setidx{\vbounded[\sigma_i]{v_i}{\sproj i E}}{i=0,1}$}
\ndUIC{$\vbounded[\sprod{\sigma_0}{\sigma_1}]{\spair{v_0}{v_1}}{E}$}
\DisplayProof
&
  \AXC{$\vbounded[\sigma_i]{v}{E_i}$}
  \AXC{$\sinj i{E_i}\szleq_{\ssum{\sigma_0}{\sigma_1}} E$}
\ndBIC{$\vbounded[\ssum{\sigma_0}{\sigma_1}]{\sinj i v}{E}$}
\DisplayProof
\\[3ex]
  \AXC{$\setst{\bounded[\sigma]
                       {\cl e{\bindin\theta x {v'}}}
                       {\capp E{E'}}}
              {\vbounded[\rho]{v'}{E'}}$}
\ndUIC{$\vbounded[\sarr\rho\sigma]{\cl*{\slam x e}{\theta}}{E}$}
\DisplayProof
&
  \AXC{$\bounded[\sigma]{\cl e\theta}{E}$}
\ndUIC{$\vbounded[\ssusp\sigma]{\cl*{\sdelay e}{\theta}}{E}$}
\DisplayProof
\\[3ex]
\multicolumn{2}{c}{
  \AXC{$\vbounded[F,\delta]{v}{E'}$}
  \AXC{$\ccons{\ptrans\delta}{E'}
        \szleq_{\ptrans{\delta}}
        E$}
\ndBIC{$\vbounded[\delta]{\scons\delta v}{E}$}
\DisplayProof
}
\end{array}
\]
\bigbreak
\hbox to\textwidth{\underline{Value bounding at type schemes:}\hfill}
\[
  \AXC{For all closed $\rho$:
       $\vbounded[\substin\tau\rho\alpha]{v}{\substin E\ptrho\alpha}$}
\ndUIC{$\vbounded[\sforall\alpha\tau]{v}{E}$}
\DisplayProof
\]
\caption{The type-indexed bounding relations.}
\label{fig:bounding-reln-type}
\end{figure}


\begin{figure}
\[
\begin{array}{cc}
\multicolumn{2}{c}{
  \AXC{$\evalin {\cl e\theta} v n$}
  \AXC{$n\szleq_{\C} \ccost E$}
  \AXC{$\vbounded[F,\rho] v {\cpot E}$}
\ndTIC{$\bounded[F,\rho]{\cl e\theta}E$}
\DisplayProof
}
\\[3ex]
  \AXC{$\vbounded[\rho]{v}{E}$}
\ndUIC{$\vbounded[t,\rho]{v}{E}$}
\DisplayProof
&
  \AXC{$\vbounded[\sigma]{v}{E}$}
\ndUIC{$\vbounded[\sigma,\rho]{v}{E}$}
\DisplayProof
\\[3ex]
  \AXC{$\setidx{\vbounded[F_i,\rho]{v_i}{\sproj i E}}{i=0,1}$}
\ndUIC{$\vbounded[\sprod{F_0}{F_1},\rho]{\spair{v_0}{v_1}}{E}$}
\DisplayProof
&
  \AXC{$\vbounded[F_i,\rho]{v}{E_i}$}
  \AXC{$\sinj i{E_i}\szleq_{(\ssum{F_0}{F_1})[\rho]} E$}
\ndBIC{$\vbounded[\ssum{F_0}{F_1},\rho]{\sinj i v}{E}$}
\DisplayProof
\\[3ex]
\multicolumn{2}{c}{
  \AXC{$\setst{\bounded[F,\rho]
                       {\cl e{\bindin\theta x {v'}}}
                       {\capp E{E'}}}
              {\vbounded[\rho_0]{v'}{E'}}$}
\ndUIC{$\vbounded[\sarr{\rho_0} F,\rho]{\cl*{\slam x e}{\theta}}{E}$}
\DisplayProof
}
\end{array}
\]
\caption{The shape-indexed bounding relations.  When writing
$\vbounded[F,\rho]{v}{E}$, $\ftv(F)\subseteq\set{t}$ and $\rho$ is closed.}
\label{fig:bounding-reln-shape}
\end{figure}

The (value) bounding relations in Figures~\ref{fig:bounding-reln-type}
and~\ref{fig:bounding-reln-shape} are really defined on typing derivations.
That is, we really define the relations
\[
\begin{aligned}
{(\typejudgeE{\cl e\theta}{\sigma})}
&\boundedby[\sigma]
{(\typejudgeE E {\ctrans\sigma})}
\\
{(\typejudgeE v \sigma)}
&\vboundedby[\sigma]
{(\typejudgeE E {\ptrans\sigma})}
\end{aligned}
\qquad
\begin{aligned}
{(\typejudgeE{\cl e\theta}{F[\rho]})}
&\boundedby[F,\rho]
{(\typejudgeE E {\ctrans{F[\rho]}})}
\\
{(\typejudgeE v {F[\rho]})}
&\vboundedby[F,\rho]
{(\typejudgeE E {\ptrans{F[\rho]}})}
\end{aligned}
\]
The following lemma acts as an inversion theorem for the bounding relation at
inductive types.

\begin{lemma}~
\label{lem:bounding-datatype-inversion}
\begin{enumerate}
\item If $\bounded[\sfsubst F\rho]{\cl e\theta}{E}$, then
$\bounded[F,\rho]{\cl e\theta}{E}$.
\item If $\vbounded[\sfsubst F\rho]{\cl e\theta}{E}$, then
$\vbounded[F,\rho]{\cl e\theta}{E}$.
\end{enumerate}
\end{lemma}
\begin{proof}
(2) implies~(1), so it suffices to prove the latter, which is done by a
straightforward induction on shape functors.
\end{proof}

The bounding relations on closures are extended to (open) terms in the
standard way for logical relations.

\begin{defn}[Bounding relation]~
\begin{enumerate}
\item Let $\theta$ be a $\sctx$-environment and $\Theta$ a
$\ptrans\sctx$-environment.  We write 
$\vbounded[\sctx]\theta\Theta$ to mean that for all~$x\in\dom\sctx$,
$\vbounded[\sctx(x)]{\theta(x)}{\Theta(x)}$ (note that
$\sctx(x)$ is a type scheme, so this relation is value bounding
at a type scheme).

\item We write
$\bounded[\sigma]
         {(\typejudge\sctx e\sigma)}
         {(\typejudge{\ptrans\sctx}{E}{\ctrans\sigma})}$
to mean that for all~$\vbounded[\sctx]\theta\Theta$,
$\bounded[\sigma]{\cl e\theta}{\esubst E \Theta}$.
\end{enumerate}
\end{defn}

The syntactic bounding theorem relies on various weakening and
substitution properties that we collect here.

\begin{lemma}[Weakening]\hfill
\label{lem:weakening}
\begin{enumerate}
\item If $\bounded e E$ and $E\szleq E'$, then $\bounded e {E'}$.
\item If $\vbounded v E$ and $E\szleq E'$, then $\vbounded v {E'}$.
\end{enumerate}
\end{lemma}

\begin{lemma}
\label{lem:cost-pot-weakening}
$c + \ccost E \szleq \ccost{(\costpluscpy c E)}$ and
$\cpot E \szleq \cpot{(\costpluscpy c E)}$.
In particular, if $\evalin {\cl e\theta} v n$, $n\leq c+\ccost E$, and
$\vbounded v {\cpot E}$, then
$\bounded{\cl e \theta} {\costpluscpy c E}$.
\end{lemma}

The main theorem is analogous to the fundamental theorem for any logical
relation:  every source language program is related (bounded by) the syntactic
recurrence extracted from it.  The proof is somewhat technically involved, but
at its core follows the reasoning typical in the proof of any such fundamental
theorem, so we delegate it to the Appendix.

\begin{thm}[Syntactic bounding theorem]
\label{thm:syn-bounding}
If $e$ is in the core language and
$\typejudgeG e \sigma$, then $\bounded[\sigma] e {\ctrans e}$.
\end{thm}
\begin{proof}
See Appendix~\ref{app:syn-bounding-thm-proof}.
\end{proof}

\section{Environment models}
\label{sec:models}


The syntactic bounding theorem tells us that the syntactic recurrences
extracted from source programs provide bounds on the evaluation cost and
potential of those programs.  However, the syntactic recurrences maintain
sufficient information about the source program to describe cost and potential
in terms of almost any notion of size.  In particular, a syntactic recurrence
extracted from a program over an inductive type maintains all the structure of
the values of that type---e.g., a syntactic recurrence over a list program
describes the bounds in terms of lists again.  It is by defining a
denotational semantics for the recurrence language that we obtain a
``traditional'' recurrence, because that permits us to abstract inductive
values to some notion of size.
We might define a semantics in which a $\ctree \sigma$ type is interpreted by the natural
numbers~$\N$, with the constructor interpreted in terms of either the
maximum function (so a tree is interpreted by its height) or the sum function
(so a tree is interpreted by its size).
So a semantic value in $\csum\cunit{\cprod\sigma{\cprod\N\N}}$,
the one-step unfolding of the interpretation of the tree type, tells
us the sizes of the data supplied to the tree constructor.  The constructor
tells us the size of the tree constructed from such data, and the destructor
tells us about the kind of data that can be used to construct a tree of a
given size.
The denotation of the recurrence extracted from a source program~$f$
is then a function~$T$ such that $T(n)$ is
(a bound on) the cost and size of the result of $f(x)$ when
$x$ has size at most~$n$.
In other words, the end goal is a
``semantic'' recurrence obtained by composing a denotation function with the
extraction function.  Soundness of the denotational semantics with respect to
the size ordering in conjunction with the syntactic bounding theorem 
ensures that the semantic recurrence also provides bounds on the cost and
potential of the source program in terms of the potentials of its arguments.

To that end, we need to define an appropriate notion of model for the
recurrence language.  We will define environment (Henkin) models following
\cite[Ch.~9.2.4]{mitchell:foundations}, which in turn follows
\cite{Bruce-et-al:Semantics}, specializing the definition to the
setting of the recurrence language.  Since the recurrence language is
characterized by the size order, we require that types be interpreted by
preorders, and what would usually be equations describing various semantic
functions will be inequalities.  This leads to a slight challenge in extending
an interpretation of inductive type constructors and destructors to a
canonical
interpretation of~$\cfoldkw_\delta$, because the interpretation of~$\delta$ is
no longer an initial algebra.  However, we shall see that it is sufficient to
have a initiality condition that is weak (requires existence, but not
uniqueness) and lax (is an inequality, not an equality), and that we can
arrange.

Applicative structures (and hence pre-models and models) are defined in terms
of preordered sets.  In such a setting, it is natural to restrict ourselves
to functions that respect the pre-order structure---i.e., monotone functions.
So in the remaining sections, when $A$ and~$B$ are preordered sets,
we write $A\to B$ for the set of monotone functions from~$A$ to~$B$, and
$A\parto B$ for the set of partial monotone functions from~$A$ to~$B$.
$A\to B$ is preordered pointwise, and $\id_A : A \to A$ is the identity
function (we drop the subscript when clear from context).  We also frequently
write $\llambda a.\dotsb$ for the semantic function that takes~$a$
to~$\dotsb$.

\subsection{Models of the recurrence language}

We start by defining the notions of type frame and
applicative structure for the recurrence language.

\begin{defn}
A \emph{type frame} is specified by the following data:
\begin{itemize}
\item A set~$\Usm$ of \emph{small semantic types} and a set $\Ulg$ of
\emph{large semantic types} with $\Usm\subseteq\Ulg$;
\item Distinguished semantic types $U_\C, U_{\cunit}\in\Usm$; 
\item Functions $\smcross:\Usm\cross\Usm\to\Usm$, $\smplus:\Usm\cross\Usm\to\Usm$,
$\smto:\Usm\cross\Usm\to\Usm$, and $\smmu:(\Usm\to\Usm)\parto\Usm$; and
\item A function $\smforall : (\Usm\to\Ulg)\parto\Ulg$.
\end{itemize}
\end{defn}

Let $\TyVar$ be the set of type variables and
let $\eta : \TyVar\to \Usm$.  The denotation of $\tau$ with respect
to~$\eta$, $\den\tau\eta$, is given in
Figure~\ref{fig:type-den}.

\begin{defn}
A type frame is a \emph{type model} if for all~$F$ and $\eta$,
$\llambda V.\den F{\extend\eta V t}\in\dom\smmu$ and for all~$\tau$
and all~$\eta$, $\llambda U.\den\tau{\extend\eta U \alpha}\in\dom\smforall$
(and hence $\den\tau\eta$ is defined for all~$\tau$ and~$\eta$).
\end{defn}

\begin{figure}
\begin{gather*}
\begin{aligned}
\den \alpha \eta &= \eta(\alpha) \\
\den\cunit\eta &= U_\cunit
\end{aligned}
\qquad
\begin{aligned}
\den{\cprod\sigma{\sigma'}}\eta &= \den\sigma\eta\smcross\den{\sigma'}\eta \\
\den{\csum\sigma{\sigma'}}\eta &= \den\sigma\eta\smplus\den{\sigma'}\eta \\
\den{\carr\sigma{\sigma'}}\eta &= \den\sigma\eta\smto\den{\sigma'}\eta \\
\den{\clfp t F}\eta &= \smmu(\llambda V.\den{F}{\extend\eta V t})
\end{aligned}
\\[\baselineskip]
\den{\forall \alpha.\tau}\eta = \smforall(\llambda U.\den\tau{\extend\eta U \alpha})
\end{gather*}
\caption{The denotation (partial) function of types and type schemes
into a type frame.}
\label{fig:type-den}
\end{figure}

\begin{defn}
An \emph{applicative structure} is specified by the following data:
\begin{itemize}
\item A type model~$(\Usm, \Ulg)$.

\item For each~$U\in\Ulg$, a preordered set~$(D^U,\leq_U)$.

\item For
each~$\Phi\in\dom\smmu$ and $U, V\in\Usm$, a function
$\Phi_{U,V}:(D^U\to D^V)\to(D^{\Phi\,U}\to D^{\Phi\,V})$.

\item Distinguished elements $0,1\in D^{U_\C}$ and an associative function
$+:D^{U_\C}\to D^{U_\C}$ such that $0$ is a right- and left identity for~$+$.

\item A distinguished element $*\in D^{U_\cunit}$

\item For each $U, V\in \Usm$, functions
\[
(D^U\to D^V) \xrightharpoonup{\smAbs[U,V]} D^{U\smto V}
\xrightarrow{\smApp[U, V]} (D^U\to D^V)
\]
such that 
$\smApp\comp\smAbs \geq \id$.  Note that $\smAbs$ is a partial function.

\item For each $U_0, U_1\in\Usm$, functions 
\[
D^{U_0}\cross D^{U_1} \xrightarrow{\smPair[U_0,U_1]} D^{U_0\smcross U_1}
\xrightarrow{\smProj[U_0,U_1]^i} D^{U_i}
\]
such that
$\smProj^i(\smPair(a_0, a_1)) \geq a_i$.

\item 
For each $U_0$, $U_1$, $V\in\Usm$, functions
\[
D^{U_i} \xrightarrow{\smInj[U_0, U_1]^i} D^{U_0\smplus U_1}
\xrightarrow{\smCase[U_0, U_1, V]} (D^{U_0}\to V)\cross (D^{U_1}\to V)\to D^V
\]
such that
$(\smCase\comp\smInj^i)\,a\,(f_0, f_1)\geq f_i\,a$.
We often write $\smCase(a, f_0, f_1)$ for $\smCase\,a\,(f_0, f_1)$.

\item For each $\Phi\in\dom\smmu$, functions
\[
D^{\Phi(\smmu\,\Phi)} \xrightarrow{\smCons[\Phi]}
D^{\smmu\,\Phi} \xrightarrow{\smDest[\Phi]}
D^{\Phi(\smmu\,\Phi)}
\]
such that
$\smDest\comp\smCons\geq\id$.

\item For each $\Phi\in\dom\smmu$ and $U\in\Usm$, functions
$\smFold[\Phi,U] : (D^{\Phi\,U}\to D^U) \to (D^{\smmu\,\Phi}\to D^U)$
such that
$(\smFold[\Phi,U]\,f)\comp \smCons[\Phi]\geq 
f\comp (\Phi_{\smmu\,\Phi,U}(\smFold[\Phi,U]\,f))$.

\item For each $\Phi\in \dom\forall$, functions
\[
\prod_{U\in\Usm} D^{\Phi(U)}
\xrightharpoonup{\smTyAbs[\Phi]}
D^{\smforall(\Phi)}
\xrightarrow{\smTyApp[\Phi]}
\prod_{U\in\Usm} D^{\Phi(U)}
\]
such that $\smTyApp\comp\smTyAbs)\geq \id$.  Note that $\smTyAbs$ is a partial
function.
\end{itemize}
Remember that when we write, e.g., $D^U\to D^V$, we mean the monotone
functions from $D^U$ to $D^V$, and hence all of the semantic functions that
make up the data of an applicative structure are monotone.
\end{defn}

We write $\U = (\Usm,\Ulg,\setidx{D^U}{U\in\Ulg})$ for a typical
applicative structure, or just $\U = \setidx{D^U}{U\in\Ulg}$ when
$\Ulg$ is clear from context.
For a context~$\cctx$ define
$\tyvar(\cctx) = \setst{\alpha}{\text{$\alpha$ occurs in 
$\ftv(\cctx(x))$ for some~$x$}}$.
Define
a \emph{$\cctx$-environment} to be a map~$\eta$ such that
\begin{itemize}
\item $\eta(\alpha)\in\Usm$ for $\alpha\in\tyvar(\cctx)$; and
\item $\eta(x)\in D^{\den{\cctx(x)}\eta}$ for $x\in\dom\cctx$.
\end{itemize}
For an applicative structure and environment~$\eta$, define
a partial denotation function $\den{\typejudge\cctx e\sigma}\eta$ as in
Figure~\ref{fig:den-good}.  The only way in which
$\den\cdot\cdot$ may fail to be total is if the arguments
to $\smAbs$ or $\smTyAbs$ are not in the corresponding domains (because we
start with a type model, we know that $\smmu$ and $\smforall$ are only applied
to functions in their domains).

\begin{defn}
Let $\U$ be an applicative structure.
\begin{enumerate}
\item $\U$ is a \emph{pre-model} if 
\begin{itemize}
\item Whenever $\typejudge\cctx e \tau$ and $\eta$ is a $\cctx$-environment,
$\den{\typejudge\cctx e\tau}\eta$ is defined and an element
of~$D^{\den{\tau}{\eta}}$; and
\item Whenever $\typejudge{\cctx,y\oftype\rho}{e'}{\sigma}$ and
$\typejudge\cctx{e}{\sfsubst{F}{\rho}}$ and $\eta$ is a $\cctx$-environment,
\[
\den{\cmap F \rho y {e'} e}\eta\leq_{\den\sigma\eta}
(\llambda V.\den F {\bindin\eta t V})_{\den\rho\eta,\den\sigma\eta}(
  \llambda a.\den{e'}{\bindin\eta y a})(\den e \eta
).
\]
\end{itemize}

\item $\U$ is a \emph{model} if $\U$ is a pre-model and whenever
$\prejudge\cctx e {e'} \tau$, and $\eta$ is a $\cctx$-environment,
$\den{\typejudge\cctx e\sigma}\eta \szleq_{\den\tau\eta}
\den{\typejudge\cctx {e'}\sigma}\eta$.
\end{enumerate}
\end{defn}

\begin{figure}
\begin{align*}
\den{\typejudge{\cctx,x\oftype\tau} x \tau}\eta &= \eta(x) \\
\den{\typejudge\cctx\ctriv\cunit}\eta &= 
  \mathord* \\
\den{\typejudge\cctx{\cpair{e_0}{e_1}}{\cprod{\sigma_0}{\sigma_1}}}\eta &= 
  \smPair(\den{e_0}\eta, \den{e_1}\eta) \\
\den{\typejudge\cctx{\cproj i e}{\sigma_i}}\eta &=
  \smProj^i(\den{e}{\eta}) \\
\den{\typejudge\cctx{\cinj i e}{\csum{\sigma_0}{\sigma_1}}}\eta &=
  \smInj^i(\den e \eta) \\
\den{\typejudge\cctx{\ccase e x {e_0} x {e_1}}{\sigma}}\eta &=
  \smCase(
    \den e\eta, 
    \llambda a.\den {e_0}{\extend\eta a x}, 
    \llambda a.\den{e_1}{\extend\eta a x}
  ) \\
\den{\typejudge\cctx{\clam x e}{\carr\sigma{\sigma'}}}\eta &=
  \smAbs(\llambda a.\den e {\extend\eta a x}) \\
\den{\typejudge\cctx{\capp e {e'}}{\sigma}}\eta &=
  \smApp(\den e\eta)\,(\den{e'}\eta) \\
\den{\typejudge\cctx
               {\ccons\delta{e}}
               {\delta}}
    {\eta}
  &= \smCons[\llambda V.\den{F}{\bindin{\eta}{t}{V}}](\den e\eta) \\
\den{\typejudge\cctx
               {\cdest\delta{e}}
               {\sfsubst F \delta}}
    {\eta}
  &= \smDest[\llambda V.\den{F}{\bindin{\eta}{t}{V}}](\den e\eta) \\
\den{\typejudge\cctx
               {\cfold\delta {e'} x e}
               {\sigma}}{\eta}
  &= \smFold[\llambda V.\den{F}{\bindin{\eta}{t}{V}},\den\sigma\eta](
       \llambda a.\den{e}{\bindin\eta x a}
     )\,(\den {e'} \eta)
     \\
\den{\typejudge\cctx {\ctylam\alpha e} {\cforall \alpha \tau}}\eta
  &=  \smTyAbs[\llambda U.\den\tau{\extend\eta U \alpha}](\llambda U.\den{\typejudge \cctx e \tau}{\extend \eta U \alpha}) \\
\den{\typejudge\cctx {\ctyapp e \sigma} {\substin\tau\sigma \alpha}}\eta
  &= \smTyApp[\llambda U.\den\tau{\extend\eta U \alpha}](\den{\typejudge\cctx e {\cforall \alpha \tau}}\eta)(\den\sigma\eta)
\end{align*}
\caption{The denotation (partial) function into an applicative
structure.  For constructors and destructors, assume $\delta = \clfp t F$
and $\fv(\delta) = \set{\alpha_0,\dots,\alpha_{n-1}}$, and define
$\eta^* = \bindin\eta{\vec\alpha}{\vec U}$.}
\label{fig:den-good}
\end{figure}

The indirection of interpreting syntactic
types by semantic types, and then 
interpreting terms of a given syntactic type as elements of a domain associated
to the corresponding semantic type is necessary, especially in our setting
of non-standard models.  This makes is much easier (seemingly,
\emph{possible}) to define things like the $\smmu$ operator.  Without
the indirection, we would have to define $\smmu$ on (functions on)
a collection of domains, some of which represent syntactic types.  That
ends up being very difficult to do.  For example, we might have to first
define a notion of polynomial function on the semantic domains in order
to define the domain of $\smmu$, and then somehow identify each semantic
polynomial function with a structure functor.  But doing so gets us into
problems with unique representation; e.g., there may be multiple
structure functors corresponding to the same semantic polynomial.  
And with non-standard models, we seem to have even more troubles, because we
end up trying to define the interpretations of inductive types simultaneously
with the $\smmu$ function.
But first interpreting the syntactic types by semantic types gives us a way
around these problems, because (if we wish) we can define the semantic types 
to be closely tied to the syntactic types.  That is exactly what we do for the 
standard type frame, so we can essentially define~$\smmu$ syntactically,
and then choose a domain corresponding to~$\clfp t F$ (which is a
semantic type as well as a syntactic one) after having
defined~$\smmu$.

\begin{lemma}
\label{lem:den-facts}
Let $\U$ be a pre-model.  Then:
\begin{enumerate}
\item $\den{\substin\tau\sigma \alpha}\eta = 
\den\tau{\bindin\eta \alpha {\den\sigma\eta}}$.  If $\alpha\notin\ftv(\tau)$
then for all~$U$, $\den\tau\eta = \den\tau{\bindin\eta\alpha U}$ and
for all term variables~$x$ and all~$a$,
$\den\tau\eta = \den\tau{\bindin\eta x a}$.

\item $\den{\substin e {e'} x}\eta =
\den e {\bindin \eta x {\den {e'} \eta}}$.  If $x\notin\fv(e)$, then
for all~$a$, $\den e \eta = \den e{\bindin\eta x a}$.

\item If $a\leq a'$, then
$\den e {\bindin\eta x a} \leq \den e {\bindin\eta x {a'}}$; in other words,
$\llambda a.\den e {\bindin \eta x a}$ is monotone.
\end{enumerate}
\end{lemma}

\begin{prop}[Environment model soundness]
\label{prop:env-model-soundness}
If $\U$ is an pre-model, then $\U$ is a model.
\end{prop}
\begin{proof}
By induction on the derivation of $\prejudge\cctx e {e'} \tau$.
\end{proof}

One might hope that a model of the fragment of the recurrence language that
omits~$\cfoldkw_\delta$ can be extended to one that does, but in our setting
this does not quite hold.
Since we only have 
directed versions
of the usual equalities, initial algebras for structure functors may not
exist.  And even if they do, they are not necessarily what we want.  
For clarity, in this discussion we will write syntactic types for semantic
types.
The point behind different semantics is to abstract inductive values to some
notion of size, and when this abstraction is non-trivial, 
$D^{\clfp t F}$ and $D^{\sfsubst F {\clfp t F}}$ are probably not isomorphic.
Instead of the usual initial algebra for 
interpreting~$\clfp t F$, we typically want an algebra 
$\smCons[F] : D^{F[\clfp t F]}\to D^{\clfp t F}$ such that for any
other algebra $s : D^{F[\sigma]}\to D^\sigma$, there is a
function~$\smFold[F,\sigma]\,s$
that makes the diagram
\[
\begin{tikzcd}
D^{F[\clfp t F]}\arrow{dd}[left]{\smMap[F](\smFold[F]\,s)}
                \arrow{rr}{\smCons[F]}
  && D^{\clfp t F}\arrow{dd}{\smFold[F]\,s}         \\
  & \leq \\
D^{F[\sigma]}\arrow{rr}{s}
  && D^\sigma
\end{tikzcd}
\]
commute, where $\smMap[F]$ is a semantic function that corresponds to
the~$F[\cdot,\cdot]$ macro.
Relative to the usual definition of initial algebra, this requirement is
\emph{weak}, in that we ask only for existence of a $\smFold{}$ function
making the diagram commute ($\beta$ reduction) and not the uniqueness of
$\smFold{}$ ($\eta$/induction), and it is \emph{lax}, in that we ask
that $\beta$-reduction holds only as an inequality, rather than an
equality.

Nonetheless, under assumptions that turn out to be relatively easy to ensure,
we can define~$\smFold[\Phi,U]$.  Given a subset $X\subseteq A$ of a
preordered set~$A$, we say that $a\in A$ is a \emph{least upper bound} of~$X$,
written $a = \bigmax X$, if for all~$x\in X$, $x\leq a$, and if $b\in A$
satisfies the condition that for all~$x\in X$, $x\leq b$, then $a\leq b$.
When $A$ is preordered, $\bigmax X$ may not exist, and when it does, it need
not be unique.  If $A$ is a \emph{partial} order (i.e., $a\leq b\leq a$
implies that $a=b$), then $\bigmax X$ is unique when it exists, and we say
that $A$ is a \emph{complete upper semi-lattice} if $A$ is a partial order and 
$\bigmax X$ exists for every $X\subseteq A$.  Though this seems like a very
strong condition, in practice it is easy to ensure.

In a model in which every~$D^U$ is a complete upper semi-lattice, we would
like to define
\[
\smFold[\Phi,U]\,s\,x =
\bigmax\setst{
         s\bigl(
           \Phi_{\smmu\,\Phi,U}(\smFold[\Phi,U] s)\,z
         \bigr)
       }
       {z\in D^{\Phi(\smmu\,\Phi)}, {\smCons[\Phi]\,z \leq x}}.
\tag{std-fold}
\label{eq:std-fold}
\]
\emph{A priori}, this definition may not be well-founded, but in fact it is, as
shown in the next proposition.

\begin{prop}
\label{prop:standard-smfold}
Suppose that $\U = \setidx{D^U}{U\in\Ulg}$ is a model
of the fragment of the recurrence language that omits~$\cfoldkw_\delta$
and each $D^U$ is a complete upper semi-lattice, and suppose that
$\smFold$ is defined by~\eqref{eq:std-fold}.  Then:
\begin{enumerate}
\item For all~$s$, $\smFold[\Phi,U]\,s$ is total and monotone.
\item $\smFold[\Phi,U]$ is total and monotone.
\end{enumerate}
\end{prop}
\begin{proof}\hfill
\begin{enumerate}
\item Fix $s$ and consider
\[
 Q = \llambda g.\llambda x.
       \bigmax\setst{s\bigl(\Phi_{\smmu\,\Phi,U}\,g\,z\bigr)}
                    {z\in D^{\Phi(\smmu\,\Phi)}, \smCons[\Phi]\,z \leq x}.
\]
$Q : (D^{\smmu\,\Phi}\to D^U)\to(D^{\smmu\,\Phi}\to D^U)$ and it is easy to
see that $Q$ is monotone.
Since $D^U$ is a complete upper semi-lattice,
$D^{\smmu\,\Phi}\to D^U$ is a complete
partial order.  So
$Q$ has a least fixed point; that is $\smFold[\Phi,U]\,s$.\footnote{%
The least fixed
point is obtained by the standard iteration of~$Q$ starting at the bottom
element.  Because we only have that $Q$ is monotone (not necessarily 
continuous on chains), the iteration may have to
be extended transfinitely---see 
\citet[Exercise~8.19]{davey-priestley:intro-lattices-order}.}  
Monotonicity
of~$\smFold[\Phi,U]\,s$ is immediate from its definition.

\item Totality follows from~(1) and monotonicity
from the fact that the
function that maps a monotone function to its least fixed point is itself
monotone.\qedhere
\end{enumerate}
\end{proof}

The proof of Prop.~\ref{prop:standard-smfold}, and hence the
interpretation of $\cfoldkw_\delta$, may seem a bit heavy-handed, making
use of general least fixed point theorems and even iterating into the
transfinite.  As we noted earlier, we are in a setting in which we do not have
(and do not want) initial algebras, but must nonetheless show an
initiality-like property of a given algebra.  Accordingly, we would expect to
use technology at least as strong as that needed for typical initial algebra
existence theorems.  The canonical such theorem 
(e.g., as described
by \citet[Thm.~7.6]{aczel:non-well-founded-sets}) verifies that the least
fixed point of a set-continuous operator is an initial algebra,
and the verification consists of constructing the equivalent of
$\smFold\,s$ by induction on the (ordinal-indexed) construction of the least
fixed point.

The reader may have noticed that an
alternative possible definition for $\smFold$ is 
\[
\smFold\,s\,x 
= s(\Phi(\smFold\,s)(\smDest\,x))
\]
and Prop.~\ref{prop:standard-smfold} would still hold.  
This fact witnesses that the initiality condition for
$\smCons[\Phi] : D^{\Phi(\smmu\,\Phi)}\to D^{\smmu\,\Phi}$ is weak, in that
functions to other algebras are not unique.
In practice, this alternative definition
yields far worse bounds for extracted recurrences, because we end up defining
$\smDest[\Phi]\,x = \bigmax\setst{z}{\smCons[\Phi](z)\leq x}$ and so
$\smFold\,s\,x 
= s(\Phi(\smFold\,s)(\bigmax\setst{z}{\smCons[\Phi](z)\leq x}))$.
Monotonicity of~$f$ only allows us to conclude
that $f(\bigmax X) \geq
\bigmax\setst{f(x)}{x\in X}$, but this putative definition
for~$\smFold\,s$ exposes a case in which this inequality is strict.

\subsection{The standard type frame}
\label{sec:standard-type-frame}

Our last step in our general discussion of models is to define the type frame
upon which all of our examples will be based.  It gives us enough data to
provide a standard definition of the functions~$\Phi_{U,V}$, which in turn
lets us use~\eqref{eq:std-fold} to define~$\smFold$ and so for most
of our examples, it will suffice to define~$\smCons[\Phi]$ (because we will
set $\smDest[\Phi] = \bigmax\setst{z}{\smCons[\Phi](z)\leq x}$).
Our examples are all based on variations of
the \emph{standard type frame}, which is defined as follows:

\begin{itemize}
\item $\Usm$ is the set of closed types and $\Ulg$ the set of closed type
schemes of the recurrence language.
\item $\smto$, $\smcross$, and $\smplus$ are the standard type constructors;
e.g., $\sigma_0\smplus\sigma_1 = \csum{\sigma_0}{\sigma_1}$.
\item $\dom\smmu = \setst{\llambda\sigma.F[\sigma]}{\fv(F) \subseteq \set{t}}$ 
and $\smmu(\llambda\sigma.F[\sigma]) = \clfp t F$ (we call a structure
functor~$F$ with $\fv(F)\subseteq\set{t}$ \emph{closed}).
\item $\dom\smforall = \setst{\llambda\sigma.\substin\tau\sigma\alpha}{\fv(\tau)=\set{\alpha}}$ and
$\smforall(\llambda\sigma.\substin\tau\sigma\alpha) = \cforall\alpha\tau$.
\end{itemize}

It is straightforward to show that if $\llambda\sigma.F[\sigma] =
\llambda\sigma.F'[\sigma]$, then $F = F'$, and if
$\llambda\sigma.\substin\tau\sigma\alpha = \llambda\sigma.\substin{\tau'}\sigma\alpha$, then $\tau = \tau'$, so $\smmu$ and $\smforall$ are well-defined.
It should be clear and occasionally helpful to observe that for any
$\tau$ and 
environment~$\eta = 
\env{\bindto{\alpha_0}{\sigma_0},\dots,\bindto{\alpha_{n-1}}{\sigma_{n-1}}}$,
$\den\tau\eta = \substin\tau{\vec\sigma}{\vec\alpha}$.
For models based on the standard type frame and any closed structure
functor~$F$, we will usually
write $F$ in place of
$\llambda\sigma.\sfsubst F\sigma$ in subscripts for readability.

\begin{prop}
\label{prop:std-type-frame-is-model}
The standard type frame is a type model.
\end{prop}

For any applicative structure based on (an extension of) the standard type
frame, define 
$F_{\rho,\sigma} : (D^\rho\to D^\sigma)\to
(D^{\sfsubst F \rho}\to D^{\sfsubst F \sigma})$ for each closed structure
functor~$F$ and closed~$\rho$ and~$\sigma$ by:
\[
\begin{aligned}[t]
t_{\rho,\sigma}\,g\,x &= g\,x \\
(\sigma_0)_{\rho,\sigma}\,g\,x &= x \\
\end{aligned}
\qquad
\begin{aligned}[t]
(\csum{F_0}{F_1})_{\rho,\sigma}\,g\,x
  &= \smCase(x, \llambda y.\smInj^0((F_0)_{\rho,\sigma}\,g\,y),
                \llambda y.\smInj^1((F_1)_{\rho,\sigma}\,g\,y)) \\
(\cprod{F_0}{F_1})_{\rho,\sigma}\,g\,x
  &= \smPair((F_0)_{\rho,\sigma}\,g\,(\smProj^0 x),
             (F_1)_{\rho,\sigma}\,g\,(\smProj^1 x)), 
  \\
(\carr{\sigma_0}{F})_{\rho,\sigma}\,g\,x
  &= \llambda y.F_{\rho,\sigma}\,g\,(x\,y)
\end{aligned}
\]

\begin{lemma}
\label{lem:map-denotation}
If $\U$ is an applicative structure based on an extension of the standard type
frame that is a model for the fragment of the recurrence language that
omits~$\cfoldkw_\delta$, $\typejudge{\cctx,y\oftype\rho}{e'}{\sigma}$,
$\typejudge\cctx e {\sfsubst F\rho}$, and $\eta$ is a $\cctx$-environment,
then
\[
 \den{\cmap F y \rho {e'} e}\eta =
 (\den F\eta)_{\den\rho\eta,\den\sigma\eta}\,(
   \llambda a.\den{e'}{\bindin\eta y a}
 )\,(\den e \eta).
\]
\end{lemma}
\begin{proof}
By induction on~$F$.
\end{proof}

Combining Prop.~\ref{prop:standard-smfold} with
Lemma~\ref{lem:map-denotation}, we conclude that
to define a model of the recurrence language, it suffices to
define an extension of the standard type frame and the following applicative
structure data:
\begin{itemize}
\item The sets~$D^\tau$, along with an argument that $D^\tau$ is a complete
upper semi-lattice;
\item The semantic functions for arrow, product, and sum types;
\item $\smCons[F]$ for each structure functor~$F$.
\end{itemize}
From this data we can define 
$\smDest[F](x) = \bigmax\setst{z}{\smCons[F]\leq x}$,
$F_{\rho,\sigma}$ as just given, and $\smFold[F,\sigma]$
by~\eqref{eq:std-fold}.
Of course, there are models that are not constructed this way; 
Section~\ref{sec:lower-bounds} gives an example that is useful for
extracting recurrences for lower bounds.

\subsection{Syntactic sugar}
\label{sec:syntactic-sugar}

We now introduce some syntactic sugar that will make our discussion of
recurrences somewhat more pleasant.  To simplify the discussion, we restrict
the details to the source language type $\stree\sigma$ and its
recurrence language potential
$\ctree\sigma$,
but we will use analogous notation for other datatypes such as $\snat$ and
$\slist\alpha$ in our examples.
Many of our source-language functions
are really structural folds over some standard datatype---that is, the step
function is a $\scasekw$ expression where the argument for each branch is
really the argument to one of the datatype constructors.  Accordingly, we
introduce notation for such $\sfoldkw$ expressions:  for 
$y\notin\fv(e_\stemp)\union\fv(e_\stnode)$,
\[
\sfoldtree \sigma e {e_\stemp} {(x, r_0, r_1).e_\stnode}
\]
is syntactic sugar for
\[
\sfold{\stree\sigma} e w
{\scase w y {e_\stemp} y {\substin {e_\stnode} {\sproj 0 y, \sproj 1 y, \sproj
2 y} {x, r_0, r_1}}}.
\]
We introduce a similar notation in the recurrence language:
\[
\cfoldtree \sigma \rho e {e_\ctemp} {(x, r_0, r_1).e_\ctnode}
\]
is syntactic sugar for
\[
\cfold{\ctree\sigma} e {(w : \sfsubst {F_{\ctree\sigma}} \rho)} {
{(\ccase w 
         y
         {e_\ctemp}
         y
         {\substin {e_{\ctnode}} {\cproj 0 y, \cproj 1 y, \cproj 2 y}
                                 {x, r_0, r_1}})}
}
\]
where $w$ and $y$ are fresh variables.

It would be nice to establish an identity of the form
$\ctrans{\sfoldkw_{\stree\sigma}\dotsb} = \cfoldkw_{\ctree\sigma}\dotsb$, but
the size-order axioms, which give us only inequalities, are too weak.
However, the models that we will consider validate many equations, so we can
set out a nice relationship.  In the following proposition, we say
``in the semantics, $e = e'$'' to mean that for any~$\eta$,
$\den e\eta = \den {e'} \eta$:

\begin{prop}
\label{prop:sugared-fold-extraction}
Suppose that we have a model such that
\begin{itemize}
\item In the semantics: if $\ccost{\ctrans {e'}} = 0$, then
$\ctrans{\substin e {e'} x} = \substin{\ctrans e} {\cpot{\ctrans{e'}}} x$;
and
\item In the semantics:  
$\costpluscpy c {\ccase* e x {e_i}} = \ccase* e x {\costpluscpy c
{e_i}}$.
\end{itemize}
If $\typejudge\sctx
              {\sfoldtree \sigma e {e_\stemp} {(x, r_0, r_1).e_\stnode}}
              {\rho}$,
then in the semantics,
\begin{multline*}
\ctrans{\sfoldtree \sigma e {e_\stemp} {(x, r_0, r_1).e_\stnode}} =
\\
\costpluscpy{c}
            {\cfoldtree \sigma 
                        {\ctrans \rho}
                        {p} 
                        {\costpluscpy 1 {\ctrans{E_\ctemp}}}
                        {(x, r_0, r_1).{\costpluscpy 1
                        {\ctrans{E_\ctnode}}}}}
\end{multline*}
where $(c, p) = \ctrans e$.
\end{prop}

While the models that we discuss in subsequent sections satisfy
the hypotheses of Prop.~\ref{prop:sugared-fold-extraction}, they are not
necessarily satisfied in an arbitrary model.  That requires
additional axioms that correspond roughly to $\eta$ axioms.

\section{Examples}
\label{sec:examples}

\subsection{The standard model}
\label{sec:std-model}

For the standard model, we first extend the standard type frame by including
the constant $\bot$ in~$\Usm$.  A semantic type (scheme) is \emph{proper} if it
has no occurrences of $\bot$.  The proper semantic types (type schemes)
correspond exactly to the closed syntactic types (type schemes).  In the
definitions of $\smmu$ and $\smforall$, we take~$F$ and~$\tau$ to be proper.
We define the sets~$A^\sigma$ by induction on~$\sigma$
as follows:\footnote{
The collection of sets~$\setidx{A^\sigma}{\sigma\in\Usm}$ must be contained
in some set that contains~$\emptyset$ and a one-element set and is closed under 
disjoint unions, products, function spaces, unions of chains, and 
products indexed by~$\Usm$.  $V^{\omega_1}$ in the standard set-theoretic
hierarchy suffices.}
\begin{itemize}
\item $A^\C = \N$, the natural numbers.
\item $A^\bot = \emptyset$.
\item $A^{\cunit} = \set{*}$, some one-element set.
\item $A^{\carr{\sigma_0}{\sigma_1}} = (A^{\sigma_1})^{A^{\sigma_0}}$, the
set of functions from~$A^{\sigma_0}$ to~$A^{\sigma_1}$.
\item $A^{\cprod{\sigma_0}{\sigma_1}} = A^{\sigma_0}\cross A^{\sigma_1}$,
where $\cross$ is the standard set-theoretic product.
\item $A^{\csum{\sigma_0}{\sigma_1}} = A^{\sigma_0}\disjunion A^{\sigma_1}$,
where $\disjunion$ is the standard set-theoretic disjoint union.
\item $A^{\clfp t F} = \bigunion_i A^{(\llambda V.\den F {\env{\bindto t V}})^i\bot}$.
\item $A^{\cforall\alpha\tau} = \prod_{\sigma\in\Usm} A^{\substin\tau \sigma \alpha}$.
\end{itemize}
Define $a\leq_{A^\sigma} b$ iff $a=b$, and let the semantic functions for
arrows, products, and sums be the identity functions.
The definitions of $\smCons[F]$, $\smDest[F]$, and $\smFold[F,\sigma]$ are
based on the standard initial-algebra semantics.
Note that we cannot use~\eqref{eq:std-fold} because
the $A^\sigma$ are not complete upper semi-lattices, and hence the
hypotheses of Prop.~\ref{prop:standard-smfold} do not hold, and hence
\infruleref{beta-fold} must be verified directly.

At first blush, this model is not particularly interesting.  There is no
abstraction of values to sizes and the ``order'' on costs is the
identity, so the recurrences extracted from source language programs
describe the exact cost of those programs in terms of the argument
values.  However, this is a standard model of (predicative)
polymorphism, and so we can hope that parametricity may have some
interesting consequences.  Free
theorems~\citep{wadler:theorems-for-free} have been used to obtain
relative cost information, and we discuss this further in
Section~\ref{sec:related-work}.  Here, we apply parametricity to the
recurrence language and sketch the argument that if $g
: \sarr{\slist\alpha}{\slist\alpha}$, then the cost of~$g(xs)$ depends
only on the length of~$xs$ (the same can be said for the length
of~$g(xs)$, but this follows from parametricity applied to the
source language).  For any~$\rho$, let us define $T_\rho(xs) =
((\den{\ctrans g}{}\,\ptrho)_p(xs))_c$, the exact cost of evaluating
$g(xs)$ (since $\ctrans\cdot$ is a monadic translation and the
interpretation of inductive types is the standard one, syntactic values
of list type in the source language are isomorphic to the semantic
values in the model).  The goal is to show that if $xs : \slist\rho$ and
$ys : \slist\sigma$ are of the same length, then $T_\rho(xs) =
T_\sigma(ys)$.  To do so, we apply parametricity 
to~$\llambda \rho.\llambda xs.T_\rho(xs) \in
    A^{\forall \rho.\clist \rho \to \C}$.  We take
the relational interpretation of $\C$ to be equality (so the cost
constants $0$ and $+$ preserve the relation).  Expanding the definition
of parametricity, this means that for any $\rho$ and $\sigma$ and relation
$R \subseteq A^{\ptrho}\cross A^{\ptsigma}$, for any $xs \in
A^{\clist{\ptrho}}$ and $ys \in A^{\clist\ptsigma}$, if $\clist R
\subseteq A^{\clist{\ptrho}}\cross A^{\clist\ptsigma}$ holds for $xs$
and $ys$, then the relational interpretation of $\C$ holds for
$T_\rho(xs)$ and $T_\sigma(ys)$.  Since the relational interpretation of
the cost type is equality, this would give the result, so it suffices to
show that there is an $R$ such that $(\clist R)(xs,ys)$ holds whenever
$xs$ and $ys$ have the same length.  However, the standard relational
lifting $\clist R$ holds whenever $xs$ and $ys$ have the same length and
$xs_i$ is related to $ys_i$ by $R$, so taking $R$ to be the total
relation achieves this.  We conclude that if $xs$ and $ys$ have the same
length, then the cost of $g(xs)$ and $g(ys)$ is the same.

\subsection{Constructor size and height}
\label{sec:constr-counting-model}

We now describe a model in which a value~$v$ of inductive type~$\delta$ is
interpreted either by the number of~$\delta$ constructors in~$v$ (constructor
size) or by the maximum nesting depth of $\delta$-constructors in~$v$
(constructor height), so that it reflects common size abstractions such as
list length, tree size, and tree height.  For example,
in this model, the interpretation of the recurrence extracted from a function
with domain~$\slist\sigma$ describes the cost in terms of the length of the
argument list.
For concreteness we will define the constructor size model.  For the
interpretation of the types, we will need two versions of the natural numbers:
$\N_0^\infty = \set{0,1,\dots,\infty}$ for costs, and
$\N_1^\infty = \set{1,2,\dots,\infty}$ for sizes of inductive values, which
must be at least~$1$ because every value contains at least one constructor.
$\N_i^\infty$ is ordered by
$x\leq_{\N_i^\infty} y$ if $y=\infty$ or $x\leq_\N y$.
The presence of~$\infty$ may be perplexing, since all programs in the source
language terminate.  However, it is not always possible to give a finite upper
bound on cost or potential in terms of the potential of the argument, because
the notion of potential used in this model may not identify all possible
sources of recursive calls.  For example, consider the function
$\ssumtree$ defined in Figure~\ref{fig:sumtree} that sums the nodes of
a $\stree\snat$.  The cost and size of $\ssumtree\,t$ depend on the size
of~$t$ and the sizes of its labels, whereas in this model, the potential 
of~$t$ only tells us the former.  Since $\ssumtree$ is
definable in our source language, its recurrence can be extracted, and hence
must have a meaning in this model; the only sensible interpretation is one
that maps every tree size to the trivial upper bound of~$\infty$ for both
cost and potential.

We start by extending the standard type frame with additional small types
$\N_0,\N_1\in\Usm$.  Then we define the sets~$V^\tau$,
observing that each~$V^\tau$ is a complete upper semi-lattice.  This allows us
to construct a model by just defining~$\smCons[F]$.  The sets~$V^\tau$ are
defined as follows:

\begin{itemize}
\item $V^{\N_i} = \N_i^\infty$.

\item $V^\C = \N_0^\infty$ with the standard interpretations for~$\smkeyw{0}$
and $\smkeyw{+}$, where $x\mathbin{\smkeyw{+}}\infty = 
\infty\mathbin{\smkeyw{+}}x = \infty$.

\item $V^\cunit = \set{*}$.

\item $V^{\carr{\sigma_0}{\sigma_1}} =$ the set of monotone functions
from $V^{\sigma_0}$ to $V^{\sigma_1}$ with the usual pointwise order, taking
$\smAbs$ and $\smApp$ to be the identity functions.

\item $V^{\cprod{\sigma_0}{\sigma_1}} = 
V^{\sigma_0}\cross V^{\sigma_1}$ with
the usual component-wise order, taking $\smPair$ and $\smProj$ to be the
standard pairing and projection functions.

\item $V^{\csum{\sigma_0}{\sigma_1}} = 
\ordideal(V^{\sigma_0}\disjunion V^{\sigma_1})$, which we define
in Section~\ref{sec:cc-sum-interp}.

\item $V^{\clfp t F} = \N_1^\infty$.  We define $\smCons[F]$ in
Section~\ref{sec:cc-datatype-functions}
(recall that we write
$\smCons[F]$ for $\smCons[\llambda V.\den F {\env{\substfor V t}}]$, etc., and
that we can define $\smDest[F]$ and $\smFold[F,\sigma]$ from it).

\item $V^{\cforall\alpha\tau} =
\prod_{\sigma\in\Usm}V^{\substin\tau\sigma\alpha}$, with the pointwise order,
taking $\smTyAbs$ and $\smTyApp$ to be the identity functions.
\end{itemize}

\noindent
Once we define the interpretation of sums and datatypes, it is straightforward
to verify that this is a model.  

\begin{prop}
\label{prop:mc-model}
$\V = \setidx{V^\tau}{\tau\in\Ulg}$ is a model of the recurrence
language.
\end{prop}
\begin{proof}
Since $\smAbs$ and $\smTyAbs$ are total, it suffices to verify the conditions
on the semantic functions.  This is trivial for arrows, products, and type
quantification; sums and inductive types are handled in the next two sections.
\end{proof}

\subsubsection{Interpretation of sums}
\label{sec:cc-sum-interp}

As we observed, we need to ensure that
all the sets~$V^\sigma$ are complete upper semi-lattices.
Preserving the complete upper semi-lattice property is
straightforward for all type constructors except sum.  We could take the usual
disjoint sum along with a new infinite element~$\infty$ that is a common upper
bound of elements on both sides, but that ends up leading to very weak bounds
in practice.  For example, recall that $\smDest[\clist\sigma](2)$ should tell
us about the data that can be used to construct a list of size~$\leq 2$
(which is a $\clcons$ list, because we count the
number of~$\cconskw_{\clist\sigma}$ constructors, so $\clnil$ has
size~$1$).
If we were to interpret sums as just proposed,
both $\smInj^0(*)$ and $\smInj^1(a, 0)$ are such values, and their least
upper bound would be~$\infty$.  It is not hard to parlay this into an argument
that if $\sltail = \slam{xs}{\scase {xs} {x} {\slnil} {x} {\sproj 1 x}}$
is the usual tail function on~$\slist\sigma$, then the recurrence extracted
from $\sltail$ gives a bound of~$\infty$ for all lists of length~$>1$.
While correct, this is hardly satisfying!

Instead, we take inspiration from abstract interpretation
\citep{cousot-cousot:popl77:ai}:  we will define~$\smDest[F](n)$ to
be the \emph{set}
of values~$x$ such that $\smCons[F](x)\leq n$.  We can arrange this for the
typical cases of interest (i.e., finitary inductive datatypes such as lists
and trees) by defining $V^{\csum{\sigma_0}{\sigma_1}}$ to be the
downward closed subsets of $V^{\sigma_0}\disjunion V^{\sigma_1}$.  We could
arrange this for \emph{all}
inductive datatypes if we were to do something similar in
the interpretation of arrows and products, but that entails some additional
notational cost in reasoning about extracted recurrences while providing
no benefits for the examples that we present.
We start with some standard order-theoretic and set-theoretic definitions:

\begin{itemize}
\item For any partially ordered set~$A$, the \emph{order ideal} of~$A$ is
\[
\ordideal(A) =_{df}
\setst{X\subseteq A}{\text{$x\in X$ and $y\leq x$}\Implies y\in X}.
\]
$\ordideal(A)$ is partially ordered by set inclusion and is a complete
upper semi-lattice; concretely, if $X\subseteq\ordideal(A)$, then
$\bigmax X = \bigunion X$.
\item For any $X\subseteq A$, 
$\downset[A] X = \setst{x\in A}{\exists y\in X.x\leq y}\in\ordideal(A)$
and for $a\in A$, $\downset[A] a = \downset\set{a}$ (we drop the superscript
when it is clear from context).
\item For any $f : A \to B$ and $X\subseteq A$, 
$f[X] = \setst{f(x)}{x\in X}$.
\item If $X_0$ and~$X_1$
are partially-ordered sets, $X_0\disjunion X_1$ 
is the usual disjoint union with injection functions 
$\smin[i]:X_i\to X_0\disjunion X_1$ partially ordered by $x\leq y$ iff
$x = \smin[i](x')$, $y = \smin[i](y')$, and $x'\leq_{X_i} y'$.
\end{itemize}

For the interpretation of sums, we define $V^{\csum{\sigma_0}{\sigma_1}} = 
\ordideal(V^{\sigma_0}\disjunion V^{\sigma_1})$ with the semantic functions
defined by

\begin{align*}
\smInj^i(x) 
  &= \smin[i][\downset[V^{\sigma_i}] x] 
     = \downset[V^{\sigma_0}\disjunion V^{\sigma_1}](\smin[i](x)) \\
\smCase(X_0\disjunion X_1, f_0, f_1) 
  &= \bigmax f_0[X_0] \bmax \bigmax f_1[X_1]
\end{align*}

\begin{lemma}
$\smCase(\smInj^i(x), f_0, f_1) \geq f_i(x)$.
\end{lemma}
\begin{proof}
\begin{align*}
\smCase(\smInj^i(x), f_0, f_1)
  &= \smCase(\smin[i][\downset x]\disjunion\emptyset, f_0, f_1) \\
  &= \bigmax f_i[\downset x] \bmax \bigmax f_{1-i}[\emptyset] \\
  &= \bigmax f_i[\downset x] \\
  &\geq f_i(x)
    & & (x\in\downset x).\qedhere
\end{align*}
\end{proof}

Note that $\ordideal A$ is a monad on the category of partially ordered
sets and monotone functions, with unit $A \to \ordideal A$ given by
$\downarrow^A$, and multiplication $\ordideal {\ordideal A} \to
\ordideal A$ given by union, and it plays the role of a powerset monad
on posets (the ordinary powerset operation, without the additional
downward closure requirement, does not have a \emph{monotone} function
$A \to \mathcal{P}(A)$, because $x \szleq_A y$ does not imply that
$\set{x}\subseteq\set{y}$.
When a partially ordered set is a complete upper
semilattice (i.e. supports the maximum operation that we use to
interpret the recursor), it is an \emph{algebra} for this monad,
i.e. there is a monotone function $\ordideal A \to A$, satisfying some
equations.  Thus, another way of understanding these models is that, for
functions and products, we build algebras $\bigmax_{A\to B} : \ordideal
(A \to B) \to (A \to B)$ and $\bigmax_{A\times B} : \ordideal(A \times
B) \to A \times B$ from algebra structures $\bigmax_{A} : \ordideal A
\to A$ and $\bigmax_B : \ordideal B \to B$, but for sums, we use the
free algebra $\ordideal(A + B)$, with $\bigmax_{\ordideal(A + B)} :
\ordideal{\ordideal (A + B)} \to \ordideal{(A+B)}$ given by union.

\subsubsection{Semantic functions for inductive datatypes}
\label{sec:cc-datatype-functions}

We define $\smCons[F]$ by first defining 
a function $\smsize{F} : V^{\sfsubst F {\clfp t F}}\to \N_0^\infty$.
For $\delta = \clfp t F$, a semantic value of 
type~${\sfsubst F {\clfp t F}}$
represents the data from which a value of type~$\delta$
is constructed, but with the inductive substructures replaced by their sizes,
and $\smsize{F}$ returns the size of the inductive value constructed from that
data.
\[
\begin{aligned}[t]
\smsize{t}(n) &= n \\
\smsize{\sigma}(x) &= 0
\end{aligned}
\qquad
\begin{aligned}[t]
\smsize{\csum{F_0}{F_1}}{}(X_0\disjunion X_1) 
  &= \bigmax\smsize{F_0}[X_0]\bmax\bigmax\smsize{F_1}[X_1]  \\
\smsize{\cprod{F_0}{F_1}}(a_0, a_1)
  &= \smsize{F_0}{}(a_0) + \smsize{F_1}(a_1) \\
\smsize{\carr{\sigma}{F}}(g)
  &= \sum\setst{\smsize{F}(g\,x)}{x\in V^\sigma}
\end{aligned}
\]
For the constructor height model, define a function $\smheight F$
analogously, replacing the sums in the 
product and arrow shapes with maximums.  
The semantic constructor and destructor are then defined by
\[
\smCons[F](a) = 1 + \smsize{F}(a)
\qquad
\smDest[F](n) = \bigmax\setst{a}{\smCons[F]\,a \leq n}.
\]
To use \eqref{eq:std-fold}, it suffices to verify
the conditions of 
Prop.~\ref{prop:standard-smfold}, which is trivial, so we have
\[
\smFold[F,\sigma]\,s\,x =
\bigmax\setst{s\bigl(F_{\delta,\sigma}(\smFold[F,\sigma]\,s)z\bigr)}
             {1 + \smsize{F}(z) \leq x}.
\]

\subsubsection{Examples: lists and trees}
\label{sec:cc-examples-lists-trees}

Referring to Figure~\ref{fig:cpy-lang-examples}, 
\begin{align*}
\den{\typejudgeE\clnil{\clist\sigma}}\eta 
  &= 1 \\
\den{\typejudge{x\oftype\sigma,xs\oftype\clist\sigma}
               {\clcons(x, xs)}{\clist\sigma}}\eta 
  &= 1 + \eta(xs) \\
\den{\typejudgeE\ctemp{\ctree\sigma}}\eta 
  &= 1 \\
\den{\typejudge
        {x\oftype\sigma, t_0\oftype\ctree\sigma, t_1\oftype\ctree\sigma}
        {\ctnode(x, t_0, t_1)}
        {\ctree\sigma}}\eta
  &= 1 + \eta(t_0) + \eta(t_1).
\end{align*}
It is not hard to see that
$\den{\typejudgeE t {\ctree\sigma}}{} = 2n + 1$, where $n$ is the usual size
of~$t$ (i.e., $\den{\typejudgeE t {\ctree\sigma}}{}$ is
the number of internal and external nodes of~$t$).  
Since this is linear in the usual
notion of size of a tree, it suffices for showing that the recurrences that we
extract have the expected $O$-behavior.  

Destructors exhibit the desired behavior; consider $\clist\sigma$ again:
\begin{align*}
\smDest[F_{\clist\sigma}](x) 
  &=
    \begin{cases}
    \set{*}\disjunion\emptyset,&x=1 \\
    \set{*}\disjunion\setst{(a, x')}
                           {a\in V^\sigma, 1+x'\leq x},&2\leq x \leq \infty
    \end{cases} \\
  &= \set{*}\disjunion (V^\sigma\cross \downset[\N_1^\infty](x-1))
\end{align*}
where we define $\downset[\N_1^\infty]0 = \emptyset$.
In other words,
a list of size~$1$ must be~$\slnil$
and a list of length at most~$x$ is either~$\slnil$
or $\slcons(x, xs)$, where $xs$ has length at most~$x-1$.  
Remember that $V^{\sfsubst F {\clist\sigma}} = 
\ordideal(\set{*}\disjunion (V^\sigma\cross\N_1^\infty))$, so if
$\smin[1](a, x)\in X\in V^{\sfsubst F {\clist\sigma}}$, then $x\geq 1$; that is
why $\smDest[F_{\clist\sigma}](1)\not=\set{*}\disjunion X$ with
$X\not=\emptyset$.
For
$\ctree\sigma$, the result is equally pleasant:
\[
\smDest[F_{\ctree\sigma}](x) =
\begin{cases}
\set{*}\disjunion\emptyset,&x=1 \\
\set{*}\disjunion\setst{(a, x_0, x_1)}
                       {a\in V^\sigma, 1+x_0+x_1\leq x},&2\leq x \leq \infty
\end{cases}
\]

Finally, we observe the following simple forms for the denotation of
recurrences over lists and trees:

\begin{prop}\hfill
\label{prop:mcfold}
\begin{enumerate}
\item If
$f\,n = \den{\cfoldlist {\sigma} {\rho}
                     {y}
                     {e_\clnil} 
                     {(x, r).e_\clcons}}{\bindin\eta y n}$, 
then in the constructor size and height models,
\begin{align*}
f\,1 &= \den{e_\clnil}\eta \\
f\,n &= 
  \den{e_\clnil}\eta \bmax
  \bigmax\setst{\den{e_\clcons}
                    {\bindin\eta
                             {x, r}
                             {
                               \infty^\sigma,
                               f\,n'
                             }
                    }
               }{n' < n} \\
  &= 
  \den{e_\clnil}\eta \bmax
  \den{e_\clcons}
                    {\bindin\eta
                             {x, r}
                             {
                               \infty^\sigma,
                               f(n-1)
                             }
                    }
  & & (n > 1).
\end{align*}
The second form for $f\,n$, $n>1$, follows from monotonicity of the denotation
function.

\item If
$f\,n = \den{\cfoldtree{\sigma} 
                       {\rho}
                       {y}
                       {e_\ctemp} 
                       {(x, r_0, r_1).e_\ctnode}}{\bindin\eta y n}$, 
then in the constructor size model,
\begin{align*}
f\,1 &= \den{e_\ctemp}\eta \\
f\,n &= 
  \den{e_\ctemp}\eta \bmax
  \bigmax\setst{\den{e_\ctnode}
                    {\bindin\eta
                             {x, r_0, r_1}
                             {
                               \infty^\sigma,
                               f\,n_0,
                               f\,n_1
                             }
                    }
               }{n_0+n_1 < n} & & (n > 1).
\end{align*}
In the constructor height model, replace $n_0+n_1 < n$ with
$n_0\bmax n_1 < n$.
\end{enumerate}
\end{prop}

\begin{proof}
The verification is a moderately tedious calculation; here it is
for~(2) with $n>1$.  Let
\begin{align*}
s &= \llambda(Z\disjunion X).\den{
            {\ccase w y {e_\ctemp}  
                      y {\substin{e_\ctnode}
                                 {\pi_0\,y,\pi_1\,y,\pi_2\,y}
                                 {x,r_0,r_1}}}}{\bindin\eta w {Z\disjunion X}} \\
  &= \llambda(Z\disjunion X).
       \den{e_\ctemp}\eta
       \bmax
       \bigmax\setst[big]{\den{e_\ctnode}{\bindin\eta{a,b_0,b_1}{x,r_0,r_1}}}
                    {(a, b_0, b_1)\in X}
\end{align*}
Observe that $f = \smFold\,s$ and let us write
$\smMap$ for 
$(\den{F_{\ctree\sigma}}{\eta})_{\den{\ctree\sigma}\eta,\den\rho\eta}$.
By definition, we have
$f\,n = \bigmax\setst{s(\smMap\,f\,(Z\disjunion X))}
                     {\smCons(Z\disjunion X) \leq n}$.
By monotonicity we need only consider $Z = \set{*}$, and by definition
of $\smCons$ we need only consider non-empty sets~$X$ such that
$(\cblank, n_0, n_1)\in X$ implies $n_0+n_1 < n$, so
\begin{align*}
f\,n
  &= \bigmax\setst{s(\smMap\,f\,(\set{*}\disjunion X))}
                  {(\cblank, n_0, n_1)\in X\Implies n_0+n_1 < n} \\
  &= \bigmax\setst{s(\set{*}\disjunion \smMap\,f\,(\smin[1][X]))}
                  {(\cblank, n_0, n_1)\in X\Implies n_0+n_1 < n} \\
  &= \bigmax
     \setst*[Big]
       {
         \den{e_\ctemp}\eta
         \bmax
         \bigmax\setst[big]
                  {\den{e_\ctnode}
                       {\bindin\eta
                               {x,r_0,r_1}
                               {b,k_0,k_1}}}
                  {(b,k_0,k_1)\in\smMap\,f\,(\smin[1][X])}
       }
       {
         (\cblank,n_0,n_1)\in X\Implies n_0+n_1<n
       } \\
  &= \den{e_\ctemp}{\eta}
     \bmax
     \bigmax
     \setst*[Big]
     {
       \bigmax
       \setst[big]
       {
         \den{e_\ctnode}
             {\bindin\eta
                     {x,r_0,r_1}
                     {b,k_0,k_1}}
       }
       {
         (b,k_0,k_1)\in\smMap\,f\,(\smin[1][X])
       }
     }
     {
       (\cblank, n_0, n_1)\in X\Implies n_0 + n_1 < n
     } \\
  &= \den{e_\ctemp}{\eta}
     \bmax
     \bigmax
     \setst*[Big]
     {
       \bigmax
       \setst*[big]
       {
         \den{e_\ctnode}
             {\bindin\eta
                     {x,r_0,r_1}
                     {b,k_0,k_1}}
       }
       {
         (b,k_0,k_1)\in\bigmax\setst
                                {\downset(a,f\,n_0,f\,n_1)}
                                {(a, n_0, n_1)\in X}
       }
     }
     {
       (\cblank, n_0, n_1)\in X\Implies n_0 + n_1 < n
     } \\
  &= \den{e_\ctemp}{\eta}
     \bmax
     \bigmax
     \setst*[Big]
     {
       \bigmax
       \setst*[big]
       {
         \den{e_\ctnode}
             {\bindin\eta
                     {x,r_0,r_1}
                     {b,k_0,k_1}}
       }
       {
         \exists(a,n_0,n_1)\in X : (b,k_0,k_1) \leq (a,f\,n_0,f\,n_1)
       }
     }
     {
       (\cblank, n_0, n_1)\in X\Implies n_0 + n_1 < n
     } \\
  &= \den{e_\ctemp}{\eta}
     \bmax
     \bigmax
         \setst[big]{\den{e_\ctnode}
                    {\bindin\eta
                      {x,r_0,r_1}
                      {\infty^\sigma,f\,n_0,f\,n_1}
                    }}
             {n_0+n_1 < n}
\end{align*}
Let us write the last equation as
\[
L = \den{e_\ctemp}\eta
\bmax
\bigmax\setst[Big]{\bigmax A_X}{(\cblank,n_0,n_1)\in X\Implies n_0+n_1 < n}
=
\den{e_\ctemp}\eta
\bmax
\bigmax B =
R
\]
First let us show that for any~$X$ such that 
$(\cblank,n_0,n_1)\in X\Implies n_0+n_1 < n$, $A_X\subseteq\downset B$, from
which we conclude that $\bigmax A_X\leq\bigmax B$, and hence
$L\leq R$.  For any such~$X$, take $(b, k_0, k_1)$ such that there is
$(a, n_0, n_1)\in X$ with $(b, k_0, k_1)\leq (a, f\,n_0, f\,n_1) \leq
(\infty,f\,n_0, f\,n_1)$.
By Prop.~\ref{prop:mc-model},
$\den{e_\ctnode}{\bindin\eta{x,r_0,r_1}{b,k_0,k_1}} \leq
\den{e_\ctnode}{\bindin\eta{x,r_0,r_1}{\infty,f\,n_0,f\,n_1}}$.  Since
$(b, k_0, k_1)$ was chosen arbitrarily, $A_X\subseteq\downset B$, as needed.
To show that $R\leq L$, suppose that $n_0+n_1<n$.  Then
$\den{e_\ctnode}{\bindin\eta{x,r_0,r_1}{\infty,f\,n_0,f\,n_1}} \in
A_{\downset(\infty,n_0, n_1)}$, from which $R\leq L$ follows.
\end{proof}

\sloppypar
Although we will primarily use Prop.~\ref{prop:mcfold},
it may be instructive to work through an example of explicitly constructing
$\smFold[F,\rho]$ from the proof 
of Prop.~\ref{prop:standard-smfold}.
Consider $F = \csum\cunit t$ (the structure functor
for $\cnat$), and set 
$s(x) = \smCase(x, \llambda u.1, \llambda u.1+x)$.  $s$ might be the step
function for the recurrence that describes the cost of the 
copy function on $\snat$.  Define~$Q$ as in the proof of
Lemma~\ref{prop:standard-smfold}.  In this setting, the bottom element at which
we start iterating~$Q$ is the function that is constantly~$0$.  Set
$f_0 = Q\,\bot$ and $f_{n+1} = Q\,f_n$.  Just as in the calculation
of $\smDest[F_{\clist\sigma}]$,
\[
\smDest[F_\cnat](x) =
\begin{cases}
\set{*}\disjunion\emptyset,&x=1 \\
\set{*}\disjunion\setst{x'}{1+x'\leq x},&2\leq x
\end{cases}
\]
and so a bit more calculation shows that
\[
f_k(n) =
\begin{cases}
n,& n\leq k+1 \\
k+1,& n>k+1.
\end{cases}
\]
It is not hard to see that
$f_\omega(n) = n$ is a fixed point of~$Q$, so we conclude that
$\den{\cfoldnat \rho x 1 {r.\cnsucc r}}{\env{\bindto x n}} = n$.

Of course, this is precisely
what we expect, though for readers familiar with how a typical recursive
function on numbers is defined by successive approximations, the route may
feel a bit different.  Usually when defining a recursive function on numbers,
one takes the flat order and starts with the everywhere-undefined function.
For a typical total function, the $k$-th
approximation is a partial function that is defined and correct
on some initial segment of the natural numbers
and undefined elsewhere.  Here we take the (more-or-less) standard order
and start with a function
that is everywhere an unlikely bound (namely, $0$).  Each successive
approximation yields a function with more likely bounds, 
terminating with a (hopefully low but) correct bound.  In the case of a
partial recursive function, the ``bad'' case is that the function is not
defined for some numbers (the value of the approximants never gets
above~$\bot$).  In our setting, the ``bad'' case is that the bound is infinite
(the value of the approximants never stops growing).
The reader may wish to compare this with the use of~$\N_0^\infty$ by 
\citet{rosendahl:auto_complexity_analysis}, where $\infty$ corresponds to the
bottom element in the usual CPO semantics for fixpoints.  We return to this in
Section~\ref{sec:recursion} when we discuss general recursion.

\subsubsection{Example:  tree copy}
\label{sec:cc-model-tree-copy}

\begin{figure}
\begin{align*}
\scopy_{\stree\sigma} &= \slam t {%
  \sfoldtree*\sigma 
            t 
            {\stemp} 
            {(x,r_0, r_1).\stnode(x, \sforce{r_0}, \sforce{r_1}).}
} \\
\ccopy_{\ctree\sigma} &= \clam t {%
  \cfoldtree*\sigma
             \rho
             t
             {\cpair 1 1}
             {(x, r_0, r_1).\cpair{1 + \ccost{r_0} + \ccost{r_1}}
                                  {\ctnode(x, \cpot{r_0}, \cpot{r_1})}}
}.
\end{align*}
\caption{The monomorphic tree copy function and its extracted recurrence.}
\label{fig:cc-model-tree-copy}
\end{figure}

For a first ``sanity check,'' let us analyze the tree copy function that is
defined in Figure~\ref{fig:cc-model-tree-copy}.  We will also describe 
some of the
main features in the analysis that are typical of all of our examples.
The first is that a source language program
$e = \slam{x,y,z}{e'}$ extracts to a recurrence of the form
$\cpair 0 {\clam x {\cpair 0 {\clam y {\cpair 0 {\clam z {\ctrans {e'}}}}}}}$.
However, we are really only interested in $\ctrans{e'}$ as a function
of the potentials~$x$, $y$, and $z$.  Accordingly, when analyzing a
program such as~$e$, we focus on the recurrence language
program $\clam{x,y,z}{\ctrans{e'}}$.  Here, this means we will analyze
the (denotation of the) 
recurrence $\ccopy_{\ctree\sigma}$ that is also shown
in Figure~\ref{fig:cc-model-tree-copy}.  
Second, we shall use 
Prop.~\ref{prop:sugared-fold-extraction} freely as though it is a theorem
about the syntax when we write our examples.  Third,
in our examples, we typically
use the identifier~$r$ in syntactic recurrences for a recursive call to the
computation of a complexity, and hence $\cpot r$ and $\ccost r$ correspond to
recursive calls that compute potential and cost, respectively.  
Finally, we remind
the reader that our goal is to show that the semantic recurrences are
essentially the same as those that we expect to arise from an informal
analysis, and so we make no attempt to solve them.  

The analysis for $\scopy_{\stree\sigma}$ proceeds as follows.
Define
$T(n) = (\den{\ccopy_{\ctree\sigma}}{}(n))_c$.  Following the definition
of the denotation function and using Prop.~\ref{prop:mcfold} and
facts about $\bmax$ and $\bigmax$ in the semantics, we have
\[
T(1) = 1
\qquad
\begin{aligned}[t]
T(n) &= 1 \bmax \bigmax\setst{1 + T(n_0) + T(n_1)}{n_0+n_1<n} \\
     &= \bigmax\setst{1 + T(n_0) + T(n_1)}{n_0+n_1<n}
\end{aligned}
\]
and we obtain a similar recurrence for
$S(n) = (\den{\ccopy_{\ctree\sigma}}{}(n))_p$.  We observe that these are
precisely the expected recurrences from an informal analysis which, if one is
careful, must consider all possible combinations of subtree sizes when
computing the cost or size of the result when the argument tree has size~$n$.

\subsubsection{Example:  binary search tree membership}
\label{sec:cc-model-bst-membership}

\begin{figure}
\begin{align*}
\sbstmem = {}
&\lambda{cmp^{\sarr{\sprod\sigma\sigma}\sorder},t^{\stree\sigma},x^{\sigma}}. \\
&     {\sfoldtree* \sigma
                 t
                 {\sfalse}
                 {
                   (y, r_0, r_1).
                   \scaseorder* {\sapp{cmp}{\spair x y}}
                               {\sforce r_0}
                               {\strue}
                               {\sforce r_1}
                 }
     }
\end{align*}
\begin{multline*}
\cbstmem = \\
\begin{aligned}
&\lambda{(cmp : \carr{\cprod\ptsigma\ptsigma}{\ctrans\corder}),
         (h : \ctree\ptsigma),
         (x : \ptsigma)}. \\
&\cfoldtree*
   \ptsigma
   {\ctrans\sbool}
   h
   {\cpair 1 \cfalse}
   {
     (y, r_0, r_1).
     \ccaseorder*
       {\cpot*{\capp{cmp}{\cpair x y}}}
       {\cpair{1 + \ccost*{\capp{cmp}{\cpair x y}} + \ccost{r_0}}
              {\cpot{r_0}}}
       {\cpair 1 \ctrue}
       {\cpair{1 + \ccost*{\capp{cmp}{\cpair x y}} + \ccost{r_1}}
              {\cpot{r_1}}}
   }
\end{aligned}
\end{multline*}
\caption{Binary search tree membership and its extracted recurrence.}
\label{fig:cc-model-bst-membership}
\end{figure}

For an interesting example, let us consider membership testing in
$\sigma$-labeled binary search trees.  First we define the type
\[
\sorder = \ssum{\ssum\sunit\sunit}\sunit
\]
and write $\scaseorder e {e_0} {e_1} {e_2}$ for
$\scasekw^\sorder\,e\,\skeyw{of}\,x.e_0; x.e_1; x.e_2$, and we assume
comparable notation in the recurrence language.  The membership test function
is given in Figure~\ref{fig:cc-model-bst-membership}.

Let us consider
an informal analysis of~$\sbstmem$, which is somewhat simpler to describe in
reference to the $\skeyw{member}$ function of
Figure~\ref{fig:susp-fold}(a).  Let $T(h)$ be
the number of calls to
$\skeyw{member}$ in terms of the height of~$t$.  We would probably argue that
$T(1) = 1$ and for $h>1$, 
\[
T(h) \leq \underbrace{1}_{\text{the call to $\skeyw{member}$}} + 
\underbrace{\bigmax\set{0, T(h_0), T(h_1)}}_{\text{cost of a $\scasekw$ is
bounded by the costs of its branches}},
\]
where
$h_0, h_1 < h$.  But since the only information we have is that $t_0$ and
$t_1$ are subtrees of some tree~$t$ of height~$h$, what we must 
really mean is that
\[
  T(h) \leq 1 + \bigmax\setst{0, T(h_0), T(h_1)}{h_0\bmax h_1<h},
\]
so this is the recurrence we expect to see in a formal analysis.

Taking the same approach as in the previous section, we analyze the
recurrence~$\cbstmem$ given in Figure~\ref{fig:cc-model-bst-membership},
this time considering its denotation in 
the constructor height model.  
The extracted recurrence makes explicit the dependence of the complexity of
$\sbstmem$ on the complexity of the comparison function~$cmp$.  Of course, a
typical analysis will make assumptions about this complexity.  The most
common such (and the one we implicitly made in our informal analysis)
is that the cost of the comparison function is independent of the
size of its arguments, which we can model here by assuming that
$\ccost*{\capp{cmp}{\cpair x y}} = 0$
for all~$x$ and~$y$ (more precisely, we only analyze $\den{\cbstmem}{}\,cmp$ 
under
the assumption that $cmp$ satisfies this condition). 
Define
$T(h) = (\den{\cbstmem}\,cmp\,h\,x)_c$ and assume that
$cmp(x, \infty) = A\disjunion B\disjunion C$.  Then making use
of Prop.~\ref{prop:mcfold},
\[
T(1) = 1
\qquad
\begin{aligned}[t]
T(h)
  &= 1
     \bmax
     \bigmax
     \setst*[Big]
     {
       \begin{aligned}[t]
       &\bigmax\setst{1 + \ccost*{mem\,cmp\,h_0\,x}}{a\in A} \bmax \\
       &\quad\bigmax\setst{1}{b\in B} \bmax \\
       &\quad\bigmax\setst{1 + \ccost*{mem\,cmp\,h_1\,x}}{c\in C}
       \end{aligned}
     }
     {
       h_0 \bmax h_1 < h
     } \\
  &= 1
     \bmax
     \bigmax
     \setst*[Big]
     {
       \begin{aligned}[t]
       &\bigmax\setst{1 + T(h_0)}{a\in A} \bmax \\
       &\quad\bigmax\setst{1}{b\in B} \bmax \\
       &\quad\bigmax\setst{1 + T(h_1)}{c\in C}
       \end{aligned}
     }
     {
       h_0 \bmax h_1 < h
     } \\
  &\leq 1 \bmax \bigmax\setst{1 + T(h_0), 1, 1 + T(h_1)}{h_0\bmax h_1 < h} \\
  &= \bigmax\setst{1 + T(h_0), 1, 1 + T(h_1)}{h_0\bmax h_1 < h}.
\end{aligned}
\]
Again, we have essentially the same recurrence as given by the informal
analysis.
The last inequality is valid because $A$, $B$, and $C$ are all subsets
of~$\set{*}$, and hence are either $\emptyset$ or $\set{*}$ itself, and we
take advantage of the fact that $\bigmax\emptyset = 0$.
The comparison
with~$\infty$ might be a bit perturbing.  In this
model, the labels do not contribute to the potential of a tree.  Since the
comparison in the recurrence arises from the comparison of~$x$ with an
arbitrary node label~$y$, the best we can say about the potential of~$y$ is
that it is at most~$\infty$.  
For another perspective, keep in mind that $cmp$ is monotone, so unless it is
a particularly odd function, 
$cmp(x,\infty) = \set{*}\disjunion\set{*}\disjunion\set{*}$, which forces the
recurrence to take all possible outcomes into account.  This is precisely what
we would expect in an informal analysis.

\subsubsection{Inductive types as an abstract interpretation}
\label{sec:ind-types-as-ai}

Our justification for the interpretation of sum types appealed to intuition
from abstract interpretation.  For datatypes
with structure functors that are sums of products (e.g., lists and trees), the
connection goes beyond just intuition, as it is easy to see that
not only do we have that
$\smDest\comp\smCons\geq \id$ \infruleref{beta-delta} but also
that $\smCons\comp\smDest = \id$.  This is precisely the kind of
Galois connection we would expect to see in an abstract interpretation, where
here we think of the datatype as being the abstract domain and its unfolding
to be the concrete domain.\footnote{
Of course, the domains here are not of finite height as in typical AI
analyses, but that is typically for the benefit of computability of those
analyses; that would correspond to computing the denotation of the bounding
recurrence, which is not our primary concern here.}
Intuitively, this is exactly how we think of
models of the recurrence language as performing a size abstraction on
datatypes.  Interpreting a datatype value (i.e., an application of the
constructor) as a size abstracts away information.  Destructing a size 
tells us how a value of that size may be constructed from other data, but that
data can only tell us the sizes of the substructures used in the construction.
In other words, the application of the destructor gives us more concrete
information about a size, namely, something about the composition of a value
of that size.

\subsection{Counting all constructors}
\label{sec:cac-model}

The cost of
some functions cannot be usefully described in terms of the ``usual'' notion of
size captured by the model~$\V$ of the previous section.  For example, to
usefully analyze the $\ssumtree$ function of Figure~\ref{fig:sumtree}, we
need a model in which the size of a $\ctree\cnat$ value measures both the
number of $\ctree\cnat$ constructors and the number of $\cnat$ constructors.
In this section, we give an example of how to construct such a model.  In it,
a value~$v$ of inductive type is interpreted by a function~$\phi$ such that
$\phi(\delta)$
is the size of the largest maximal subtree of~$v$
that contains only $\delta$-constructors.  
For $v : \ctree\cnat$, this means that
$\den v {}(\ctree\cnat)$ is the usual size of~$v$, $\den v {}(\cnat)$ is 
the maximum label size
of~$v$, and $\den v {}(\delta) = 0$ for $\delta\notin\set{\ctree\cnat,\cnat}$.

Because we want to distinguish between constructors for different inductive
types, 
it is convenient to use the following alternative
grammar for types and structure functors, which just spells out the closed
type production for structure functors:
\[
\begin{aligned}
\sigma &::= \alpha
       \mid \C
       \mid \cunit
       \mid \csum \sigma\sigma
       \mid \cprod\sigma\sigma
       \mid \carr\sigma\sigma
       \mid \clfp t F \\
F &::= t
  \mid \alpha
  \mid \C
  \mid \cunit
  \mid \clfp t F
  \mid \csum F F
  \mid \cprod F F
  \mid \carr \sigma F.
\end{aligned}
\tag{$*$}
\]

The content of the next proposition is just that the grammar~($*$) defines
the same words as that of Figure~\ref{fig:rec-lang-full}.

\begin{prop}\hfill
\label{prop:sf-equiv}
\begin{enumerate}
\item If $\sigma$ is a type by the grammar~($*$), then $\sigma$ is a structure
functor by the grammar~($*$).
\item $\sigma$ is a type by the grammar of Figure~\ref{fig:rec-lang-full}
iff $\sigma$ is a type by the grammar~($*$), and $F$ is a structure functor
by the grammar of Figure~\ref{fig:rec-lang-full} iff $F$ is a structure
functor by the grammar~($*$).
\end{enumerate}
\end{prop}
\begin{proof}\hfill
  \begin{enumerate}
\item Induction on~$\sigma$.
\item Induction on the $\mu$-nesting depth of $\sigma$ and~$F$.
  The main idea is that we treat $t$ as a fixed symbol, rather than a
  meta-variable ranging over a class of variables, so inside the $\mu
  t.F$ production of $F$, it is no longer possible to refer to the
  ``outer'' $t$, and the $\mu t.F$ production of $F$ always corresponds
  to a constant shape functor.\qedhere
  \end{enumerate}
\end{proof}

The type frame is the same as for the constructor-counting model
of Section~\ref{sec:constr-counting-model}; for the current model,
we write $W^\sigma$ for the interpretation of~$\sigma$.  Except for
inductive types, the clauses for $W^\sigma$ are the same as those 
for~$V^\sigma$ from
Section~\ref{sec:constr-counting-model}.   Set
$\mathcal D = \setst{\clfp t F}{\text{$F$ closed}}$ and
\begin{itemize}
\item $W^{\clfp t F} = \setst{\phi \in D \to \N_0^\infty}
                      {\phi(\clfp t F) \geq 1,
                      \text{$\delta$ not a syntactic subtype of $F$}\Implies
                      \phi(\delta) = 0}$.
\end{itemize}
To define $\smCons[F]$ and $\smDest[F]$, we define
$\smsize{F,\delta} : W^{\sfsubst F \delta}
\to (\mathcal D\to\N_0^\infty)$ 
similarly to the previous section.  The additional subscript
enables us to track which datatype is the ``main'' datatype, as the counting
is different for products for the main datatype and others.
The definition is as
follows:
\[
\begin{aligned}[t]
\smsize{t,\delta}(\phi) &= \phi \\
\smsize{\C,\delta}(x) &= \llambda\delta.0 \\
\smsize{\cunit,\delta}(*) &= \llambda\delta.0 \\
\smsize{\clfp t F,\delta}(\phi) &= \phi
\end{aligned}
\qquad
\begin{aligned}[t]
\smsize{\csum{F_0}{F_1},\delta}(X_0\disjunion X_1)
  &= \bigmax\smsize{F_0,\delta}[X_0] \bmax \bigmax\smsize{F_1,\delta}[X_1] \\
\smsize{\cprod{F_0}{F_1},\delta}(x_0, x_1)
  &= 
  \llambda\delta'.
  \begin{cases}
  \smsize{F_0,\delta}(x_0)(\delta') + \smsize{F_1,\delta}(x_1)(\delta'),
    &\delta' = \delta \\
  \smsize{F_0,\delta}(x_0)(\delta') \bmax \smsize{F_1,\delta}(x_1)(\delta'),
    &\delta' \not= \delta \\
  \end{cases}
  \\
\smsize{\carr\sigma F,\delta}(f)
  &= 
  \llambda\delta'.
  \begin{cases}
  \sum\setst{\smsize{F,\delta}(f\,x)(\delta')}{x\in D^\sigma},
    &\delta' = \delta \\
  \bigmax\setst{\smsize{F,\delta}(f\,x)(\delta')}{x\in D^\sigma},
    &\delta' \not= \delta \\
  \end{cases}
\end{aligned}
\]
Set
\[
\smCons[F](a) = \llambda\delta.
                \chi_F(\delta) + \smsize{F,\clfp t F}(a)(\delta)
\qquad
\smDest[F](\phi) = \bigmax\setst{a}{\smCons[F](a)\leq\phi}
\]
where $\chi_F(\delta) = 1$ if $\delta=\clfp t F$, $\chi_F(\delta) = 0$ 
otherwise.  Notice that for $\delta = \clfp t F$,
$\smCons[F](a)(\delta) =
1 + \smsize{F,\delta}(a)(\delta) \geq 1$, so
$\smCons[F](a)\in W^\delta$.  $\cfoldkw_\delta$ is
interpreted by~\eqref{eq:std-fold} as usual.

\subsubsection{Example:  the potential of $\cnattree$}
\label{sec:cac-nat-tree-potential}

Although we could prove a general theorem to show that $\smsize{F,\delta}$
encapsulates the description given above, seeing the details of the specific
case of $\ctree\cnat$ is more illuminating.  
To start, some notation is helpful:
set $\phi^\cnat_n\in W^\cnat$ and $\phi^\cnattree_{n,k}\in W^\cnattree$
to be the functions
\[
\begin{aligned}[t]
\phi^\cnat_n(\cnat) &= n \\
\phi^\cnat_n(\cblank) &= 0
\end{aligned}
\qquad
\begin{aligned}[t]
\phi^\cnattree_{n,k}(\cnat) &= n \\
\phi^\cnattree_{n,k}(\cnattree) &= k \\
\phi^\cnattree_{n,k}(\cblank) &= 0
\end{aligned}
\]

First we start with a useful lemma:

\begin{lemma}
\label{lem:cac-S-interp}
$\den \cnsucc {}(\phi) = \chi_{F_\cnat} + \phi$, and in particular,
$\den\cnsucc{}(\phi^\cnat_k) = \phi^\cnat_{k+1}$.
\end{lemma}
\begin{proof}
\begin{multline*}
\den{\cnsucc}{}(\phi)
  = \chi_{F_\cnat} + 
     \smsize{F_\cnat,\cnat}(\smInj^1(\phi))
  = \chi_{F_\cnat} +
     \smsize{F_\cnat,\cnat}(\emptyset\disjunion\downset\phi) = \\
    \chi_{F_\cnat} +
     (\bigmax\smsize{t,\cnat}[\downset\phi])
  = \chi_{F_\cnat} +
     (\bigmax(\downset\phi))
  = \chi_{F_\cnat} + \phi.\qedhere
\end{multline*}
\end{proof}

Now set
$\cnnum n = \cnsucc(\dotsc(\cnsucc\,\cnzero)\dotsc) \oftype\cnat$ 
($n$ $\cnsucc$s).  
We will show that $\den{\cnnum n}{} = \phi^\cnat_{n+1}$
by induction on $n$.  For $n=0$,
\begin{multline*}
\den{\cnnum 0}{}
  = \smCons[F_\cnat](\smInj^0 *)
  = \smCons[F_\cnat](\set{*}\disjunion\emptyset) = \\
    \chi_{F_\cnat} + 
     \smsize{F_\cnat,\cnat}(\set{*}\disjunion\emptyset)
  = \chi_{F_\cnat} + (\llambda\delta.0)
  = \chi_{F_\cnat}
  = \phi^\cnat_1.
\end{multline*}
And for $n\geq 0$, we use Lemma~\ref{lem:cac-S-interp} to show that
\[
\den{\cnnum{n+1}}{}
  = (\den{\cnsucc}{}\,\den{\cnnum n}{})
  = \den{\cnsucc}{}(\phi^\cnat_{n+1})
  = \phi^\cnat_{n + 2}.
\]

Now let us consider closed $\ctree\cnat$ expressions built up using only
$\ctree\cnat$ and $\cnat$ constructors---i.e., $\cnat$-labeled binary trees.
We show that if $t$ is such a tree, then
$\den{t}{} = \phi^\cnattree_{m,k}$, where 
$m =\bigmax\setst{1+n}{\text{$n$ a label in $t$}}$ and $k$ is the number of
$\cnattree$ constructors in~$t$.
In the following calculations, we will save a bit of space by writing
$\smsize{F}$ for $\smsize{F,\ctree\cnat}$.
For $\ctemp$, the argument is essentially the same as for the analysis
of $\den{\cnnum 0}{}$, noting that
$\bigmax\setst{1+n}{\text{$n$ a label in $\ctemp$}} = \bigmax\emptyset = 0$.  
For the inductive step, assume that
$\den{t_i}{} = \phi^\cnattree_{n_i,k_i}$, so our goal is to show that
$\den{\ctnode(\cnnum n, t_0, t_1)}{} = 
\phi^\cnattree_{(n+1)\bmax n_0\bmax n_1,1+k_0+k_1}$:
\begin{align*}
\den{\ctnode(\cnnum n, t_0, t_1)}{}
  &= \smCons[F_{\ctree\cnat}](\smInj^1(
       \den{\cnnum n}{}, \den{t_0}{}, \den{t_1}{}))  \\
  &= \smCons[F_{\ctree\cnat}](
       \emptyset\disjunion\downset(\phi^\cnat_{n+1}, 
                                   \phi^\cnattree_{n_0,k_0}, 
                                   \phi^\cnattree_{n_1,k_1})) \\
  &= \chi_{F_{\ctree\cnat}} +
     \smsize{F_{\ctree\cnat}}(
       \emptyset\disjunion\downset(\phi^\cnat_{n+1}, 
                                   \phi^\cnattree_{n_0,k_0}, 
                                   \phi^\cnattree_{n_1,k_1})) \\
  &= \chi_{F_{\ctree\cnat}} +
     \bigmax\smsize{\cprod\cnat{\cprod t t}}[
       \downset(\phi^\cnat_{n+1}, 
                                   \phi^\cnattree_{n_0,k_0}, 
                                   \phi^\cnattree_{n_1,k_1})
     ].
\end{align*}
If $(\phi',\phi_0',\phi_1')\in\downset(\phi^\cnat_{n+1}, 
                                       \phi^\cnattree_{n_0,k_0}, 
                                       \phi^\cnattree_{n_1,k_1})$, then
\begin{align*}
\smsize{{\cprod\cnat{\cprod t t}}}(\phi',\phi_0',\phi_1')(\delta)
  &= \begin{cases}
     \smsize{\cnat}(\phi')(\delta) +
     \smsize{t}(\phi_0')(\delta) +
     \smsize{t}(\phi_1')(\delta),
       &\delta=\cnattree \\
     \smsize{\cnat}(\phi')(\delta) \bmax
     \smsize{t}(\phi_0')(\delta) \bmax
     \smsize{t}(\phi_1')(\delta),
       &\delta\not=\cnattree
     \end{cases} \\
  &= \begin{cases}
     \phi'(\cnattree) + \phi_0'(\cnattree) + \phi_1'(\cnattree),
       &\delta = \cnattree \\
     \phi'(\delta) \bmax \phi_0'(\delta) \bmax \phi_1'(\delta),
       &\delta \not= \cnattree
     \end{cases}
\end{align*}
and hence the computation of 
$\den{\ctnode(\cnnum n, t_0, t_1)}{}(\delta)$ proceeds as:
\begin{align*}
{}
  &= \begin{cases}
     1 +
     \bigmax
     \Biggl\{
     \begin{aligned}
     &\phi'(\cnattree) + \phi_0'(\cnattree) + \phi_1'(\cnattree)\setseparator
     \\
     &\qquad{(\phi',\phi_0',\phi_1')\in\downset(\phi^\cnat_{n+1},\phi^\cnattree_{n_0,k_0},\phi^\cnattree_{n_1,k_1})}
     \end{aligned}
     \Biggr\}, &\delta = \cnattree \\
     \bigmax
     \Biggl\{
     \begin{aligned}
     &\phi'(\delta) \bmax \phi_0'(\delta) \bmax \phi_1'(\delta)\setseparator
     \\
     &{(\phi',\phi_0',\phi_1')\in\downset(\phi^\cnat_{n+1},\phi^\cnattree_{n_0,k_0},\phi^\cnattree_{n_1,k_1})}
     \end{aligned}
     \Biggr\}, &\delta \not= \cnattree
     \end{cases} \\
  &= \begin{cases}
     1 + k_0 + k_1,&\delta = \cnattree \\
     (n+1) \bmax n_0 \bmax n_1,&\delta = \cnat \\
     0,&\text{otherwise}
     \end{cases} \\
  &= \phi^{\ctree\cnat}_{(n+1)\bmax n_0 \bmax n_1, 1 + k_0 + k_1}(\delta).
\end{align*}

We have simplified descriptions of recurrences that are analogous to those of
Prop.~\ref{prop:mcfold}:

\begin{prop}\hfill
\label{prop:ac-fold}
\begin{enumerate}
\item If
$f\,\phi = \den{\cfoldnat\rho{x}{e_{\cnzero}}{r.e_{\cnsucc}}}
               {\bindin\eta{x}{\phi}}$, then
\[
f\,\phi^\cnat_1
  = \den{e_\cnzero}\eta
\qquad
\begin{aligned}[t]
f\,\phi^\cnat_n
  &= \den{e_\cnzero}\eta
     \bmax
     \bigmax
     \setst{\den{e_\cnsucc}{\bindin\eta{r}{\phi^\cnat_j}}}{j<n} \\
  &= \den{e_\cnzero}\eta
     \bmax
     \den{e_\cnsucc}{\bindin\eta{r}{\phi^\cnat_{n-1}}}
     & & (n>1)
\end{aligned}
\]
\item If
$f\,\phi =
\den{\cfoldtree{\cnat}{\rho}{x}{e_\ctemp}{(x,r_0,r_1).e_\ctnode}}
    {\bindin\eta x \phi}$,
then
\begin{align*}
f\,\phi^\cnattree_{n,1}
  &= \den{e_\ctemp}\eta \\
f\,\phi^\cnattree_{n,k}
  &= \den{e_\ctemp}{\eta} \bmax
     \bigmax
     \setst*[Big]
     {
       \den{e_\ctnode}
           {\bindin\eta
                    {x,r_0,r_1}
                    {\phi^\cnat_{n'},f\,\phi^{\cnattree}_{n_0,k_0},
                                     f\,\phi^{\cnattree}_{n_1,k_1}}}
     }
     {
       n'\bmax n_0\bmax n_1\leq n, 1 + k_0 + k_1 \leq k
     }
     & & (n>1)
\end{align*}
\end{enumerate}
\end{prop}

\subsubsection{Example:  summing the nodes of a $\snattree$}
\label{sec:cac-summing-nodes}

\begin{figure}
\begin{align*}
\splus &=
  \slam{x,y}{\sfoldnat x y {(r).\snsucc\,(\sforce r)}} \\
\ssumtree &=
  \slam 
    t
    {
    \sfoldtree* 
      \snat
      t
      \snzero
      {(x, r_0, r_1).\splus\,x\,{(\splus\,{(\sforce{r_0})}\,{(\sforce{r_1})})}}
    }
\\
\cplus &=
  \clam
    {x,y}
    { \cfoldnat {\ctrans\snat} x {\cpair 1 y} {(r).\cpair{1+\ccost r}{\cnsucc\,\cpot r}}} \\
\csumtree &=
  \clam
    {t}
    {
      \cfoldtree*
        \cnat
        {\ctrans\snat}
        t
        {\cpair 1 \cnzero}
        {%
          \begin{aligned}[t]
          (x,r_0, r_1).%
            & (1 + \ccost{r_0} + \ccost{r_1} +
            \ccost*{\cplus\,\cpot{r_0}\,\cpot{r_1}}) +_c {} \\
            & \cplus\,x\,\cpot*{\cplus\,\cpot{r_0}\,\cpot{r_1}}
          \end{aligned}
        }
    }
\end{align*}
\caption{A function that sums the nodes of a $\snattree$.}
\label{fig:sumtree}
\end{figure}

\bgroup
\newcommand{\sst}{\skeyw{st}}
\newcommand{\cst}{\ckeyw{st}}
Let us use this model to analyze the function
$\ssumtree : \sarr\snattree\snat$ that sums the nodes of a $\snattree$.
Its definition is given in Figure~\ref{fig:sumtree}, along the relevant
extracted recurrences.  An informal analysis might proceed as follows.
Because the cost of $\ssumtree$ depends on both the cost
and size of the result of $\splus$ as well as the size of the results of
the recursive calls, we must extract recurrences for all of these.  If
$S_\splus(m, n)$ and $T_\splus(m, n)$ are the size of the result and the cost 
of $\splus(\snn {m-1}, \snn {n-1})$, respectively (recall from
Figure~\ref{fig:src-lang-examples} that $\snn n$ is the source language numeral
for~$n$), then an informal analysis yields the recurrences
\[
\begin{aligned}[t]
S_\splus(1, n) &= n \\
S_\splus(m, n) &= 1 + S_\splus(m-1, n)
\end{aligned}
\qquad
\begin{aligned}[t]
T_\splus(1, n) &= 1 \\
T_\splus(m, n) &= 1 + T_\splus(m-1, n).
\end{aligned}
\]
Similarly, if $S_\sst(n, k)$ and $T_\sst(n, k)$ are the size of the result and
the cost of $\ssumtree(t)$ when $t$ has maximum label size~$n$ and size~$k$, 
we end up with the recurrences
\begin{align*}
S_\sst(n, 1) &= 1 \\
S_\sst(n, k) 
  &= \bigmax
     \setst{S_\splus(n, S_\splus(S_\sst(n, k_0), S_\sst(n, k_1)))}
           {k_0+k_1 < k}
\end{align*}
and
\begin{align*}
T_\sst(n, 1) &= 1 \\
T_\sst(n, k)
  &= 
     \begin{aligned}[t]
     \bigmax
     \{
     &T_\sst(n, k_0) + T_\sst(n, k_1) + 
     T_\splus(S_\sst(n, k_0), S_\sst(n, k_1)) +  \\
     &\qquad T_\splus(n, S_\splus(S_\sst(n, k_0), S_\sst(n, k_1)))
     \mid k_0 + k_1 < k
     \}.
     \end{aligned}
\end{align*}
To solve these recurrences, one would first use any standard technique to
conclude that $S_\splus(m, n) = m+n-1$ and
$T_\splus(m, n) = m$ to simplify the recurrence clauses for~$S_\sst$
and~$T_\sst$, then establish bounds on the latter by induction.  However, the
solution of the recurrences is not our focus here, but rather the justified
extraction of them.

Now let us turn to our formal analysis.  Set
$\tilde S_\cplus(\phi,\phi') = (\den{\cplus}{}\,\phi\,\phi')_p$.
Then making use of Prop.~\ref{prop:ac-fold},
$\tilde S_\cplus(\phi^\cnat_1,\phi') = \phi'$ and for $m>1$,
\begin{align*}
\tilde S_\cplus(\phi^\cnat_m,\phi')
  &= \phi'
     \bmax
     \den{\cnsucc(\cpot r)}
         {\env{\bindto r {\den{\cplus}{}\,\phi^\cnat_{m-1}\,\phi'}}} \\
  &= \phi'
     \bmax
     (\chi_{F_\cnat} + \cpot*{{\den{\cplus}{}\,\phi^\cnat_{m-1}\,\phi'}})
     & & \text{(Prop.~\ref{lem:cac-S-interp})} \\
  &= \phi'
     \bmax
     (\chi_{F_\cnat} + \tilde S_\cplus(\phi^\cnat_{m-1},\phi'))
\end{align*}
This recursive description of~$\tilde S_\cplus$ is sufficient to prove that
$\tilde S_\cplus(\phi^\cnat_m,\phi') \geq \phi'$, and so we can conclude the
reasoning with
\[
\tilde S_\cplus(\phi^\cnat_m,\phi') =
\chi_{F_\cnat} + \tilde S_\cplus(\phi^\cnat_{m-1},\phi')
\]
and so in particular
\[
\tilde S_\cplus(\phi^\cnat_1,\phi^\cnat_n) = \phi^\cnat_n
\qquad
\tilde S_\cplus(\phi^\cnat_m,\phi^\cnat_n) =
\chi_{F_\cnat} + \tilde S_\cplus(\phi^\cnat_{m-1},\phi^\cnat_n),
\]
recurrences that are equivalent to those derived informally.
The analysis of
$\tilde T_\cplus(\phi,\phi') = (\den{\cplus}{}\,\phi\,\phi')_c$ is similar and
results in the recurrence
\[
\tilde T_\cplus(\phi^\cnat_1,\phi^\cnat_n) = 1
\qquad
\tilde T_\cplus(\phi^\cnat_m,\phi^\cnat_n) =
1 + \tilde T_\cplus(\phi^\cnat_{m-1},\phi^\cnat_n).
\]

Now set $\tilde S_\cst(\phi) = (\den{\csumtree}{}\,\phi)_p$.  Making
use of Prop.~\ref{prop:ac-fold}, 
$\tilde S_\cst(\phi^\cnattree_{1,k}) = \phi^\cnat_1$ and for
$k>1$,
\newcommand{\tS}{\tilde S}
\newcommand{\tT}{\tilde T}
\begin{align*}
\tS_\cst(\phi^\cnattree_{n,k})
  &= \phi^\cnat_1
     \bmax
     \bigmax
     \begin{aligned}[t]
     & \bigl\{(\den{\cplus\,x\,\cpot*{\cplus\,\cpot{r_0}\,\cpot{r_1}}}
            {\env{\bindto
                    {x,r_i}
                    {
                      \phi^\cnat_{n'},
                      \den{\csumtree}{}\,\phi^\cnattree_{n_i,k_i}
                    }
                  }
            })_p \\
     & \qquad \setseparator n'\bmax n_0 \bmax n_1 \leq n, k_0+k_1 < k \bigr\}
     \end{aligned} \\
  &= \phi^\cnat_1
     \bmax
     \bigmax
     \begin{aligned}[t]
     & \bigl\{
         (\den{\cplus}{}\,\phi^\cnat_{n'}(
           \den{\cplus}{}\,(\tS_\cst(\phi^\cnattree_{n_0,k_0}))\,
                           (\tS_\cst(\phi^\cnattree_{n_1,k_1}))
         )_p)_p \\
     & \qquad \setseparator n'\bmax n_0 \bmax n_1 \leq n, k_0+k_1 < k \bigr\}
     \end{aligned} \\
  &= \phi^\cnat_1
     \bmax
     \bigmax
     \begin{aligned}[t]
     & \bigl\{
         \tS_\cplus(
           \phi^\cnat_{n'},
           \tS_\cplus(
             \tS_\cst(\phi^\cnattree_{n_0,k_0}),
             \tS_\cst(\phi^\cnattree_{n_1,k_1})
           )
         ) \\
     & \qquad \setseparator n'\bmax n_0 \bmax n_1 \leq n, k_0+k_1 < k \bigr\}
     \end{aligned}
\end{align*}
Since $\phi^\cnat_1$ is the bottom element of $W^\cnat$ and we can prove from
this recurrence that $\tS_\cst(\phi^\cnattree_{n,k})$ is monotone with respect
to~$n$, we can conclude this reasoning with
\[
\tS_\cst(\phi^\cnattree_{n,k}) =
\bigmax\setst{
         \tS_\cplus(
           \phi^\cnat_{n},
           \tS_\cplus(
             \tS_\cst(\phi^\cnattree_{n,k_0}),
             \tS_\cst(\phi^\cnattree_{n,k_1})
           )
         )}{k_0+k_1 < k},
\]
which is analogous to the recurrence we derived informally.  The analysis
of $\tilde T_\cst(\phi^\cnattree_{n,k}) = 
(\den{\csumtree}{}\,\phi^\cnattree_{n,k})_c$ is similar and leads to
\[
\tilde T_\cst(\phi^\cnattree_{1,k}) = 1
\qquad
\tilde T_\cst(\phi^\cnattree_{n,k}) =
\bigmax
\setst*[Big]
  {
    \begin{aligned}[t]
    1 + 
    & \tT_\cst(\phi^\cnattree_{n,k_0}) +
    \tT_\cst(\phi^\cnattree_{n,k_1}) + \\
  & \tT_\cplus(\tS_\cst(\phi^\cnattree_{n,k_0}),
                     \tS_\cst(\phi^\cnattree_{n,k_1})) + \\
  & \tT_\cplus(\phi^\cnat_n,
                     \tS_\cplus(\tS_\cst(\phi^\cnattree_{n,k_0}),
                                \tS_\cst(\phi^\cnattree_{n,k_1}))) \\
    \end{aligned}
  }
  {
    k_0 + k_1 < k
  }
\]
As a final note, in order to obtain the desired final form, we sometimes had
to do some reasoning about the function on the basis of its recurrence, such
as proving that the function is monotone.  In fact, such reasoning is almost
always required in the informal analysis as well, even though we typically
gloss over such points when analyzing algorithms.
\egroup

In may be helpful to contrast this analysis with the
interpretation of $\cplus$ and $\csumtree$ in the model of 
Section~\ref{sec:constr-counting-model}.  Since $\cnat$ values involve no
other datatype constructors, the interpretation of $\cplus$ is essentially
just the same, only requiring less notation to write down.  However, the
cost component 
of~$\den{\csumtree}{\env{\substfor n t}}$ is less helpful.  Because 
the model of Section~\ref{sec:constr-counting-model} only accounts for the
tree constructors, it does not account for the
sizes of the node labels, and so
this computation includes
the cost component
of~$\den{\cplus\,x\,\cpot*{\cplus\,\cpot{r_0}\,\cpot{r_1}}}
        {\env{\substfor{\infty,\dotsc}{x,r_0, r_1}}}$
and this will result in a bound of~$\infty$
(cf.\ to the occurrence of~$\infty$ in the analysis of $\sbstmem$ in the
previous section, which did no harm there).  
This is correct as a bound.  It reflects a cost analysis in which we
have decided that we are counting each recursive call as a computation step,
but then analyze a program in which data values whose size we ignore is the
source of some recursive calls.  However, this rather poor choice of size for
this particular context yields a very weak bound, and so shows more generally
that the choice of model does really matter.

\subsection{Size abstraction and polymorphism: merging the
constructor-counting models} 
\label{sec:merged-model}

Let us make a couple of observations about the previous two sections.  It
seems at least intuitive that counting only the main constructors is a more
abstract notion of size than counting all constructors.  And it also seems
that even if we are working in the model of Section~\ref{sec:cac-model}, if we
have a polymorphic function in hand, it ought to be analyzable by just
counting main constructors.  This leads to the idea that if we have a model in
hand (such as counting all constructors), then at least in some cases, it
ought to be possible to interpret polymorphic recurrences so that the
potentials arise from a more abstract notion of size than that given by the
model.  We give an example of how that might be done now.

\begin{defn}
Suppose $\U = (\Usm, \Ulg, \set{D^\sigma})$ and $\U' = (\Usm, \Ulg,
\set{D'}^\sigma)$ are two models of the recurrence language, both based
on the (same extension of the) standard type frame.  We say that 
\emph{$\U'$ is an abstraction of~$\U$}, or \emph{$\U$ is a concretization
of~$\U'$}, if for every $\sigma\in\Usm$ there are functions
\[
\begin{tikzcd}
D^\sigma \arrow[bend left]{r}{\smabs_\sigma}      
  & {D'}^\sigma \arrow[bend left]{l}{\smconc_\sigma}
\end{tikzcd}
\]
such that for all~$\sigma$, $\smconc_\sigma$ is monotone,
$\smconc_\sigma\comp\smabs_\sigma\geq\id_{D^\sigma}$ and
$\smabs_\sigma\comp\smconc_\sigma = \id_{D'^\sigma}$.
\end{defn}

\begin{defn}
Suppose $\U' = (\Usm, \Ulg, \set{D'}^\sigma)$ is an abstraction of
$\U = (\Usm, \Ulg, \set{D^\sigma})$.
The \emph{polymorphic abstraction of~$\U$ relative to~$\U'$} is the 
model~$\U\to\U' = (\Usm, \Ulg, \set{B^\sigma})$
that is defined as follows:
\begin{itemize}
\item For $\sigma\in\Usm$, $B^\sigma = D^\sigma$, with the semantic functions
for small types taken from~$\U$.
\item For $\tau\in\Ulg\setminus\Usm$, $B^\tau = {D'}^\tau$, where:
\begin{itemize}
\item 
If $\rho$ is quantifier-free and $\fv(\rho)\subseteq\set{\alpha}$, then
\begin{align*}
\dom(\smTyAbs[\llambda\sigma.\substin\rho\sigma\alpha]^{\U\to\U'}) &=
\setst{f\in\prod_{\sigma\in\Usm} (D')^{\substin\rho\sigma\alpha}}
      {\llambda\sigma.\smabs_{\substin\rho\sigma\alpha}(f\,\sigma)\in\dom(\smTyAbs[\llambda\sigma.\substin\rho\sigma\alpha]^{\U'})}
      \\
\smTyAbs[\llambda\sigma.\substin\rho\sigma\alpha](f)
  &= \smTyAbs[\llambda\sigma.\substin\rho\sigma\alpha]^{\U'}(
       \llambda\sigma.\smabs_{\substin\rho\sigma\alpha}(
         f\,\sigma
       )
     ) \\
\smTyApp[\llambda\sigma.\substin\rho\sigma\alpha](f)
  &= \llambda\sigma.\smconc_{\substin\rho\sigma\alpha}(
       \smTyApp[\llambda\sigma.\substin\rho\sigma\alpha]^{\U'}\,f\,\sigma
     )
\end{align*}
\item If $\tau$ is not quantifier-free and $\fv(\tau)\subseteq\set{\alpha}$, 
then we take
$\smTyAbs[\llambda\sigma.\substin\tau\sigma\alpha] =
\smTyAbs[\llambda\sigma.\substin\tau\sigma\alpha]^{\U'}$ and
$\smTyApp[\llambda\sigma.\substin\tau\sigma\alpha] =
\smTyApp[\llambda\sigma.\substin\tau\sigma\alpha]^{\U'}$.
\end{itemize}
\end{itemize}
\end{defn}

\begin{prop}\hfill
\label{prop:poly-abs-is-model}
\begin{enumerate}
\item If $\U$ and~$\U'$ are applicative structures, then $\U\to\U'$ is an
applicative structure.
\item If $\U$ and $\U'$ are pre-models such that
whenever $\typejudge\cctx e\rho$ and $\eta$ is a $\cctx$-environment,
$\llambda\sigma.\den{\smabs_{\substin\rho\sigma\alpha}(e)}
                    {\bindin\eta\alpha\sigma} \in\dom\smTyAbs^{\U'}$, 
then $\U\to\U'$ is a pre-model.
\end{enumerate}
\end{prop}
\begin{proof}
The only non-trivial verification is that when $\rho$ is quantifier-free and
$\fv(\rho)\subseteq\set{\alpha}$, 
$\smTyApp[\llambda\sigma.\substin\rho\sigma\alpha](
   \smTyAbs[\llambda\sigma.\substin\rho\sigma\alpha]\,f
 )\geq f$:
\[
\begin{split}
\smTyApp[\llambda\sigma.\substin\rho\sigma\alpha]&(
  \smTyAbs[\llambda\sigma.\substin\rho\sigma\alpha]\,f
)\,\sigma \\
  &= \smTyApp[\llambda\sigma.\substin\rho\sigma\alpha](
       \smTyAbs[\llambda\sigma.\substin\rho\sigma\alpha]^{\U'}(
         \llambda\sigma.\smabs_{\substin\rho\sigma\alpha}(
           f\,\sigma
         )
       )
     )\,\sigma \\
  &= (\llambda\sigma.\smconc_{\substin\rho\sigma\alpha}(
       \smTyApp[\llambda\sigma.\substin\rho\sigma\alpha]^{\U'}(
         \smTyAbs[\llambda\sigma.\substin\rho\sigma\alpha]^{\U'}(
           \llambda\sigma.\smabs_{\substin\rho\sigma\alpha}(
             f\,\sigma
           )
         )
       )\,\sigma
     ))\,\sigma \\
  &= \smconc_{\substin\rho\sigma\alpha}(
       \smTyApp[\llambda\sigma.\substin\rho\sigma\alpha]^{\U'}(
         \smTyAbs[\llambda\sigma.\substin\rho\sigma\alpha]^{\U'}(
           \llambda\sigma.\smabs_{\substin\rho\sigma\alpha}(
             f\,\sigma
           )
         )
       )\,\sigma
     ) \\
  &\geq 
     \smconc_{\substin\rho\sigma\alpha}((
       \llambda\sigma.\smabs_{\substin\rho\sigma\alpha}(
         f\,\sigma
       )
     )\,\sigma) \\
  &\geq
     \smconc_{\substin\rho\sigma\alpha}(
       \smabs_{\substin\rho\sigma\alpha}(
         f\,\sigma
       )
     ) \\
  &\geq f\,\sigma. \qedhere
\end{split}
\]
\end{proof}

As an example, we define abstraction and concretization functions in
Figure~\ref{fig:abs-conc-functions} that show that the main constructor
counting model~$\V$ from Section~\ref{sec:constr-counting-model} is an
abstraction of the all-constructor counting model~$\W$ from
Section~\ref{sec:cac-model}.

\begin{figure}
\begin{align*}
\smabs_\sigma &: W^\sigma\to V^\sigma
    & \smconc_\sigma &: V^\sigma\to W^\sigma \\
\smabs_\cunit(*) &= \mathord*
    & \smconc_\cunit(*) &= \mathord* \\
\smabs_\C(n) &= n
    & \smconc_\C(n) &= n \\
\smabs_{\clfp t F}(\phi) &= \phi(\clfp t F)
    & \smconc_{\clfp t F}(n) &= \llambda\delta.\begin{cases}
                                              n,&\delta = \clfp t F \\
                                              \infty,&\delta\not=\clfp t F
                                              \end{cases} \\
\smabs_{\csum{\sigma_0}{\sigma_1}}(X_0\disjunion X_1)
  &= \downset\smabs_{\sigma_0}[X_0] \disjunion \downset\smabs_{\sigma_1}[X_1]
    & \smconc_{\csum{\sigma_0}{\sigma_1}}(Y_0\disjunion Y_1)
      &= \downset\smconc_{\sigma_0}[Y_0] \disjunion 
         \downset\smconc_{\sigma_1}[Y_1] \\
\smabs_{\cprod{\sigma_0}{\sigma_1}}(x_0, x_1)
  &= (\smabs_{\sigma_0}(x_0), \smabs_{\sigma_1}(x_1))
    & \smconc_{\cprod{\sigma_0}{\sigma_1}}(y_0, y_1)
      &= (\smconc_{\sigma_0}(y_0), \smconc_{\sigma_1}(y_1)) \\
\smabs_{\carr\rho\sigma}(f) &= \smabs_\sigma\comp f\comp\smconc_\rho
    & \smconc_{\carr\rho\sigma}(f) &= \smconc_\sigma\comp f \comp \smabs_\rho
\end{align*}
\caption{Abstraction and concretization functions that relate the 
all-constructor (concrete) and main-constructor (abstract) models.}
\label{fig:abs-conc-functions}
\end{figure}

\begin{prop}\hfill
\label{prop:ai-relation}
\begin{enumerate}
\item $\smabs_\sigma$ and $\smconc_\sigma$ are monotone for all~$\sigma$.
\item $\smabs_\sigma\comp\smconc_\sigma = \id$ and
$\smconc_\sigma\comp\smabs_\sigma \geq\id$.
\end{enumerate}
\end{prop}
\begin{proof}\hfill
\begin{enumerate}
\item By induction on~$\sigma$.
\item By induction on~$\sigma$; we just do $\sigma = 
\csum{\sigma_0}{\sigma_1}$.
Let us write $\smabs$ for $\smabs_{\csum{\sigma_0}{\sigma_1}}$,
$\smabs_i$ for $\smabs_{\sigma_i}$, and similarly for~$\smconc$.
To see that $\smabs\comp\smconc = \id$, notice that
$(\smabs\comp\smconc)(Y_0\disjunion Y_1) =
\downset\smabs_0[\downset\smconc_0[Y_0]]\disjunion
\downset\smabs_1[\downset\smconc_1[Y_1]]$, so
if $a'\in(\smabs\comp\smconc)(Y_0\disjunion Y_1)$, then there are~$i$, $b$, and
$a\in Y_i$ such that
$a'\leq\smabs_i(b)$ and $b\leq \smconc_i(a)$, and hence
$a'\leq\smabs_i(\smconc_i(a)) = a$ (by monotonicity
and the induction hypothesis).  But since~$Y_i$ is
downward closed, $a'\in Y_i$, so 
$(\smabs\comp\smconc)(Y_0\disjunion Y_1) \subseteq Y_0\disjunion Y_1$.
To see that $\smabs\comp\smconc\geq\id$, notice that if
$a\in Y_0\disjunion Y_1$, then $a\in Y_i$ for some~$i$, and hence
$a = \smabs_i(\smconc_i(a))\in
\downset\smabs_0[\downset\smconc_0[Y_0]]\disjunion
\downset\smabs_1[\downset\smconc_1[Y_1]] = 
(\smabs\comp\smconc)(Y_0\disjunion Y_1)$, so
$Y_0\disjunion Y_1 \subseteq (\smabs\comp\smconc)(Y_0\disjunion Y_1)$.

To see that $\smconc\comp\smabs\geq\id$, suppose
$b\in X_i$.  Then by the induction hypothesis
$b\leq(\smconc_i\comp\smabs_i)(b)$, and by unraveling the definition,
$(\smconc_i\comp\smabs_i)(b)\in\downset\smconc_i[\downset\smabs_i[X_i]]$.
Since $\downset\smconc_i[\downset\smabs_i[X_i]]$ is downward-closed,
$b\in\downset\smconc_i[\downset\smabs_i[X_i]]
  \subseteq(\smconc\comp\smabs)(X_0\disjunion X_1)$.  \qedhere
\end{enumerate}
\end{proof}

\begin{prop}
\label{prop:w-to-v-is-model}
$\W\to\V$ is a model.
\end{prop}
\begin{proof}
From Props.~\ref{prop:poly-abs-is-model} and~\ref{prop:ai-relation} 
and the fact that $\smTyAbs^\V$ is total.
\end{proof}

The definition of the abstraction and concretization functions in
Figure~\ref{fig:abs-conc-functions} looks fairly canonical, so a natural
question is whether for any two models of the recurrence language one
can extend given functions on the interpretations of base types to all
small types.  In fact these definitions are an instance of a general
pattern, but to state the pattern we will need a few definitions.  A
2-category is a generalization of a category with a notion of
morphism-between-morphism: if $X$ and $Y$ are objects, and $f,g : X
\longrightarrow Y$ are morphisms, then we will write $f
\le_{\mathcal{C}} g : X \longrightarrow Y$ for a 2-cell from $f$ to $g$.
We will mainly consider the 2-category $\mathbf{Preorder}$, whose
objects $X,Y$ are preordered sets, whose morphisms $f : X
\longrightarrow Y$ are monotone functions, and whose 2-cells $f \le g :
X \longrightarrow Y$ are bounds $\forall x:X. f(x) \le_Y g(x)$.  We will
also need $\mathbf{Preorder}^{op}$ (the 1-cell dual of
$\mathbf{Preorder}$): the objects are again preorders, a 1-cell $X
\longrightarrow_{\mathbf{Preorder}^{op}} Y$ in $\mathbf{Preorder}^{op}$
is a 1-cell in $Y \longrightarrow_{\mathbf{Preorder}} X$, i.e. a
monotone function $Y \to X$, but the 2-cells $f \le_{Preorder^{op}} g :
X \longrightarrow_{Preorder^{op}} Y$ are still the 2-cells $f
\le_{\mathbf{Preorder}} g : Y \longrightarrow_{\mathbf{Preorder}} X$,
i.e. $\forall y:Y. f(y) \le_X g(y)$.  A standard construction is to take
the cartesian product of two 2-categories, where the objects, 1-cells,
and 2-cells are given pointwise; in particular we will consider
$\mathbf{Preorder} \times \mathbf{Preorder}$ and $\mathbf{Preorder}^{op}
\times \mathbf{Preorder}$.  A 2-functor $F : \mathcal{C} \to
\mathcal{D}$ between 2-categories acts on objects, 1-cells (preserving
identity and composition either strictly or up to 2-cell isomorphism),
and 2-cells. For example, a (strict) 2-functor $F : \mathbf{Preorder}
\to \mathbf{Preorder}$ consists of (0) for each preorder $X$, a preorder
$F(X)$; (1) for each monotone function $f : X \to Y$, a monotone
function $F(f) : F(X) \to F(Y)$ such that $F(id) = id$ and $F(g \circ f)
= F(g) \circ F(f)$; (2) if $\forall x:X. f(x) \le_Y g(x)$ then $\forall
w:F(X). F(f) w \le_{F(Y)} F(g) w$.  I.e. $F$ sends preorders to
preorders and monotone functions to monotone functions, in such a way
that if $g$ bounds $f$ then $F(g)$ bounds $F(f)$.  

An abstract interpretation in the sense above is often called a
\emph{Galois insertion}, which is a \emph{reflection in
  $\mathbf{Preorder}$}: a (strict) reflection of $A$ into $C$ consists
of a pair of 1-cells $\smabs \dashv \smconc$ where $\smabs : C \to A$ and
$\smconc : A \to C$, with an equality $\smabs \circ \smconc = id_A$ and a
2-cell $id_C \le_C \smconc \circ \smabs$.  A standard observation is that
\emph{any} 2-functor $F : \mathcal{C} \to \mathcal{D}$ preserves
reflections (this is used, for example, in domain
theory~\citep{smythplotkin82recursive}): if $\smabs \dashv \smconc$ is a reflection
then $F(\smabs) \dashv F(\smconc)$ is a reflection between $F(C)$ and
$F(A)$.  Applying $F$ to the equality $\smabs \circ \smconc = id_A$ and
using strict preservation of identity and composition gives $F(\smabs)
\circ F(\smconc) = id_{F(A)}$, and using the action on 2-cells of $F$ on
$id_C \le_C \smconc \circ \smabs$ (and again preservation of identity and
composition) gives $id_{F(C)} \le_{F(C)} F(\smconc) \circ F(\smabs)$.

This all means that we can lift the abstraction and concretization from
base types to any type constructor that extends to a 2-functor.
The product of preorders $X \times Y$ is the action on objects of a
functor $\mathbf{Preorder} \times \mathbf{Preorder} \to
\mathbf{Preorder}$, where the action on maps $f_0 : X_0 \to X_0'$ and $g
: X_1 \to X_1'$ is given by
\[
f_0 \times f_1 : X_0 \times X_1 \to X_0' \times X_1' := z \mapsto \langle f_0 (\pi_0 z), f_1(\pi_1 z) \rangle
\]
This acts on 2-cells (preserves bounds) because pairing and application
are monotone operations.  To show that it preserves composition, we need
a full $\beta$-reduction equation, and to
show that it preserves identity, we also need the corresponding
$\eta$/surjective pairing equation.  However, these are true for the
standard cartesian product of preorders.  A reflection in
$\mathbf{Preorder} \times \mathbf{Preorder}$ is a pair of reflections
for each component.  Unwinding these definitions gives the definitions
of $\smabs_{\sigma_0 \times \sigma_1}$ and  $\smconc_{\sigma_0 \times
  \sigma_1}$ in Figure~\ref{fig:abs-conc-functions}.

The case of sums is more interesting.  The standard coproduct of
preorders $X + Y$ is the disjoint union $X \sqcup Y$ ordered as defined
above.  This extends to a 2-functor $\mathbf{Preorder} \times
\mathbf{Preorder} \to \mathbf{Preorder}$ with $f_0 + f_1$ defined via
case-analysis.  This is bound-preserving because the branches of a
case-analysis (on the standard coproduct in preorders) are a monotone
position, and preserves identity/composition if we have $\beta\eta$
equations for case-analysis, which $X+Y$ does.

In the models under consideration, we do not define
$D^{\sigma_0+\sigma_1}$ to be $D^{\sigma_0} + D^{\sigma_1}$, but
$\ordideal{(D^{\sigma_0} + D^{\sigma_1})}$.  However, it is also the
case that $\ordideal$ is a 2-functor $\mathbf{Preorder} \to
\mathbf{Preorder}$: $\ordideal{f} : \ordideal{X} \to \ordideal{Y}$ is
$\downarrow \{ f(x) : x \in X \}$, which preserves bounds and identities
and compositions.  The composition of 2-functors is again a 2-functor,
so $\ordideal{(- + -)} : \mathbf{Preorder} \times \mathbf{Preorder} \to
\mathbf{Preorder}$
is as well, 
and unwinding definitions gives $\smabs_{\sigma_0+\sigma_1}$
and $\smconc_{\sigma_0+\sigma_1}$ from
Figure~\ref{fig:abs-conc-functions}.  

For functions, the preorder of pointwise-ordered monotone maps $X \to Y$
extends to a mixed-variance 2-functor $\mathbf{Preorder}^{op} \times
\mathbf{Preorder} \to \mathbf{Preorder}$, with functorial action given
by pre- and post-composition.  Moreover, a reflection $\smabs \dashv
\smconc$ in $\mathbf{Preorder}$ is a reflection $\smconc \dashv \smabs$ in
$\mathbf{Preorder}^{op}$ with the roles of concretization and
abstraction exchanged.  This unpacks to the definitions of
$\smabs_{\rho \to \sigma}$ and $\smconc_{\rho \to \sigma}$ in
Figure~\ref{fig:abs-conc-functions}, where abstraction precomposes with
concretization, and vice versa.    

Thus, while our general definition of model does not require types to be
interpreted as 2-functors---for example, being a model does not
require the $\eta$ law for pairs that ensures preservation of
identities---a number of more specific models will have this form, and
thus admit the same definition of relativized model, given abstraction
and concretization for base/inductive types.  For example, we may freely
apply $\mathord{}$ in the interpretation of any type constructor,
e.g.\ defining $D^{\sigma_0\times\sigma_1}$ to be
$\ordideal{(D^{\sigma_0} \times D^{\sigma_1})}$ for more precision.

\subsubsection{Example:  list reverse}
\label{sec:mm-model-reverse}

\begin{figure}
\begin{align*}
\slrev
  &= \lambda xs.%
     \slet* 
       {\slrev'}
       {\slam{xs}{\sfoldlist*
                    {\alpha}
                    {xs}
                    {\slam {zs} {zs}}
                    {(x, r).\slam{zs}{\sapp*{\sforce r}{(\slcons(x, zs))}}}}}
       {\slrev'\,xs\,\slnil} \\
\clrev'
  &= \ctylam{\alpha}{
       \clam{xs}{
         \cfoldlist*
           \alpha
           {\clist\alpha}
           {xs}
           {\cpair 1 {\clam{zs}{\cpair 0 {zs}}}}
           {(x,r).\cpair 1 {\clam{zs}
                                 {\costpluscpy{\ccost r}
                                              {\capp{\cpot r}
                                                    {(\clcons(x, zs))}}}}}}}
\end{align*}
\caption{Linear-time list reversal and its extracted recurrences.}
\label{fig:mm-model-reverse}
\end{figure}

To get a sense of how polymorphic abstraction behaves, 
let us analyze the polymorphic
linear-time list reverse function given in Figure~\ref{fig:mm-model-reverse}
in the model $\W\to\V$.  We choose this model because on the one hand $\W$
provides enough information for analyzing monomorphic functions like
$\ssumtree$ that depend on more than just the usual notion of size, yet we
still want to analyze a polymorphic function like list reversal in terms of
list length, ignoring any information about the elements of the argument list.
Since polymorphism in the source language arises only via let-bindings, the
recurrence for $\clrev'$ that is given is the recurrence
that is substituted for for $\clrev'$ according to the definition of
extraction for $\sletkw$-expressions.  A typical informal analysis of
$\slrev$ would really analyze~$\slrev'$, and might define $S(n, m)$ and
$T(n, m)$ to be the size and cost of $\slrev'\,xs\,ys$ when $xs$
and~$ys$ have length~$n$ and~$m$, respectively.  One would then observe that
$S$ and $T$ satisfy the recurrences
\[
\begin{aligned}[t]
S(1, m) &= m \\
S(n, m) &= S(n-1, m+1)
\end{aligned}
\qquad
\begin{aligned}[t]
T(1, m) &= 1 \\
T(n, m) &= 1 + T(n-1, m)
\end{aligned}
\]
from which one establishes the $O(n)$ bound on cost.

Just as with our other models, to analyze~$\clrev$, we must consider its
instantiation at some arbitrary small type~$\sigma$.  In the model~$\W$, this
would entail understanding how to compute $\smFold^\W\,s\,\phi$ for
arbitrary~$\phi$, which would be defined in terms of all~$\phi'\leq\phi$.  The
key point of $\W\to\V$ is that while we cannot avoid considering the
instantiation of $\clrev$ at arbitrary~$\sigma$, we only need to know how to
compute $\smFold^\W\,s\,\phi$ for those~$\phi$ that are the concretizations of
values in $V^{\clist\sigma}$.  
To see this, let us define
$\phi^{\clist\sigma}_n = \smconc_{\clist\sigma}(n)$---observe that
$\phi^{\clist\sigma}_n$ maps $\clist\sigma$ to~$n$ and all other datatypes
to~$\infty$---and then compute $\clrev'$, where we write~$f_\sigma\,\phi$
for~$\den{\cfoldlist{\sigma}
                    {\clist\sigma}
                    {xs}
                    {\cdots}
                    {\cdots}}
         {\bindin\eta{xs}{\phi}}$:

\begin{align*}
\den{\clrev'}{}
  &= \smTyAbs(
       \llambda\sigma.\llambda\phi^{W^{\clist\sigma}}.
         \den{\cfoldlist{\alpha}
                        {\clist\alpha}
                        {xs}
                        {\cdots}
                        {\cdots}}
             {\env{\bindto\alpha\sigma,\bindto{xs}{\phi}}}
     ) \\
  &= \smTyAbs(
       \llambda\sigma.\llambda\phi^{W^{\clist\sigma}}.
         \den{\cfoldlist{\sigma}
                        {\clist\sigma}
                        {xs}
                        {\cdots}
                        {\cdots}}
             {\env{\bind {xs} {\phi}}}
     ) \\
  &= \smTyAbs(
       \llambda\sigma.\llambda\phi^{W^{\clist\sigma}}.
         f_\sigma\,\phi
     ) \\
  &= \llambda\sigma.\smabs(
       \llambda\phi^{W^{\clist\sigma}}.
         f_\sigma\,\phi
     ) \\
  &= \llambda\sigma.\llambda n^{V^{\clist\sigma}}.\smabs(
         f_\sigma\,\phi^{\clist\sigma}_n
     ) \\
  &= \llambda\sigma.\llambda n^{V^{\clist\sigma}}.(
       \cblank,
       \smabs(
         f_\sigma\,\phi^{\clist\sigma}_n
       )_p
     ) \\
  &= \llambda\sigma.\llambda n^{V^{\clist\sigma}}.(
       \cblank,
       \llambda m^{V^{\clist\sigma}}.\smabs(
         (f_\sigma\,\phi^{\clist\sigma}_n
         )_p\,\phi^{\clist\sigma}_m)
     ) \\
  &= \llambda\sigma.\llambda n^{V^{\clist\sigma}}.(
       \cblank,
       \llambda m^{V^{\clist\sigma}}.(
         \cblank,
         \smabs(
           (f_\sigma\,\phi^{\clist\sigma}_n
           )_p\,\phi^{\clist\sigma}_m)_p
       )
     )
\end{align*}

When restricted to concretizations of abstract values, $\smFold^\W$ is
straightforward to compute.

\begin{prop}
If
$f\,n = \den{\cfoldlist {\sigma} {\rho}
                     {y}
                     {e_\clnil} 
                     {(x, r).e_\clcons}}
            {\bindin\eta y {\phi^{\clist\sigma}_n}}^\W$, 
then 
\begin{align*}
f\,1 &= \den{e_\clnil}\eta \\
f\,n &= 
  \den{e_\clnil}\eta \bmax
  \den{e_\clcons}
                    {\bindin\eta
                             {x, r}
                             {
                               \infty^\sigma,
                               f(n-1)
                             }
                    }
  & & (n > 1).
\end{align*}
\end{prop}

With this in mind, set 
$\tilde S(n, m) = \smabs(
           (f_\sigma\,\phi^{\clist\sigma}_n
           )_p\,\phi^{\clist\sigma}_m)_p$.
Our goal is to write a recurrence for $\tilde S(n, m)$.  We start with
\begin{align*}
\tilde S(1, m)
  &= \smabs((f_\sigma\,\phi^{\clist\sigma}_1)_p\,\phi^{\clist\sigma}_m)_p \\
  &=
  \smabs((\den{\cpair{1}{\clam{zs}{\cpair{0}{zs}}}}{})_p\,
              \phi^{\clist\sigma}_m)_p \\
  &= \smabs(\llambda\phi.(0, \phi)\,\phi^{\clist\sigma}_m)_p \\
  &= \smabs(0,\phi^{\clist\sigma}_m)_p \\
  &= \smabs(\phi^{\clist\sigma}_m) \\
  &= m.
\end{align*}
To compute $\tilde S(n, m)$ for $n>1$, we first compute
\begin{align*}
f\,\phi^{\clist\sigma}_n
  &=     \den{\cpair{1}{\clam{zs}{\cpair 0 {zs}}}}{}
         \bmax
         \den{\cpair{1}
                    {\clam{zs}
                          {\costpluscpy
                             {\ccost r}
                             {\capp{\cpot r}{(\clcons(\cpair{x}{zs}))}}}}}
             {\lenv\bindto{x,r}{\infty,f\,\phi^{\clist\sigma}_{n-1}}\renv} \\
  &= 
       (1, 
       \llambda\phi.(0,\phi)
       \bmax
       \llambda\phi.(f\,\phi^{\clist\sigma}_{n-1})_c +_c
                    (f\,\phi^{\clist\sigma}_{n-1})_p(
                      \chi_{F_{\clist\sigma}} + \phi
                    )
       ) \\
  &=
       (1, \llambda\phi.
       (0,\phi)
       \bmax
       (f\,\phi^{\clist\sigma}_{n-1})_c +_c
       (f\,\phi^{\clist\sigma}_{n-1})_p(
       \chi_{F_{\clist\sigma}} + \phi
       )
       ) \\
\intertext{and so}
\bigl((f\,\phi^{\clist\sigma}_n)_p\,\phi^{\clist\sigma}_m\bigr)_p
  &= \bigl(
     (0,\phi^{\clist\sigma}_m)
     \bmax
       (f\,\phi^{\clist\sigma}_{n-1})_c +_c
       (f\,\phi^{\clist\sigma}_{n-1})_p(
         \chi_{F_{\clist\sigma}} + \phi^{\clist\sigma}_m
       )
     \bigr)_p \\
  &= \bigl(
     (0,\phi^{\clist\sigma}_m)
     \bmax
       (f\,\phi^{\clist\sigma}_{n-1})_c +_c
       (f\,\phi^{\clist\sigma}_{n-1})_p\,\phi^{\clist\sigma}_{m+1}
     \bigr)_p \\
  &= \phi^{\clist\sigma}_m
     \bmax
     ((f\,\phi^{\clist\sigma}_{n-1})_p\,\phi^{\clist\sigma}_{m+1})_p \\
\intertext{and hence in the end we have}
\tilde S(n, m)
  &= \smabs\bigl((f\,\phi^{\clist\sigma})_p\,\phi^{\clist\sigma}_m\bigr)_p \\
  &= \smabs\,\phi^{\clist\sigma}_m
     \bmax
     \smabs(
       (f\,\phi^{\clist\sigma}_{n-1})_p\,\phi^{\clist\sigma}_{m+1}
     )_p \\
  &= m \bmax \tilde S(n-1, m+1) \\
  &= \tilde S(n-1, m+1).
\end{align*}
Analysis of cost proceeds in a similar manner.  We have again extracted the
recurrences we expect from an informal analysis, but instead of those
recurrences being in terms of arbitrary values in~$W^{\clist\sigma}$, they are
in terms of the length of the argument list.

Stepping back a bit, recall from Section~\ref{sec:std-model} that we can apply
parametricity to the standard model to reason about the cost
of~$\slrev\,xs$, which seems comparable to what we have just done.  But there
is a difference.  The result from parametricity tells us that the cost of
the result is determined by the length of the argument, but it does not tell
us how to compute the former in terms of the latter.  What we have done here
is to formally justify the recurrence that does just that.

\subsection{Lower bounds and an application to map fusion}
\label{sec:lower-bounds}

So far we have focused on extracting recurrences for upper bounds.  However,
the syntactic bounding theorem is agnostic with respect to the actual
interpretation of the size order.  We take advantage of this to derive
recurrences for upper and lower bounds in the main constructor counting model
of Section~\ref{sec:constr-counting-model}.  Let us consider the $\slmap$
function given in Figure~\ref{fig:cc-model-map}.  By reasoning that is by now
hopefully somewhat mundane, if we set
$T_{\slmap\,f}(n) = (\den{\clmap}{}\,f\,n)_c$, then we obtain the recurrence
\[
T_{\slmap\,f}(1) = 1
\qquad
T_{\slmap\,f}(n) = 1 + (f\,\infty)_c + T_{\slmap\,f}(n-1).
\]
Solving this recurrence yields an upper bound of
$T_{\slmap\,f}(n) = n(1 + (f\,\infty)_c)$.  Now let us apply this to the two
sides of the usual map fusion law
\[
\slmap\,f\,(\slmap\,g\,xs) = \slmap\,(f\comp g)\,xs.
\]
We hope to show that the right-hand side is less costly than the left.
Working through the recurrence extractions,
we conclude that the cost of the left-hand side is bounded
by $T_{\slmap\,f\comp\slmap\,g}(n) = 2n(1 + (g\,\infty)_c + (f\,\infty)_c)$,
whereas the right-hand side is bounded by
$T_{\slmap(f\comp g)}(n) = n(1 + (g\,\infty)_c + (f(g\,\infty)_p)_c)$.  Even
under the assumption that the costs of $f$ and $g$ are independent of their
arguments does not result in the desired conclusion, because we only know that
these recurrences yield \emph{upper bounds}, and the fact that one upper bound
is larger than another tells us nothing about the actual costs.  What we would
like to know is that these recurrences are tight, and for that we need lower
bounds as well.

\begin{figure}
\begin{align*}
\slmap =
&\lambda{f^{\sarr\rho\sigma},xs^{\slist\rho}}. \\
&     {\sfoldlist* \rho
                 {xs}
                 {\slnil}
                 {(x, r).\slcons(\sapp f x, r)}
     } \\
\clmap =
&\lambda{(f : \carr\ptrho{\ctrans\sigma}), (xs : \clist\ptrho)} \\
&\cfoldlist*
   \ptrho
   {\clist\ptsigma}
   {xs}
   {\cpair 1 \clnil}
   {
     (x, r).
     \cpair{1 + \ccost*{\capp f x} + \ccost r}
           {\clcons(\cpot*{\capp f x}, \cpot r)}
   }
\end{align*}
\caption{List map and its extracted recurrence.}
\label{fig:cc-model-map}
\end{figure}

As we already mentioned, as long as we have a model of the recurrence language
in which the interpretation of the size order
satisfies the axioms of Figure~\ref{fig:synrec-lang-preorder-full}, the
bounding theorem holds.  So to obtain lower bounds, we would want a model in
which the order on the interpretation of $\C$ is the reverse of the usual
order.  That means we would have two models in hand, one that gives us upper
bounds, and one that gives us lower bounds; we would then have to ensure that
the recurrences in each model can be sensibly compared.  As it turns out, we
can arrange that by using the model in
Section~\ref{sec:constr-counting-model}, because the interpretations of the
types are all complete upper semi-lattices.  We take advantage of the fact
that a complete upper semi-lattice is in fact a complete lattice, where
greatest lower bounds are defined by 
$\bigmeet X = \bigmax\setst{x}{\forall y\in X: x\szleq y}$.  This permits us
to define the dual interpretation of the model
$(\Usm,\Ulg,\setidx{V^\sigma}{\sigma})$ to be
$(\Usm,\Ulg,\setidx{(V^*)^\sigma}{\sigma})$, where
$(V^*)^\sigma = (V^\sigma, \szleq^*_\sigma)$ and
$x\szleq^*_\sigma y$ iff $y\szleq_\sigma x$.  Because all of the size-order
axioms except~\infruleref{beta-delta} and~\infruleref{beta-fold} are witnessed
by identities in~$\V$ (i.e., the left- and right-hand sides of the axioms have
the same denotation), we can take the semantic functions in~$\V^*$
not related to datatypes to be those of~$\V$.  For datatype-related functions,
it is
unnecessary to change either $\smsize{F}$ or $\smCons[F]$; the only change
needed is that we define
\[
\smDest[F]^*(n) = \bigmeet\setst{a}{\smCons[F](a)\geq n}.
\]
We can verify that~\infruleref{beta-delta} holds by observing that
\[
\smDest[F]^*(\smCons[F](a))
  = \bigmeet\setst{a'}{\smCons[F](a')\geq \smCons[F](a)}
  \leq a
\]
and hence
$\smDest[F]^*(\smCons[F](a)) \geq^* a$ as required.  Of course, the value of
the destructor is different in this model, but not by much; a routine
calculation shows that
\[
\smDest[F_{\clist\sigma}](x) 
  = \emptyset\disjunion(\set{\bot_\sigma}\cross\downset[\N_1^\infty](x-1));
\]
compare this to the calculation in Section~\ref{sec:constr-counting-model}.

We likewise can define the semantic fold function in this model by
\begin{align*}
\smFold[F]^*\,s\,x
  &= \bigmeet\setst{s(\smMap(\smFold\,s)\,z)}{\smCons[F] z\geq x}
\end{align*}
\sloppypar Similar to the computation of $\smDest[F]$, we have an analogue of
Proposition~\ref{prop:mcfold}:  if
$f\,n = \den{\cfoldlist\sigma {\rho} {y} {e_\clnil} {(x, r).e_\clcons}}
            {\bindin\eta y n}$,
then
\begin{align*}
f\,1 &= \bot_\sigma \\
f\,n &= \den{e_\clcons}{\bindin\eta{x,r}{\bot_\sigma,f(n-1)}}.
\end{align*}

Returning to our discussion of comparing the costs of
$\slmap\,f \comp \slmap\,g$ and $\slmap(f\comp g)$ we now conclude that
$T^\ell_{\slmap\,f\comp\,\slmap\,g}(n) = 2n(1+(g\,\bot)_c + (f\,\bot)_c)$ is a
\emph{lower} bound on the cost of $\slmap\,f\comp\slmap\,g$, so to show that
$\slmap\,(f\comp g)$ is the more efficient alternative, it suffices to show
that 
\[
n(1 + (g\infty)_c + (f(g\,\infty)_p)_c)
\szleq_\C
2n(1+(g\,\bot)_c + (f\,\bot)_c),
\]
which is trivial when the costs of $f$ and $g$ are independent of their
arguments.

\section{Recursion}
\label{sec:recursion}

We have not included general recursion in our languages in order to focus on
the key idea that different models formally justify various informal cost
analyses.  The presence of recursion does not change this perspective, but it
does complicate the model descriptions in ways orthogonal to our main thrust.
We sketch the approach of \citet{kavvos-et-al:popl2020} here.

For the syntax, we add recursive definitions to the source language with a
standard~$\sletreckw$ construct and to the recurrence language with a standard
$\cfixkw$ constructor, corresponding to the usual approach for call-by-value
and call-by-name languages.  The details are given in 
Figure~\ref{fig:recursion-syntax}, where we also give two new size-order rules
to replace~\infruleref{beta-fold}.  In these new rules, $\mathcal E$ is an
elimination context and $\cfixn n x e$ is defined by
\[
\cfixn 0 x e = \cfix x x
\qquad
\cfixn {n+1} x e = \substin e {\cfixn n x e} {x}.
\]
The two rules codify the relation between the size order and the information
order that is implicit in the presence of $\cfixkw$:  a more defined bound is
a better (i.e., smaller) bound.  In the presence of non-termination, the
bounding relation requires a slight adjustment:  $\bounded e E$ provided:  if
$E$ terminates, then $\evalin{\cl e\theta}{v}{n}$, where $n\leq\ccost E$ and
$\vbounded v {\cpot E}$.  This is the only place a (standard) operational
semantics is needed in the recurrence language, and we are investigating how
to eliminate its use.

\begin{figure}
\hbox to\textwidth{Source language:\hfill}
\[
\begin{array}{c}
  \AXC{$\typejudge
          {\cctx,f\oftype\carr\rho{\rho'}}
          {\slam{x}{e'}}
          {\carr\rho{\rho'}}$}
  \AXC{$\typejudge
          {\cctx,f\oftype\forall{(\carr\rho{\rho'})}}
          {e}
          {\sigma}$}
\ndBIC{$\typejudge\cctx{\sletrec f {\slam x {e'}} e}{\sigma}$}
\DisplayProof
\\[3ex]
  \AXC{$\evalin
          {\cl
             {e}
               {\bindin
                \theta
                {f}
                {\cl{\slam x {\sletrec f {\slam x {e'}} {e'}}}{\theta}}}}
          {v}
          {n}$}
\ndUIC{$\evalin{\cl{\sletrec f {\slam x {e'}} {e}}{\theta}}{v}{n+1}$}
\DisplayProof
\end{array}
\]
\hbox to\textwidth{Recurrence language:\hfill}
\[
\begin{array}{c}
  \AXC{$\typejudge{\cctx,x\oftype\sigma}{e}{\sigma}$}
\ndUIC{$\typejudge\cctx{\cfix x e}{\sigma}$}
\DisplayProof
\\[3ex]
  \AXC{$\setidx{\prejudge\cctx e {\elimctx{\cfixn n x e}} {\sigma}}
               {n=0,1,\dotsc}$}
\ndUIC{$\prejudge\cctx e {\elimctx{\cfix x e}} {\sigma}$}
\DisplayProof
\qquad
\ndAXC{$\prejudge\cctx{\cfixn{n+1} x e}{\cfixn n x e}{\sigma}$}
\DisplayProof
\end{array}
\]
\hbox to\textwidth{Recurrence extraction:\hfill}
\[
\ctrans{\sletrec f {\slam x {e'}} {e}}
  = \clet f {\cfix f {\clam x {\costpluscpy 1 {\ctrans {e'}}}}} {\ctrans e}
\]
\caption{Adding general recursion to the source and recurrence languages.}
\label{fig:recursion-syntax}
\end{figure}

For the semantics of the recurrence language, we impose additional structure
on our applicative structures.  We call the new structures \emph{sized
domains} and they are defined just like applicative structures, except that
for each $U\in\Usm$, $D^U = (D^U,\szleq_U,\defleq_U, \bot_U)$, where
$(D^U,\szleq_U)$ is a preorder as before, and 
$(D^U,\defleq_U,\bot_U)$ is a complete partial order.  The semantic domains
must satisfy two additional constraints:
\begin{itemize}
\item If $x\defleq_U y$, then $y\szleq_U x$; and
\item If $y_0\defleq_U y_1\defleq_U\dotsb$ and for all~$i$, $x\szleq_U y_i$,
then $x\szleq\bigdefmax y_i$.
\end{itemize}

That leaves us with verifying that the models that we presented in 
Section~\ref{sec:examples} are sized domains.  For each of the models, we take
$\defleq_{\N_i^\infty}$ to be the usual flat order with
$\bot_{\N_i^\infty} = \infty$ (again, 
cf.~\citep{rosendahl:auto_complexity_analysis}) extended pointwise and
componentwise for functions and products.  For sums, set
$X\defleq Y$ if $Y\subseteq X$.  It is a straightforward exercise to show that 
$D^{\csum\rho\sigma}$ is a CPO that satisfies the constraints just given.  
To show that we have a model, it suffices to
verify that the semantic functions are simultaneously monotone with respect to
$\szleq$ and continuous with respect to~$\defleq$, after which 
Prop.~\ref{prop:mc-model} can be extended with the clause that
$\llambda a.\den{\typejudge\cctx e \sigma}{\bindin\eta x a}$ is
continuous with respect to~$\defleq$.
Verification of continuity for $\smCase$ relies on two facts that hold in
these models at all types:
\begin{itemize}
\item If $a\defleq a'$ and $b\defleq b'$, then 
$(a\bmax b)\defleq(a'\bmax b')$; and
\item If $a_0\defleq a_1\dotsb$ and $b_0\defleq b_1\dotsb$, then
$\bigdefmax\set{a_i\bmax b_i} = (\bigdefmax a_i)\bmax(\bigdefmax b_i)$.
\end{itemize}

Extracting syntactic recurrences from general recursive functions
and interpreting them in our models follows the same pattern we have already
seen several times.  But now the recurrences may have more complex solutions
(such as poly-log solutions).
For example, \citet{kavvos-et-al:popl2020} analyze the standard
implementation of merge-sort and interpret it in the model of
Section~\ref{sec:constr-counting-model}.  Under the usual assumption that the
cost of the comparison function is constant the recurrence
clause of the semantic recurrence is
$T(n) = c + dn + T(n/2)$ for some constants~$c$ and~$d$ (that arise from the
analyses of the functions that divide a list in two and merge two sorted
lists), just as expected.  Now one may reason in the semantics to
establish the $O(n\lg n)$ cost from this recurrence.

Quick-sort provides an interesting example of how more complex models can be
used to capture subtle information that may be necessary for an asymptotic
analysis.  Quick-sort relies on a partitioning function
$\skeyw{part} :
\sarr\alpha{\sarr{\slist\alpha}{\cprod{\slist\alpha}{\slist\alpha}}}$
such that $\skeyw{part}\,x\,xs = (ys, zs)$, where $ys$ consists of the elements
of $xs$ that are $< x$ and $zs$ those elements that are $\geq x$.  A key part
of the analysis of quick-sort is the fact that the sum of the lengths of~$ys$
and~$zs$ is the length of~$xs$.  In the models we have presented in
Section~\ref{sec:examples}, the extracted recurrence will not yield such a
bound.  For example, in the main constructor-counting model,
the best we can conclude about the extracted recurrence is
that in the semantics,
$\ckeyw{part}\,x\,n = \cpair n n$.  The problem is that the
interpretation of products requires that we choose some specific pair that is
a bound on all pairs $(k,\ell)$ such that $k+\ell = n$, and $(n, n)$ is the
least such bound.
But we have seen this
situation before when it came to interpreting sums, and the solution is the
same:  instead of taking $V^{\cprod\rho\sigma} = V^\rho\cross V^\sigma$, we
can instead take $V^{\cprod\rho\sigma} = \ordideal(V^\rho\cross V^\sigma)$.
While the calculations become more tedious, in such a model we can show that
$\ckeyw{part}\,x\,n = \setst{(k,\ell)}{k+\ell \leq n}$.  However, it turns out
this is not quite enough.  Both the source and recurrence languages have
negative products, which means that projections must be used to extract~$ys$
and~$zs$.  In the interpretation of the extracted recurrence, projection of a
set of pairs maximizes over the corresponding component, and so
$\cproj i {(\ckeyw{part}\,x\,n)} = n$ (because $n+0 = 0 + n = n$), 
which again leads to a weak bound.
Instead, we must use positive products with an elimination of the
form~$\skeyw{split}\,(x, y) = e^{\sprod\rho\sigma}\,\sinkw\,e'$.
The corresponding elimination form in the recurrence language can be
interpreted by maximizing~$\den{e'}{}$ over all pairs in~$\den{e}{}$, 
which is precisely what
is needed to carry out the rest of the usual analysis of quick-sort.

\section{Related work}
\label{sec:related-work}

We first expand upon a couple of observations that we made earlier and mention
some motivating history behind some technical details.
Then we address how our work fits into the literature on cost analysis.

We touched on an application of parametricity in Section~\ref{sec:std-model}.
\citet{seidel-voigtlander:qapl11} have interpreted free
theorems~\citep{wadler:theorems-for-free} to obtain relative complexity
information.  Their work can be viewed as applying parametricity to the
standard model, but in a somewhat more general setting of a recurrence
language that has a monadic type constructor~$\ckeyw{C}(\sigma)$ 
for ``complexity of~$\sigma$,'' with projections for cost and potential.
They define a notion of lifting relations to complexities
(much as relations are lifted to inductive types), which
allows them to interpret a free theorem such as
$f(\skeyw{hd}\,xs) = \skeyw{hd}(\slmap\,f\,xs)$ in such a way that the
interpretations of both sides yield complexity information, and the identity
then allows them to conclude, e.g., that the cost of the left-hand side is no
greater than that of the right-hand side.  With our approach, we would simply
extract recurrences from the left- and right-hand sides and reason about them
as in Section~\ref{sec:lower-bounds}.  
While on the topic of relative cost information, we would be remiss to not
mention the type-and-effect system of
\citet{cicek-et-al:popl17}, which permits a very precise analysis of the
relative cost of different algorithms on the same arguments or the same
algorithm on different arguments.
We have not investigated whether
our techniques can be adapted to provide comparable analyses.

We drew an analogy with abstract interpretation (AI) in
Section~\ref{sec:ind-types-as-ai} and made use of the existence of a Galois
connection of the sort that arises in AI in Section~\ref{sec:merged-model}.
\citet{rosendahl:auto_complexity_analysis} uses AI to extract cost bounds
directly from a first-order fragment of Lisp.  She first defines a program
translation similar to our syntactic extraction and interprets it in the
standard model~$D$ of $S$-expressions.  She then defines an AI
from~$\powerset D$ into a finite-height lattice of ``partial structures,''
whose values are essentially truncated standard values.  Given a notion of
size $s : D\to\N$ and a computable bound on $\llambda n.\alpha(\setst{x}{s(x)
= n})$, the interpretation of the syntactic recurrence in the abstract domain
is a computable upper bound on the cost of the original program.  
This work is restricted to first-order programs and does not handle branching
data structures well (e.g., if $s(t)$ is the number of nodes in the tree~$t$,
then for $n>1$, $\alpha(\setst{x}{s(x) = n})$ is a node structure that is
truncated at its children, so the bounds are all trivial).  But these ideas
may provide an approach to computing bounds on semantic recurrences in models
where the semantic recurrence itself is not computable (a situation that does
not arise in the models we have presented).

While our notion of potential is
drawn most directly from \citet{danner-royer:ats-lmcs}, it
traces back at least to \citet{shultis:complexity}, who defines a denotational
semantics for a simple higher-order language that models both the value and
the cost of an expression.  He develops a system of ``tolls,'' which play a
role similar to that of our potentials.  The tolls and the semantics are not
used directly in calculations, but rather as components in a logic for
reasoning about them.  
\citet{sands:thesis} defines a translation scheme in
which each identifier~$f$ in the source language is associated to a \emph{cost
closure} that incorporates information about the value $f$ takes on its
arguments, the cost of applying~$f$ to arguments, and arity.  Cost closures
record information about the future cost of a partially-applied function, just
as our potentials do.  
The idea of using
denotational semantics to captures cost information has been seen before.
We have already mentioned \citet{rosendahl:auto_complexity_analysis}
and \citet{shultis:complexity}.
\citet{van-stone:thesis} defines a category-theoretic denotational
semantics that uses ``cost structures'' (these include the $\C \times -$ writer
monads we use here) to capture cost information and shows that it is
sound with respect to a cost-annotated operational semantics for a
higher-order language.  Our bounding theorem is roughly analogous to Van
Stone's soundness theorem, but is a bit more general because we show an
inequality (using the
size order on the complexity language) instead of an equality, which
allows the bounding theorem to apply to models with size abstraction.  

Turning now to the literature on cost analysis, constructing resource
bounds from source code has a long history in Programming Languages.
The earliest work known to the authors is that of \citet{cohen:cacm74},
which extracts programs that describe costs from an ALGOL60-like
language that are intended to be manipulated in an interactive system,
and \citeauthor{wegbreit:cacm75}'s \citeyearpar{wegbreit:cacm75} METRIC
system, which extracts recurrences from simple first-order recursive
Lisp programs.  An interesting aspect of the latter system is that it is
possible to describe probability distributions on the input domain
(e.g., the probability that the head of an input list will be some
specified value), and the generated bounds incorporate this information.
\citeauthor{lematayer:toplas88}'s \citeyearpar{lematayer:toplas88} ACE
system converts FP programs \citep{backus:fp} (under a strict
operational semantics) to FP programs (under a non-strict semantics)
describing the number of recursive calls of the source program.  The
first phase is comparable to the cost projection of our recurrence
extraction; the potential projection is the original program.  Both
METRIC and ACE yield non-recursive upper bounds on the generated cost
functions (this is the bulk of the work for ACE).  These systems are
restricted in their datatypes and compute costs in terms of syntactic
values; the notion of ``size'' is somewhat ad-hoc and second class.
Many approaches to cost analysis rely on the idea that the cost can be
treated as an additional output of the program, or as a piece of program
state; \cite{wadler:popl92:essence} observed that this can be represented by a
monadic translation --- though in our case we use the writer monad
rather than the state monad, since we do not give programs access to
their cost.

There are many approaches to type-based cost
analysis~\citep{%
crary-weirich:popl00,
hofmann-jost:popl03,
jost-et-al:popl10,%
hoffmann-hofmann:esop10,%
hoffmann-et-al:toplas12:multivariate-amortized,%
hoffman-et-al:popl17,%
jost+17lazy,%
knoth+19synthesis,
knoth+20liquidresource,
avanzini-dal-lago:icfp17,
cicek-et-al:popl17,
wang-et-al:oopsla17,
dallagogaboardi11cost,
handley+20liquid,
rajani+21lambdaamor
}.
At a high level, these systems include special-purpose judgements or types that track
cost, indexed or refinement types that track the size
of values, and a type checking or inference mechanism that can
automatically determine some resource bounds.  
For example, 
the Automatic Amortized Resource Analysis
(AARA) technique of 
\citet{hoffmann-et-al:toplas12:multivariate-amortized,hoffman-et-al:popl17,jost+17lazy,hofmann-jost:popl03,jost-et-al:popl10,hoffmann-hofmann:esop10}, with an implementation at
\citet{website:raml}, computes cost bounds by
introducing a type system with size information that is parameterized by an
integer degree, and then performing type
inference.  
If inference is successful, then the program cost can be bounded by a
polynomial of at most that degree (and a bound is reported); otherwise it
cannot.  As its name suggests, AARA automatically incorporates amortization,
resulting in tighter bounds for some programs than our extracted recurrences
yield (but see \citep{cutler-et-al:icpf20} for an extension of our approach to
amortized analysis).  
The basic AARA technique has been extended in numerous ways, e.g. with 
refinement types~\cite{knoth+19synthesis,knoth+20liquidresource} for 
synthesizing programs with desired resource bounds, and 
for more precise tracking of
potential in values.
The Timed~ML system of
\citet{wang-et-al:oopsla17} also uses refinement types 
(indexed types in the style of DML~\citep{xi-pfenning:popl99:dml}) that permit the user to
define datatypes with their own notion of size and to include cost information
in the program type.  Type inference produces verification conditions which,
if solvable, validate the cost information.
That cost information may be very concrete, or left more open-ended, in which
case the verification conditions end up synthesizing (recurrence) relations
that must be satisfied.    
\citet{avanzini-dal-lago:icfp17} develop a non-amortized type-based analysis,
which uses a translation similar to our
recurrence extraction to explicitly represent the cost as a unary numeral.
As a result, the evaluation cost of the original program is reflected in the
size of the cost component of the translated program.  They then make use of an extension of
sized types~\citep{hughes-et-al:popl96}
to infer a type for the translated program, which therefore
includes a bound on the cost in terms of the size of the arguments.

All of these type-based approaches are impressive in the breadth of successful
analyses and/or automation thereof.  However, we believe it is
nonetheless worth studying cost analysis by recurrence extraction for several
reasons.  First, the process of inferring bounds using these specialized
type systems and their associated solvers is not, in our opinion, very
easy for a person to do, while our focus is on formalizing the method
that we readily teach students to do.
Second, automated approaches necessarily impose some limits on the kinds
of bounds that can be inferred and the notions of size that are
supported to facilitate inference (though \cite{handley+20liquid} also
allows explicit proofs; see the discussion of techniques in proof assistants
below).  For example, AARA infers polynomial bounds, while our approach
(adapted to the setting of general recursion) can produce recurrences
with non-polynomial solutions.
Third, type based approaches make the size and cost an \emph{intrinsic}
feature of the code: in approaches based on refinement types, one must,
for example, define one tree type where size means number of nodes, and
a different tree type where size means height, which causes code
duplication if both are necessary; in amortized approaches, one must
choose the potential annotations when defining a type (though sometimes
this can be mitigated by parametrizing the
datatypes~\cite{knoth+20liquidresource}).  In our approach, cost and
size are an \emph{extrinisic} property of the code, so the same function
can be interpreted in different models with different notions of size
for different analyses, which can be useful e.g. for a library function
that is used in two different programs by other functions that require
two different notions of size.  That said, this does not address
situations where two different notions of size for a type are needed in
a single program --- one possible solution is a model in which the
potential is the pair of these sizes, but this would have similar reuse
problems to changing a refinement type to include additional
information, in that all existing analyses would formally need to be modified.  

Let us now consider work that, like ours,
externalizes cost from programs that are typed
in a more-or-less standard type system.  \citet{avanzini-et-al:icfp15}
carefully defunctionalize higher-order programs to first-order programs
in order to take advantage of existing techniques from first-order
rewrite systems.  This leverages existing technologies to great effect,
but does not match the kind of recurrence extraction that we are aiming
for in this work.  The COSTA
project~\citep{albert-et-al:tcs12:cost-analysis} extracts cost
recurrences from Java bytecode; \citet{albert-et-al:tocl13:inference}
provides techniques for constructing closed forms for both lower and
upper bounds on these recurrences.  This group has also pushed forward
on parallel cost~\citep{albert-et-al:tocl18}, something that
\citet{raymond:thesis} has looked into in our setting, but the COSTA
work has focused on first-order, low-level languages.

\citet{cutler-et-al:icpf20} adapt our technique to
handle amortized analysis.  Reinforcing our goal of formalizing informal
approaches, the source language there includes constructions for describing a
credit allocation policy (the banker's method) and extraction of an
\emph{amortized} cost recurrence, to which a general theorem applies that
total amortized cost bounds total actual cost.  The language is sufficient for
describing structures like splay trees in which the number of credits
allocated to different parts of the structure is not constant, and the source
language type system ensures that credits are not misused.  The key point is
that the amortized cost recurrence is extracted into essentially the same
recurrence language as we have presented here, reflecting the fact that the
recurrences that we use to describe amortized cost do not themselves
refer to credits.

\citet{kavvos-et-al:popl2020} give an
approach to extending our technique to handle general (as opposed to
structural) recursion by using call-by-push-value (CBPV)
\citep{levy:cbpv} as an intermediate source language into which both
call-by-value and call-by-name can be embedded.  While CBPV includes a fine
stratification of types into computational and value types, analyzing 
a program still really just relies on notions of size and cost.  Thus the
syntactic recurrence language differs from the one just described only in
replacing primitive recursion with a general fixpoint operator, along with
corresponding axioms for the size order, thereby changing it
from a version of System~$T$ with inductive types to a version
of~$\mathrm{PCF}$ with inductive types.

\citet{atkey11separation,gueneau-et-al:esop18,chargueraud-pottier:jar19,zhanhaslbeck19verifying} develop
imperative program logics for reasoning about cost based on separation
logic, essentially by treating the number of timesteps taken as part of
the heap.  A Coq or Isabelle implementation of these logics allows for reasoning
about code, and the subgoals that arise during verification result
in synthesizing recurrence relations, which play the role of our
syntactic recurrences.
While quite sophisticated algorithms
and data structures can be analyzed this way, including imperative ones,
for analyzing functional programs we find it more congruous to use (and
teach to students) standard functional program verification techniques
like inductive reasoning about outputs, as opposed to imperative program
verification techniques like weakest precondition/characteristic formula
generation.  And as we note in Section~\ref{sec:conclusions}, we conjecture that
our approach extends to the analysis of many imperative programs, because the
description of cost itself is frequently a functional description.

Turning now to semi-automated/manual reasoning in a functional
style, \citet{danielsson:popl08} verifies a number of lazy
functional programs in Agda using a dependent type tracking the number
of steps a program takes.  \citet{mccarthy-et-al:soc18} investigate a
variant, implemented in Coq, using a monad parametrized
by both the number of steps and a specification, given as a relation
between the cost and value.  The specifications are used both for
functional correctness and for reasoning about cost, and this design
allows Coq's extraction to OCaml to erase all costs and reasoning about
them.  The library also provides a source-to-source translation that
translates simply-typed code into the monad, inserting appropriate
ticks, which is analogous to our recurrence extraction.
\citet{radicek+18monadicrefinements} define a specification logic for
reasoning about monadic costs as an extension of higher-order logic.

\citeauthor{benzinger:tcs04}'s \citeyearpar{benzinger:tcs04} 
ACA system might be the closest in philosophy to ours, in that it
extracts (higher-order) recurrences
from call-by-name 
\textsc{NuPrl} programs that bound the cost of those programs.  There
we find (moderately complex) expressions that correspond to applying
higher-order functions to arguments (necessarily
alternating with projections) to describe the cost of a fully applied function
argument, corresponding to our notion of higher-order potential.  But this
does not address more realistic call-by-value or call-by-need evaluation.

Since these
approaches~\citep{danielsson:popl08,mccarthy-et-al:soc18,radicek+18monadicrefinements,benzinger:tcs04}
take place inside of a general-purpose logic or proof assistant, one can
express costs in terms of the sizes of inputs by explicitly referring to
an appropriate size function and proving how operations transform the
size.  Relative to this, a main contribution of our approach is to
systematize and partially automate the reasoning about size, in the
sense that our semantic interpretation of the potential of a function
$f$ gives a direct inductive definition of the fused ``size of the
result of $f$ on inputs of size --'' function. This is possible because
we step outside of the programming language into a denotational setting
where e.g. arbitrary maximums exist.
We claim that this corresponds better to informal analyses than using
the full power of a proof assistant to carefully prove how functions act
on sizes, because the fused size-to-size function will simplify in ways
that the original function does not.  For example, because in these
models most or all contexts are monotone in the size order, one can
freely ignore branches whose size is dominated by another.

\section{Conclusions and further work}
\label{sec:conclusions}

We have presented a technique for extracting cost-and-size recurrences from
higher-order functional programs that provably bound the operational cost in
terms of user-definable notions of size, thereby giving a formal account of
the process of many informal cost analyses.
The technique applies to the pure fragment of strict languages such as ML and
OCaml.  Although we have not investigated the question carefully, it also
seems that it applies to much reasoning about imperative programs.  The reason
is that such analysis often consists of extracting \emph{functional} cost
recurrences whose validity only depends on the fact that certain imperative
operations have certain costs.  For example, the analysis of many functions on
an arrays depends on the fact that indexed access and update is constant time.
But the analysis does not typically result in a recurrence that even refers to
an array, much less destructively updates one.  In our setting, we would
either hard-code the costs of access and update in the syntactic recurrence
extraction, or we would leave those functions as identifiers and analyze the
semantic recurrence under the assumption that those identifiers are
interpreted by constant-time functions.
The \emph{de facto}
standard for such reasoning is Separation Logic, and the work that ours seems
closest to in spirit is that of \citet{zhanhaslbeck19verifying}.  
Our goal would be to provide relatively simple approaches to formalizing
reasoning about many imperative programs.
This is certainly speculative, and we have not investigated how far one can
push this idea before requiring the machinery of something comparable to
Separation Logic.

A natural direction to extend our work would be to handle cost analysis of
lazy languages.  \citet{okasaki:purely-functional-data-structures} describes a
technique of amortized analysis in which costs are split into ``shared'' and
``unshared'' costs in order to correctly account for the memoization of
computations, and we believe our approach can be adapted to formalize this
technique.  \citet{hackett-hutton:icfp19} show that lazy evaluation is a form
of ``clairvoyant'' call-by-value and that cost can be described
non-deterministically rather than in terms of shared and unshared costs.  We
hope to adapt our approach to yield
corresponding recurrences, especially as they actually compute costs via an
interpretation in a denotational model
that appears to mesh nicely with our approach.

We have presented several models making use of different notions of size.  It
is no surprise that it is easier to work in models with simpler notions of
size, and we saw in Section~\ref{sec:merged-model} that a simpler notion of
size corresponds to a more abstract model.  Formalizing the connection between
more abstract and more concrete models so that information from the latter may
be pulled into the former, would improve the usefulness of this sort of
reasoning.  This sounds like an analogy with safety and liveness theorems from
abstract interpretation, and this is probably a fruitful direction for further
study.  More complex models should enable more sophisticated analysis.  For
example, the average case complexity of deterministic quick-sort can be
described by assuming a (uniform) probability distribution on the inputs.
That would seem to correspond to interpreting the usually extracted recurrence
in a model in which inductive types are interpreted by probability
distributions or random variables.  \citet{barnaby:thesis} has made
preliminary progress in this direction, which indicates that it is probably
necessary to have at least limited forms of dependent typing in the recurrence
language.

We have focused on the extraction of semantic recurrences to show that they
are the ones that are expected from informal analysis.  We have not studied
techniques for solving the semantic recurrences, which in general are
higher-order functions. \citet{benzinger:tcs04} discusses techniques for
solving them by reducing them to first-order recurrence equations
and then using off-the-shelf
solvers such as \textsc{Mathematica} and
OCRS~\citep{kincaid-et-al:popl18:ocrs}.  Another fruitful direction would be
to formalize the extracted semantic recurrences in proof assistants and make
use of the formalization of standard theorems like the Master Theorem
and of asymptotic reasoning as in 
\citet{gueneau-et-al:esop18}.  This would permit a formal development in a
setting where complete automation is not possible.

The extraction of the syntactic recurrence is straightforward to implement,
and a future project is to produce an end-to-end tool from source code to
semantic recurrence to solution.  
We know that automated cost analysis is a complex project that many have
attempted, and so this goal as stated is probably too ambitious, and we warn
the reader that our thoughts here are pies in the sky at the time of writing.
Our vision is more along the lines of an interactive system, in which
recurrences are extracted and ``easy'' ones solved, but allowing the user to
step in to provide assertions (hopefully proved!) about the solutions to
difficult ones.  Familiarity with recurrence extraction as a cost analysis
technique would hopefully lower the entry barrier of such a tool.  We could
also hope that that same familiarity would enable users to 
work backward from an unexpectedly poor recurrence to the code from
which it results (cf.\ \citet{benzinger:tcs04}).
\citet{wang-hoffmann:popl19}
adapt AARA to provide worst-case inputs that validate the tightness of the
produced bounds, which could be used to similar effect.  Another direction
such a project could take would be to pull either the syntactic or the
semantic information back as additional interface-level components of a
language library, so as to modularize cost reasoning and take advantage of the
compositionality of our approach.  However, this is not so straightforward.
One issue that arises is that the denotation a type that is appropriate for
analyzing an algorithm is not necessarily the one that is appropriate for
using it.  For example, the recurrence extraction approach works best to
analyze binary search tree algorithms in terms of their heights, but a client
who uses a binary search tree implementation is probably more interested in
understanding the cost in terms of the size.  This is a setting in which
composing recurrences does not work as smoothly as we might hope.
Understanding how to mesh them
together, and more generally how to hide analyses that possibly require
more complex types (such as those by \citet{cutler-et-al:icpf20})
behind an interface, is ongoing work.

\bibliographystyle{abbrvnat}
\bibliography{refs}

\appendix

\section{Type preservation for the source language}
\label{app:src-lang-type-preservation}

Type preservation depends on the usual substitution lemmas.

\begin{lemma}
\label{lem:typing-value-subst}
If $\typejudge {\Gamma,x\oftype\rho} {\cl e {(\theta-x)}} \sigma$ and
$\typejudgeV v \rho$, then 
$\typejudge\Gamma {\cl e {\bindin\theta x v}} \sigma$.
\end{lemma}

\begin{lemma}
\label{lem:typing-val-subst}
If $\typejudge{y\oftype\rho}{v'}{\sigma}$ and $\typejudgeV v\rho$,
then $\substin{v'}{v}{y}$ is a value and
\mbox{$\typejudgeV{\substin{v'}{v}{y}}{\sigma}$}.
\end{lemma}

We now have the type preservation theorem.


\begin{thm*}[Type preservation, Theorem~\ref{thm:type-preservation}]\hfill
\begin{enumerate}
\item If $\typejudgeE {\cl e\theta}{\sigma}$ and $\eval {\cl e\theta}{v}$,
then $\typejudgeV v \sigma$.
\item 
If $\typejudgeE{\cl*{\smap F y {v''} {v'}}{\theta}} {\sfsubst F \sigma}$ 
and
$\eval {\smap F y {v''} {v'}} {v}$, then
$\typejudgeV v {\sfsubst F{\sigma}}$.
\item 
If $\typejudgeE{\smapv F y {v''} {v'}} {\sfsubst F{\sigma}}$ and
$\eval {\smapv F y {v''} {v'}} {v}$, then
$\typejudgeV v {\sfsubst F{\sigma}}$.
\end{enumerate}
\end{thm*}
\begin{proof}
The proof is a simultaneous induction on the height of the
derivation that referred to in each part.
We give just a few of the more interesting cases, starting with part~(1).

\begin{proofcases}
\item[$\eval{\cl x\theta}{\theta(x)}$]
By the hypothesis, $\typejudgeE {\cl x\theta}{\sigma}$, so by the typing
rules for closures, there must be some~$\sctx'$ such that
$\sctx'(x) = \sforall{\vec\alpha}{\rho}$ and
$\sigma = \substin\rho{\vec\sigma}{\vec\alpha}$, and
$\theta$ is a $\sctx'$-environment.  But that means that in particular,
$\typejudgeV{\theta(x)}{\substin\rho{\vec\sigma}{\vec\alpha}}$, as
required.

\item[$\eval{\cl*{\scase* e x e}{\theta}}{v}$]
The typing must have the form
\[
    \AXC{$\typejudge\sctx e {\ssum{\sigma_0}{\sigma_1}}$}
    \AXC{$\setidx{\typejudge {\sctx,x\oftype\sigma_i}{e_i}{\sigma}}{i=0,1}$}
  \ndBIC{$\typejudge{\sctx}{\scase* e x e}{\sigma}$}
  \AXC{$\theta$ a $\sctx$-environment}
\ndBIC{$\typejudgeE{\cl*{\scase* e x e}{\theta}}{\sigma}$}
\DisplayProof
\]
and the evaluation must have the form
\[
  \AXC{$\eval{\cl e\theta}{\sinj i {(v_i)}}$}
  \AXC{$\eval{\cl{e_i}{\bindin\theta{x}{v_i}}}{v}$}
\ndBIC{$\eval{\cl*{\scase* e x e}{\theta}}{v}$}
\DisplayProof
\]
By definition $\typejudgeE{\cl e\theta}{\ssum{\sigma_0}{\sigma_1}}$,
so by the induction hypothesis, 
$\typejudgeV{\sinj i {v_i}}{\ssum{\sigma_0}{\sigma_1}}$, and hence by
inversion, $\typejudgeV{v_i}{\sigma_i}$.  That means that
$\bindin\theta{x}{v_i}$ is a $(\sctx,x\oftype\sigma_i)$-environment, and
hence $\typejudgeE{\cl{e_i}{\bindin\theta{x}{v_i}}}{\sigma}$.  So by
the induction hypothesis, $\typejudgeV v \sigma$, as required.

\item[$\eval{\cl*{\slam x e}\theta}{\cl*{\slam x e}\theta}$]
If $\typejudgeE {\cl*{\slam x e}\theta} {\sarr{\sigma}{\sigma'}}$, then
we must show that $\typejudgeE {\cl*{\slam x e}\theta} {\sarr{\sigma}{\sigma'}}$
as a value.  For this we must show that
$\typejudgeE {\cl*{\slam x e}\theta} {\sarr{\sigma}{\sigma'}}$ as a closure,
which is precisely the hypothesis we started with.

\item[$\eval{\cl*{\sapp{e_0}{e_1}}\theta}{v}$]
The typing has the form
\begin{prooftree}
    \AXC{$\typejudge\sctx{e_0}{\sarr\rho\sigma}$}
    \AXC{$\typejudge\sctx{e_1}{\rho}$}
  \ndBIC{$\typejudge\sctx{\sapp{e_0}{e_1}}{\sigma}$}
  \AXC{$\theta$ a $\sctx$-environment}
\ndBIC{$\typejudgeE{\cl*{\sapp{e_0}{e_1}}\theta}{\sigma}$}
\end{prooftree}
and the evaluation has the form
\begin{prooftree}
  \AXC{$\eval{\cl{e_0}{\theta}}{\cl*{\slam x {e_0'}}{\theta_0'}}$}
  \AXC{$\eval{\cl{e_1}{\theta}}{v_1}$}
  \AXC{$\eval{\cl{e_0'}{\bindin{\theta_0'}{x}{v_1}}}{v}$}
\ndTIC{$\eval{\cl*{\sapp{e_0}{e_1}}{\theta}}{v}$}
\end{prooftree}
Since $\theta$ is a $\sctx$-environment, 
$\typejudgeE{\cl{e_0}{\theta}}{\sarr\rho\sigma}$, so by the induction
hypothesis,
$\typejudgeV{\cl*{\slam x {e_0'}}{\theta_0'}}{\sarr\rho\sigma}$ and similarly
$\typejudgeV{v_1}{\rho}$.  By definition we have that there is some~$\sctx'$
such that $\typejudge{\sctx'}{\slam x {e_0'}}{\sarr\rho\sigma}$ and
$\theta_0'$ is a $\sctx'$-environment; by inversion we have that
$\typejudge{\sctx',x\oftype\rho}{e_0'}{\sigma}$.
Since $\typejudgeV{v_1}{\rho}$, 
$\bindin{\theta_0'}{x}{v_1}$ is a
$(\sctx',x\oftype\rho)$-environment, and so
$\typejudgeE{\cl{e_0}{\bindin{\theta_0'}{x}{v_1}}}{\sigma}$, and so by
the induction hypothesis,
$\typejudgeV v \sigma$, as required.

\item[$\eval{\cl*{\sfold\delta{e'}{x}{e}}\theta}{v}$]
The typing must have the form
\begin{prooftree}
    \AXC{$\typejudge\sctx {e'} {\delta}$}
    \AXC{$\typejudge{\sctx,x\oftype\sfsubst F {\ssusp\sigma}}{e}{\sigma}$}
  \ndBIC{$\typejudge\sctx{\sfold\delta{e'} x e}{\sigma}$}
  \AXC{$\theta$ a $\sctx$-environment}
\ndBIC{$\typejudgeE{\cl*{\sfold\delta{e'} x e}{\theta}}{\sigma}$}
\end{prooftree}
and the evaluation must have the form
\begin{prooftree}
  \AXC{$\eval {\cl{e'}\theta} {\scons\delta{v'}} $}
  \AXC{$\eval
    {\smapv F y {\cl*{\sdelay*{\sfold\delta y x e}}{\theta}} {v'}} 
    {v''}$}
  \AXC{$\eval {\cl e {\bindin\theta x {v''}}} {v}$}
\ndTIC{$\eval {\cl*{\sfold\delta{e'} x e}{\theta}} v $}
\end{prooftree}
where without loss of generality
we assume $y\notin\dom\sctx$ and $y\notin\dom\theta$.
By the assumptions and induction hypothesis,
$\typejudgeV{\scons\delta {v'}}{\delta}$, and so by
inversion, $\typejudgeV{v'}{\sfsubst F {\delta}}$.
From $\typejudge{\sctx,x\oftype\sfsubst F {\ssusp\sigma}}{e}{\sigma}$
we conclude that
$\typejudge{\sctx,y\oftype\delta}
   {\sdelay*{\sfold\delta y x e}}
   {\ssusp\sigma}$.
Since $\theta$ is a $\sctx$-environment, we conclude that
$\typejudge{y\oftype\delta}
  {\cl*{\sdelay*{\sfold\delta y x e}}{\theta}}
  {\ssusp\sigma}$.
These two judgments allow us to conclude that
$\typejudgeE
  {\smapv F y {\cl*{\sdelay*{\sfold\delta y x e}}{\theta}} {v'}}
  {\sfsubst F{\ssusp\sigma}}$,
so by the induction hypothesis applied to the evaluation of the
$\smapvkw$ expression,
$\typejudgeV{v''}{\sfsubst F{\ssusp\sigma}}$.  That means that
$\bindin\theta x {v''}$ is a
$(\sctx,x\oftype\sfsubst F{\ssusp\sigma})$-environment, and so by
the induction hypothesis applied to the evaluation of
$\cl e {\bindin\theta x {v''}}$,
$\typejudgeV{v}{\sigma}$, as required.
\end{proofcases}

For~(2), suppose
$\eval{\cl*{\smap F y {v'} e}\theta}{v}$.
The typing must have the form
\begin{prooftree}
    \AXC{$\typejudge{y\oftype\rho}{v'}{\sigma}$}
    \AXC{$\typejudge\sctx{e}{\sfsubst F\rho}$}
  \ndBIC{$\typejudge\sctx{\smap F y {v'} e} {\sfsubst F\sigma}$}
  \AXC{$\theta$ a $\sctx$-environment}
\ndBIC{$\typejudgeE{\cl*{\smap F y {v'} e}\theta}{\sfsubst F\sigma}$}
\end{prooftree}
and the evaluation the form
\begin{prooftree}
  \AXC{$\evalin{\cl e\theta}{v''}{n}$}
  \AXC{$\eval{\smapv F y {v'} {v''}}{v}$}
\ndBIC{$\evalin{\cl*{\smap F y {v'} e}\theta}{v}{n}$}
\end{prooftree}
As in previous cases, $\typejudgeV{v''}{\sfsubst F \rho}$, and so
$\typejudgeE{\smapv F y {v'} {v''}}{\sfsubst F\sigma}$, and so
the result follows from the induction hypothesis applied to this
$\smapvkw$ expression.

We now prove~(3).

\begin{proofcases}
\item[$\eval{\smapv t y {v'} v}{\substin{v'}{v}{y}}$]
From the typing assumption we have that
$\typejudge{y\oftype\rho}{v'}{\sigma}$ and
$\typejudgeV v \rho$, so the result follows from
Lemma~\ref{lem:typing-val-subst}.

\item[$\eval{\smapv {\sarr\rho F} y {v'} {\cl*{\slam x e}\theta}}%
            {\cl*{\slam x {\smap F y {v'} e}}{\theta}}$]
The typing must have the form
\begin{prooftree}
  \AXC{$\typejudge{y\oftype\rho}{v'}{\sigma}$}
      \AXC{$\typejudge{\sctx,x\oftype\rho}{e}{F[\rho']}$}
    \ndUIC{$\typejudge\sctx{\slam x e}{\sarr\rho{F[\rho']}}$}
    \AXC{$\theta$ a $\sctx$-environment}
  \ndBIC{$\typejudgeV{\cl*{\slam x e}\theta}{\sarr\rho{F[\rho']}}$}
\ndBIC{$\typejudgeE{\smapv {\sarr\rho F} y {v'} {\cl*{\slam x e}\theta}} 
                   {\sarr\rho{F[\sigma']}}$}
\end{prooftree}
Thus we obtain a typing of the value as
\begin{prooftree}
      \AXC{$\typejudge{y\oftype\rho'}{v'}{\sigma}$}
      \AXC{$\typejudge{\sctx,x\oftype\rho}{e}{F[\rho']}$}
    \ndBIC{$\typejudge{\sctx, x\oftype\rho}{\smap F y {v'} e}{F[\sigma]}$}
  \ndUIC{$\typejudge\sctx{\slam x {\smap F y {v'} e}}{\sarr\rho{F[\sigma]}}$}
  \AXC{$\theta$ a $\sctx$-environment}
\ndBIC{$\typejudgeE{\cl*{\slam x {\smap F y {v'} e}}{\theta}}
                   {\sarr\rho{F[\sigma]}}$}
\end{prooftree}
\end{proofcases}
\end{proof}

\section{Typeability of extracted recurrences}
\label{app:typeability-extracted-recurrences}

In this appendix we prove that extracted recurrences are typeable.
It is worth remembering that 
$\ptrans{\sarr\rho\sigma} = 
\carr{\ptrans\rho}{\ctrans \sigma}$, so
extraction ``commutes'' with type substitution in the
expected way.

\begin{lemma}~
\label{lem:extraction-commutes-with-subst}
$\ptrans{\substin\rho{\vec\sigma}{\vec\alpha}} =
\substin{\ptrans\rho}{\vec{\ptrans\sigma}}{\vec\alpha}$.  Since
shape functors are a subset of types, this implies that
$\ptrans{\substin F{\vec\sigma}{\vec\alpha}} =
\substin{\ptrans F}{\vec{\ptrans\sigma}}{\vec\alpha}$ and
$\ptrans{\sfsubst F\rho} = \sfsubst{\ptrans F}{\ptrans\rho}$
\end{lemma}

\begin{lemma}
\label{lem:rec-term-subst}
If $\typejudge{\cctx, x\oftype\tau'}{e}{\tau}$ and
$\typejudge{\cctx}{e'}{\tau'}$, then
$\typejudge\cctx{\substin e {e'} x}{\tau}$.
\end{lemma}

\begin{lemma}
\label{lem:rec-costpluscpy-type}
If $\typejudge\cctx {c} {\C}$ and 
$\typejudge\cctx e{\ctrans\sigma}$,
then $\typejudge\cctx{\costpluscpy{c}{e}}{\ctrans\sigma}$.
\end{lemma}

\begin{prop*}[Typeability of extracted recurrences,
Prop.~\ref{prop:extracted-recurrences-typeable}]
If $\typejudge\sctx e\sigma$ is in the core language, then 
$\typejudge{\ptrans\sctx}{\ctrans e}{\ctrans\sigma}$.
\end{prop*}
\begin{proof}
The proof is by induction on the derivation of $\typejudge\sctx e\sigma$; we
just do a few of the cases, since they are all fairly routine.

\begin{proofcases}

\item[$\typejudge{\sctx,x\oftype\sforall{\vec\alpha}{\rho}}{x}{\substin\rho{\vec\sigma}{\vec\alpha}}$]
$\ctrans x = \cpair{0,x}$ and 
$\ctrans{\substin\rho{\vec\sigma}{\vec\alpha}} =
\cprod\C{\ptrans{\substin\rho{\vec\sigma}{\vec\alpha}}} =
\cprod\C{\substin{\ptrans\rho}{\vec{\ptrans\sigma}}{\vec\alpha}}$;
the recurrence language typing is
\begin{prooftree}
  \ndAXC{$\typejudge{\ptrans\sctx,x\oftype\sforall{\vec\alpha}{\ptrans\rho}}
                  {0}
                  {\C}$}
    \ndAXC{$\typejudge{\ptrans\sctx,x\oftype\sforall{\vec\alpha}{\ptrans\rho}}
                      {x}
                      {\sforall{\vec\alpha}{\ptrans\rho}}$}
  \ndUIC{$\typejudge{\ptrans\sctx,x\oftype\sforall{\vec\alpha}{\ptrans\rho}}
                    {x}
                    {\substin{\ptrans\rho}{\vec{\ptrans\sigma}}{\vec\alpha}}$}
\ndBIC{$\typejudge{\ptrans\sctx,x\oftype\sforall{\vec\alpha}{\ptrans\rho}}
          {\cpair 0 x}
          {\cprod\C{\substin{\ptrans\rho}{\vec{\ptrans\sigma}}{\vec\alpha}}}$}
\end{prooftree}

\item[$\typejudge\sctx{\slam x e}{\sarr\rho\sigma}$]
$\ctrans{\sarr\rho\sigma} = \cprod\C{(\sarr{\ptrans\rho}{\ctrans\sigma})}$
and we have $\typejudge{\sctx,x\oftype\rho}{e}{\sigma}$, so by the induction
hypothesis,
$\typejudge{\ptrans\sctx,x\oftype\ptrans\rho}
           {\ctrans e}
           {\ctrans\sigma}$ and hence
\begin{prooftree}
  \AXC{$\typejudge\sctx 0 \C$}
    \AXC{$\typejudge{\ptrans\sctx,x\oftype\ptrans\rho}
                    {\ctrans e}
                    {\ctrans\sigma}$}
  \ndUIC{$\typejudge{\sctx}
                    {\clam {(x\oftype\ptrans\rho)} {\ctrans e}}
                    {\carr{\ptrans\rho}{\ctrans\sigma}}$}
\ndBIC{$\typejudge{\ptrans\sctx}
                  {\cpair{0}{\clam {(x\oftype\ptrans\rho)} {\ctrans e}}}
                  {\cprod\C{(\carr{\ptrans\rho}{\ctrans\sigma})}}$}
\end{prooftree}

\item[$\typejudge\sctx{\sfold\delta {e'} x {e}} {\sigma}$]  The typing
derivation has the form
\begin{prooftree}
  \AXC{$\typejudge\sctx{e'}{\delta}$}
  \AXC{$\typejudge{\sctx,x\oftype\sfsubst F{\ssusp\sigma}}
                  {e}
                  {\sigma}$}
\ndBIC{$\typejudge\sctx{\sfold\delta {e'} x {e}} {\sigma}$}
\end{prooftree}
so by the induction hypothesis
$\typejudge{\ptrans\sctx}
           {\ctrans{e'}}
           {\cprod\C{\ptrans{\delta}}}$
and
$\typejudge{\ptrans\sctx,x\oftype{\sfsubst{\ptrans{F}}{\ctrans\sigma}}}
           {\ctrans e}
           {\ctrans\sigma}$.
Writing $(c', p')$ for $\ctrans{e'}$, by inversion we have that
$\typejudge{\ptrans\sctx}{c'}{\C}$ and
$\typejudge{\ptrans\sctx}{p'}{\ptrans\delta}$.  We must show that
$\typejudge{\ptrans\sctx}
   {
\costpluscpy{c'}
       {\cfold{\ptrans\delta}
                {p'}
                {(x : {\sfsubst {\ptrans F}{\ctrans\sigma}})}
                {\costpluscpy{1}{\ctrans{e}}}}}
           {\ctrans\sigma}$.
This follows directly from the typings given by the induction hypothesis,
making use of the fact that
$\ptrans{\tysubst F\eta} = 
\substin{\ptrans F}{\vec{\ptrans\sigma}}{\vec\alpha}$ and 
Lemma~\ref{lem:rec-costpluscpy-type}.

\item[$\typejudge\sctx{\slet x {e'} e}{\sigma}$]
The typing derivation has the form
\begin{prooftree}
  \AXC{$\typejudge\sctx{e'}{\rho}$}
  \AXC{$\typejudge{\sctx,x\oftype\sforall{\vec\alpha}\rho}{e}{\sigma}$}
  \AXC{$\vec\alpha\notin\ftv(\sctx)$}
\ndTIC{$\typejudge\sctx{\slet x {e'} e}{\sigma}$}
\end{prooftree}
The induction hypothesis tells us that
$\typejudge{\ptrans\sctx}{\ctrans{e'}}{\ctrans\rho}$, 
so if $\ctrans e' = \cpair{c'}{p'}$, then
$\typejudge{\ptrans\sctx}{c'}{\C}$ and
$\typejudge{\ptrans\sctx}{p'}{\ptrho}$.
From the latter we conclude that
$\typejudge{\ptrans\sctx}
           {\ctylam{\vec\alpha}{p'}}
           {\cforall{\vec\alpha}{\ptrans\rho}}$
because $\vec\alpha\notin\ftv(\sctx)$ implies that
$\vec\alpha\notin\ftv(\ptrans\sctx)$.
The induction hypothesis also tells us that
$\typejudge{\ptrans\sctx,x\oftype\cforall{\vec\alpha}{\ptrans\rho}}
           {\ctrans e}
           {\ctrans\sigma}$.
Together with Lemma~\ref{lem:rec-term-subst} we conclude that
$\typejudge{\ptrans\sctx}
  {\substin{\ctrans e}{\ctylam{\vec\alpha}{p'}}{x}}
  {\ctrans\sigma}$
and so Lemma~\ref{lem:rec-costpluscpy-type} yields the desired conclusion.
\qedhere
\end{proofcases}
\end{proof}

\section{The syntactic bounding theorem}
\label{app:syn-bounding-thm-proof}

In this appendix we prove the syntactic bounding theorem 
(Theorem~\ref{thm:syn-bounding}).  
The proof relies on two lemmas that describe
bounding for $\smapvkw$ and $\sfoldkw$ expressions.


\begin{lemma}[Syntactic bounding for $\smapvkw$]
\label{lem:bounding-map}
Suppose $\ftv(F)\subseteq\set{t}$ and that the following all hold:
\begin{enumerate}
\item $\typejudge{y\oftype\rho} {v'} \sigma$ and
$\typejudge{y\oftype\ptrans\rho}{E'}{\ptrans\sigma}$.
\item $\typejudgeE{v}{\sfsubst{F}\rho}$ and
$\derives \derE {\vbounded[\sfsubst F{\rho}]{v}{E}}$;
\item If $\typejudgeE{w_0}{\rho}$ and
$\derives{\derE_0}{\vbounded[\rho]{w_0}{E_0}}$ is a subderivation
of~$\derE$ then 
$\vbounded[\sigma]{\substin {v'} {w_0}{y}}{\substin {E'} {E_0} y}$;
\item $\eval{\smapv F y {v'} {v}} {v''}$.
\end{enumerate}
Then 
$\vbounded[\sfsubst F{\sigma}]
          {v''}
          {\cmap*{\ptrans F} \ptrho y {E'} E}$.
\end{lemma}
\begin{proof}
The proof is by induction on~$F$.

\begin{proofcases}

\item[$F = t$]
Assumption (4) tells us that
$v'' = \substin {v'} v y$, so
we must show that $\vbounded[\sigma]{\substin {v'} v y}{\substin {E'} {E} y}$,
which follows from assumption~(3), taking~$v_0$ and~$E_0$ to be~$v$ and~$E$,
respectively.

\item[$F = \tau_0$]
Assumption (4) tells us that
$v'' = v$, so we must show that
$\vbounded[\tau_0]{v}{E}$, which follows from assumption~(2).

\item[$F = \sprod{F_0}{F_1}$]
Assumption~(2) and inversion tells us that $v = \spair{v_0}{v_1}$, and
assumption~(4) tells us that $v'' = \spair{v_0''}{v_1''}$, where
\[
  \AXC{$\setidx{\eval{\smapv{F_i} y {v'} {v_i}}{v_i''}}{i=0,1}$}
\ndUIC{$\eval{\smapv{\sprod{F_0}{F_1}} y {v'} {\spair{v_0}{v_1}}}
             {\spair{v_0''}{v_1''}}$}
\DisplayProof
\]
We must show that 
$\vbounded{\spair{v_0''}{v_1''}}
   {\cpair{\cmap {\ptrans{F_0}} \ptrho y {E'} {\cproj 0 E}}
          {\cmap {\ptrans{F_1}} \ptrho y {E'} {\cproj 1 E}}}$,
for which it suffices to show that
$\vbounded{v_i''}{\cmap{\ptrans{F_i}} \ptrho y {E'} {\cproj i E}}$ 
for $i=0,1$.  
To do so we apply the induction hypothesis taking~$F_i$ for~$F$,
$v_i$ for~$v$, $\cproj i E$ for~$E$, and $v_i''$ for~$v''$.
Verifying the assumptions is straightforward, noting that
(3) follows because the derivation that
$\vbounded{v_i}{\cproj i E}$ is a subderivation of
$\vbounded{\spair{v_0}{v_1}}{E}$.

\item[$F = \ssum{F_0}{F_1}$]
Assumption~(2) and inversion tells us that 
$v = \sinj i {v_i}$, where there is~$E_i$ such that
$\vbounded{v_i}{E_i}$ and 
$\cinj i {E_i} \szleq_{\sfsubst{\ptrans{F_i}}\rho} E$.
Assumption~(4) tells us that $v'' = \sinj i {v_i''}$, where
\[
  \AXC{$\eval{\smapv {F_i} y {v'} {v_i}}{v_i''}$}
\ndUIC{$\eval{\smapv{\ssum{F_0}{F_1}} y {v'} {\sinj i {v_i}}}
             {\sinj i{v_i''}}$}
\DisplayProof
\]
We must show that
\[
\vbounded{\sinj i {v_i''}}
  {\ccaset* E x {\sfsubst{\ptrans{F_i}}{\ptrans\rho}}
            {\cinj i {(\cmap{\ptrans{F_i}}
                            {\ptrans\rho}
                            y {E'} x}}}.
\]
Let us write~$E^*$ for the right-hand side.
Now we must show that there is~$E_i''$ such that
$\vbounded{v_i''}{E_i''}$ and $\cinj i{E_i''}\szleq E^*$.
We apply the induction hypothesis taking $F_i$ for~$F$,
$v_i$ for~$v$, $E_i$ for~$E$, and $v_i''$ for~$v''$ to conclude
that $\vbounded{v_i''}{E_i''}$ where
$E_i'' = \cmap{\ptrans{F_i}}\ptrho y {E'} {E_i}$, and we
notice that
\begin{align*}
\cinj i {E_i''}
  &= \cinj i {(\cmap{\ptrans{F_i}}\ptrho y {E'} {E_i})} \\
  &\leq 
     {\ccaset* {\cinj i {E_i}} x {\sfsubst{\ptrans{F_i}}{\ptrans\rho}}
            {\cinj i {(\cmap{\ptrans{F_i}}
                            {\ptrans\rho}
                            y {E'} x}}} \\
  &\leq E^*
\end{align*}
as required.  The assumptions for the induction hypothesis are straightforward
to verify, noting that~(3) follows because the derivation that
$\vbounded{v_i}{E_i}$ is a subderivation of
$\vbounded{\sinj i {v_i}}{E}$.

\item[$F = \sarr{\tau_0}{F_0}$]
Assumption~(2) and inversion tells us that $v = \cl*{\slam x e}\theta$, and
assumption~(4) tells us that 
$v'' = \cl*{\slam x {\smap F y {v'} e}}\theta$.
We must show that
\begin{align*}
\cl*{\slam x {\smap F y {v'} e}}\theta
  &\vboundedby \cmap{\ptrans{\carr{\tau_0}{F_0}}}\ptrho y {E'} E \\
  &=\clam* x {\ptrans{\tau_0}} {\cpair{\ccost{(\capp E x)}}
                                      {\cmap {\ptrans{F_0}}
                                             \ptrho
                                             y
                                             {E'}
                                             {\cpot{(\capp E x)}}}}.
\end{align*}
To do so, fix $\vbounded[\tau_0]{v_1}{E_1}$; it suffices to show that
$\bounded[\sfsubst F \rho]
         {\cl*{\smap F y {v'} e}{\bindin\theta x {v_1}}}
         {\cpair{\ccost{(\capp E {E_1})}}
                {\cmap{\ptrans F}
                   \ptrho y {E'} {\cpot{(\capp E {E_1})}}}}$.
The evaluation of the left-hand side has the form
\[
  \AXC{$\evalin{\cl e {\bindin\theta x {v_1}}} {w'} n$}
  \AXC{$\eval{\smapv F y {v'} {w'}}{w}$}
\ndBIC{$\evalin{\cl*{\smap F y {v'} e}{\bindin\theta x {v_1}}}{w}{n}$}
\DisplayProof
\]
so by Lemma~\ref{lem:cost-pot-weakening} it suffices to show that
$n\leq \ccost{(\capp E {E_1})}$ and
$\vbounded w 
   {\cmap{\ptrans{F}}\ptrho y {E'} {\cpot{(\capp E {E_1})}}}$.
Recalling that $v = \cl*{\slam x e}\theta$ and
$\vbounded{v}{E}$ by assumption~(2), we have that
$\bounded{\cl e {\bindin\theta x {v_1}}}{\capp E {E_1}}$, and hence
$n\leq \ccost{(\capp E {E_1})}$ (our first obligation) and
$\vbounded {w'} {\cpot{(\capp E {E_1})}}$.
To show that
$\vbounded w 
   {\cmap {\ptrans{F}} \ptrho y {E'} {\cpot{(\capp E {E_1})}}}$ we
apply the induction hypothesis taking $F_0$ for~$F$, $w'$ for~$v$,
$\cpot{(\capp E {E_1})}$ for~$E$, and $w$ for $v''$.  Assumptions~(1),
(2), and~(4) are straightforward to verify.  For assumption~(3), suppose that
$\typejudgeE{w_0}{\rho}$ and
$\derives{\derE_0}{\vbounded{w_0}{E_0}}$ is a subderivation of
$\derives{\derE'}{\vbounded{w'}{\cpot{(\capp E {E_1})}}}$.  
We need to show that
$\vbounded{\substin {v'} {w_0} y}{\substin{E'}{E_0}{y}}$.  To do so,
it suffices to show that $\derE_0$ is a subderivation of 
$\derives{\derE}{\vbounded{v}{E}}$, and for this it suffices
to show that $\derE'$ is a subderivation of $\derE$.
This follows from examining $\derE$:
\[
  \AXC{$\dotsb$}
    \AXC{$\evalin {\cl e {\bindin\theta x {v_1}}} {w'} n$}
    \AXC{$n\leq\ccost{(\capp E {E_1})}$}
      \AXC{$\derE'$}
    \noLine
    \ndUIC{$\vbounded{w'}{\cpot{(\capp E {E_1})}}$}
  \ndTIC{$\bounded{\cl e{\bindin\theta x {v_1}}}{\capp E {E_1}}$}
  \AXC{$\dotsb$}
\ndTIC{$\vbounded{\cl*{\slam x e}\theta}{E}$}
\DisplayProof
\]
\end{proofcases}
\end{proof}


\begin{lemma}[Syntactic bounding for $\sfoldkw$]
\label{lem:bounding-fold}
Suppose the following all hold:
\begin{enumerate}
\item 
$\bounded[\sigma]
   {(\typejudge{\sctx,x\oftype \sfsubst{F}{\ssusp\sigma}}
               e
               \sigma)}
   {(\typejudge{\ptrans\sctx,
                x\oftype\ptrans{\sfsubst{F}{\ssusp\sigma}}}
               {E}
               {\ctrans\sigma}}$;
\item $\vbounded[\sctx-x]\theta\Theta$ (w.l.o.g., $x\notin\dom\Theta$);
\item $\vbounded[\delta]{v'}{E'}$.
\end{enumerate}
Then
$\bounded[\sigma]
   {\cl*{\sfold\delta y x e}{\bindin\theta y {v'}}}
   {\cfold{\ptrans\delta}
            {E'}
            {(x : {\sfsubst{\ptrans{F}}{\ctrans\sigma}})}
            {\costpluscpy 1 {\esubst E\Theta}}}
$.
\end{lemma}
\begin{proof}
The proof is by induction on the derivation of assumption~(3), which
necessarily ends with the rule
\[
  \AXC{$\vbounded[F,\delta]{v'}{E''}$}
  \AXC{$\ccons*{\ptrans\delta}{E''}
        \szleq_{\ptrans{\delta}} E'$}
\ndBIC{$\vbounded[\delta]{\scons\delta {v'}}{E'}$}
\DisplayProof
\]
To reduce notational clutter, we will write
$E^*[z]$ for
${\cfold{\ptrans\delta}
       z
       {(x : {\sfsubst{\ptrans{F}}{\ctrans\sigma}})}
       {\costpluscpy 1 {\esubst E\Theta}}}$,
so we must show that
$\bounded{\cl*{\sfold\delta y x e}{\bindin\theta y {\scons\delta v'}}}
         {E^*[E']}$.
Using the axioms for~$\szleq$, we have that
$E^*[E']
  \geq E^*[\ccons{\ptrans\delta}{E''}]
  \geq \costpluscpy 1
        {\substin
          {\esubst E \Theta}
          {\cmap{\ptrans{F}}
                {\ptrans{\delta}}
                y
                {E^*[y]}
                {E''}} x}$.
The evaluation of interest has the form
\[
  \AXC{$\evalin{\cl y {\bindin\theta y {\scons\delta{v'}}}}
               {\scons\delta v'}
               0$}
  \AXC{$\eval{\smapv F y {\cl*{\sdelay*{\sfold\delta y x e}}{\theta}} {v'}}
             {v''}$}
  \AXC{$\evalin{\cl e {\bindin\theta x {v''}}}{v}{n}$}
\ndTIC{$\evalin{\cl*{\sfold\delta y x e}{\bindin\theta y {\scons\delta v'}}}
               {v}
               {n+1}$}
\DisplayProof
\]
We apply Lemma~\ref{lem:bounding-map} 
by taking $F$ for~$F$,
$\delta$ for~$\rho$,
$\cl*{\sdelay*{\sfold\delta y x e}}{\theta}$ for~$v'$,
$E^*[y]$ for~$E'$, $v'$ for~$v$, $E''$ for~$E$, and $v''$ for~$v''$
(we verify the assumptions momentarily)
to conclude that
$\vbounded{v''}
          {\cmap{\ptrans{F}} 
                {\ptrans{\delta}} y {E^*[y]} {E''}}$,
so by~(2),
$\vbounded[\sctx]{\bindin\theta x {v''}}
                 {\substin\Theta{\cmap{\ptrans{F}} 
                                      {\ptrans{\delta}} 
                                      y {E^*[y]} {E''}} x}$
and so by~(1),
$\bounded{\cl e {\bindin\theta x {v''}}}
         {\esubst E {\substin\Theta{\cmap{\ptrans{F}} 
                                      {\ptrans{\delta}} 
                                      y {E^*[y]} {E''}} x}}$.
This tells us that
\begin{align*}
1 + n 
&\leq 
1 + \ccost{(\esubst E {\substin\Theta{\cmap{\ptrans{F}} 
                                          {\ptrans{\delta}} 
                                          y {E^*[y]} {E''}} x})} \\
&=
\ccost{(\costpluscpy 1 
         {\esubst E {\substin\Theta{\cmap{\ptrans{F}} 
                                         {\ptrans{\delta}} 
                                         y {E^*[y]} {E''}} x}})} \\
&\leq
\ccost{(E^*[E'])}
\end{align*}
and
\begin{align*}
v
&\vboundedby
\cpot{(\esubst E {\substin\Theta{\cmap{\ptrans{F}} 
                                          {\ptrans{\delta}} 
                                          y {E^*[y]} {E''}} x})} \\
&\leq \cpot{(E^*[E'])}
\end{align*}
as needed.

We just need to verify the assumptions of Lemma~\ref{lem:bounding-map}:
\begin{enumerate}
\item $\typejudge{y\oftype\delta}
                 {\cl*{\sdelay*{\sfold\delta y x e}}\theta}
                 {\ssusp\sigma}$ and
      $\typejudge{y\oftype\ptrans{\delta}}
                 {E^*[y]}
                 {\ptrans{\ssusp\sigma}}$.

\item $\typejudgeE{v'}{\sfsubst{F}{\delta}}$ and
      $\vbounded{v'}{E''}$ with derivation~$\derE$.

\item If $\typejudgeE{w_0}{\delta}$ and
      $\derives{\derE_0}{\vbounded{w_0}{E_0}}$ is a subderivation 
      of~$\derE$, then
      $\vbounded{\cl*{\sdelay*{\sfold \delta y x e}}{\bindin\theta y {w_0}}}
                {\substin{E^*}{E_0}{y}}$.

\item $\eval{\smapv F y 
               {\cl*{\sdelay*{\sfold \delta y x e}}{\theta}}
               {v'}}
            {v''}$
\end{enumerate}
(1), (2), and (4) are immediate.  Under the assumptions of~(3), we
must show that 
$\bounded{\cl*{\sfold\delta y x e}{\bindin\theta y {w_0}}}
         {\substin{E^*}{E_0}{y}}$.  Since $\derE_0$ is a subderivation
of~$\derE$, the main induction hypothesis applies.
\end{proof}


\begin{thm*}[Syntactic bounding theorem, Thm.~\ref{thm:syn-bounding}]
If $\typejudge\sctx e\sigma$ is in the core language, then
$\bounded[\sigma] 
 {(\typejudge\sctx e \sigma)} 
 {(\typejudge{\ptrans\sctx} {\ctrans e} {\ctrans\sigma})}$.
\end{thm*}
\begin{proof}
The proof is by induction on~$\typejudgeG e \tau$.
Most cases proceed by showing that $\bounded e {\cpair c p}$ for some $c$
and $p$, where
$\evalin e v n$.  By~\infruleref{beta-times}, $c\leq\ccost{\cpair c p}$ and
$p\leq \cpot{\cpair c p}$, so it suffices to show that
$n\leq c$ and $\vbounded v p$,
and we take advantage of this fact silently.
\begin{proofcases}

\item[$\typejudge {\sctx,x\oftype\sforall{\vec\alpha}{\sigma}} x {\tysubst*\sigma{\vec\sigma}{\vec\alpha}}$]
Fix $\vbounded[\sctx,x\oftype\sforall{\vec\alpha}\sigma]\theta\Theta$; 
we must show that 
$\cl x\theta\boundedby[\substin\sigma{\vec\sigma}{\vec\alpha}]
  \esubst{\cpair 0 x}{\Theta} = {\cpair{0}{\Theta(x)}}$.  
The evaluation 
of~$\cl x\theta$ has the form
\[
\ndAXC{$\evalin{\cl x\theta}{\theta(x)}{0}$}
\DisplayProof
\]
The cost bound is immediate.  For the value bound we must show that
$\vbounded[\substin\sigma{\vec\sigma}{\vec\alpha}]{\theta(x)}{\Theta(x)}$.
This follows from the definition of
$\vbounded[\sctx,x\oftype\sforall{\vec\alpha}\sigma]\theta\Theta$.

\item[$\typejudge\sctx\striv\sunit$]
    Fix $\vbounded[\Gamma]\theta\Theta$; we must show that
    $\cl\striv\theta \boundedby[\sunit]{\esubst{\cpair{0}{\ctriv}}{\Theta}} =
    \cpair 0 \ctriv$.  The evaluation of $\cl\striv\theta$ has the form
    \begin{prooftree}
    \ndAXC{$\evalin {\cl \striv\theta} {\ctriv} 0$}
    \end{prooftree}
    and we have that
    (cost) $0\leq 0$ and
    (value) $\vbounded\striv\ctriv$ by the definition of $\vboundedby[\sunit]$.

\item[$\typejudge\sctx{\spair{e_0}{e_1}}{\sprod{\sigma_0}{\sigma_1}}$]
    Fix $\vbounded[\sctx]\theta\Theta$; we must show that
    ${\cl{\spair{e_0}{e_1}}\theta}%
    \boundedby%
    {\esubst{\cpair{c_0+c_1}{\cpair{p_0}{p_1}}}\Theta} =
    \cpair{\esubst*{c_0+c_1}\Theta}{\esubst{\cpair{p_0}{p_1}}\Theta}$, 
    where $\ctrans{e_i} = (c_i, p_i)$.  The evaluation of
    $\cl{\spair{e_0}{e_1}}\theta$ has the form
    \begin{prooftree}
        \AXC{$\evalin{\cl{e_0}{\theta}}{v_0}{n_0}$}
        \AXC{$\evalin{\cl{e_1}{\theta}}{v_1}{n_1}$}
        \ndBIC{$\evalin{\cl{\spair{e_0}{e_1}}{\theta}}%
                       {\spair{v_0}{v_1}}{n_0+n_1}$}
    \end{prooftree}
    \begin{description}
        \item[Cost] $n_i\leq \esubst{c_i}\Theta$ by the IH so 
        $n_0+n_1\leq \esubst{c_0}\Theta + \esubst{c_1}\Theta =
        \esubst*{c_0+c_1}\Theta$.
        \item[Value] $\vbounded{v_i}{\esubst{p_i}{\Theta}}$ by the IH so
            ${\spair{v_0}{v_1}}\vboundedby{\cpair{\esubst{p_0}\Theta}{\esubst{p_1}\Theta}} = \esubst{\cpair{p_0}{p_1}}\Theta$ by the IH.
    \end{description}

\item[$\typejudge\sctx{\sproj i e}{\sigma_i}$]
    Fix $\vbounded[\sctx]\theta\Theta$; we must show that
    ${\cl*{\sproj i e}\theta}%
    \boundedby%
    \esubst{\cpair{c}{\cproj i p}}\Theta =
    \cpair{\esubst c \Theta}{\esubst*{\cproj i p}\Theta}$, where
    ${\ctrans{e}} = \cpair c p$.
    The evaluation of $\cl*{\sproj i e}\theta$
    has the form
    \begin{prooftree}
        \AXC{$\evalin{\cl e\theta}{\cpair{v_0}{v_1}}{n}$}
        \ndUIC{$\evalin{\cl*{\sproj i e}\theta}{v_i}{n}$}
    \end{prooftree}
    \begin{description}
        \item[Cost] $n\leq \esubst* c \Theta$ by the IH.
        \item[Value] $\vbounded{\spair{v_0}{v_1}}{\esubst p \Theta}$ 
            by the IH, so
            ${v_i}\vboundedby 
            {\cproj i {(\esubst p \Theta)}} = \esubst*{\cproj i p}\Theta$ 
            by the definition of
            $\vboundedby[\cprod{\sigma_0}{\sigma_1}]$.
    \end{description}

\item[$\typejudge\sctx{\sinj i e}{\ssum{\sigma_0}{\sigma_1}}$]
    Fix $\vbounded[\sctx]\theta\Theta$; we must show that
    ${\cl*{\sinj i e}{\theta}}%
    \boundedby%
    \esubst{\cpair{c}{\cinj i p}}\Theta =
    \cpair{\esubst c \Theta}{\esubst*{\cinj i p}\Theta}$, where
    ${\ctrans e} = (c, p)$.
    The evaluation of $\cl*{\sinj i e}\theta$
    has the form
    \begin{prooftree}
          \AXC{$\evalin{\cl e\theta}{v}{n}$}
        \ndUIC{$\evalin{\cl*{\sinj i e}\theta}{\cinj i v}{n}$}
    \end{prooftree}
    \begin{description}
        \item[Cost] $n\leq \esubst c\Theta$ by the IH.
        \item[Value] $\vbounded v {\esubst p \Theta}$ by the IH, and
            $\cinj i {(\esubst p \Theta)}\szleq \cinj i {(\esubst p \Theta)}$,
            so
            ${\sinj i v}\vboundedby
            {\cinj i {(\esubst p \Theta)}}=
            \esubst*{\cinj i p}\Theta$ by the definition of
            $\vboundedby[\ssum{\sigma_0}{\sigma_1}]$.
    \end{description}

\item[$\typejudge\sctx{\scase* e x {e}}{\sigma}$]
    Fix $\vbounded[\sctx]\theta\Theta$; we must show that
    ${\cl*{\scase* e x e}\theta}%
    \boundedby%
    \esubst*{\costpluscpy c {\ccaset* p x {\ptrans{\sigma_i}} 
                                     {\cpair {c_i}{p_i}}}}\Theta =
    {\costpluscpy {\esubst c \Theta} 
                  {\ccaset*
                    {\esubst p\Theta} 
                    x 
                    {\ptrans{\sigma_i}}
                    {\esubst{\cpair{c_i}{p_i}}{\Theta-x}}}}$,
    where
    ${\ctrans e} = \cpair c p$ and
    $\ctrans{e_i} = \cpair{c_i}{p_i}$.
    The evaluation of
    ${\cl*{\scase* e x {e}}\theta}$
    has the form
    \begin{prooftree}
          \AXC{$\evalin {\cl e \theta} {\sinj i v} {n}$}
          \AXC{$\evalin {\cl{e_i}{\subst{\theta} v x}} {v_i} {n_i}$}
        \ndBIC{$\evalin{\cl*{\scase* e x {e}}\theta}{v_i}{n+n_i}$}
    \end{prooftree}
    By the IH for~$e$, $\vbounded{\sinj i v}{\esubst p \Theta}$, 
    so there is
    some $E'$ such that 
    $\vbounded v E'$
    and
    $\sinj i{E'}\leq \esubst p\Theta$.
    If we set $\theta' = \bindin{\theta}{v}{x}$ and
    $\Theta' = \substin{\Theta}{E'}{x}$, then
    $\vbounded{\theta'}{\Theta'}$, so by the IH for~$e_i$,
    $\bounded{\cl{e_i}{\theta'}}
             {\cpair{\esubst{c_i}{\Theta'}}{\esubst{p_i}{\Theta'}}}$.
    Since $\cinj i {E'} \szleq \esubst p\Theta$, we have
    \begin{align*}
    \cpair{\esubst{c_i}{\Theta'}}{\esubst{p_i}{\Theta'}}
      &= \esubst{\cpair{c_i}{p_i}}{\Theta'}  \\
      &\leq \ccaset* {\cinj i {E'}} x {\ptrans{\sigma_i}} {\esubst{\cpair{c_i}{p_i}}{\Theta-x}} \\
      &\leq \ccaset* {\esubst p\Theta} x {\ptrans{\sigma_i}} {\esubst{\cpair{c_i}{p_i}}{\Theta-x}}
    \end{align*}
    and so
    \begin{align*}
    \cpair{\esubst c \Theta + \esubst{c_i}{\Theta'}}{\esubst{p_i}{\Theta'}}
      &= \costpluscpy{\esubst c \Theta}{\cpair{\esubst{c_i}{\Theta'}}{\esubst{p_i}{\Theta'}}} \\
      &\szleq \costpluscpy
                {\esubst c \Theta}
                {\ccaset* 
                   {\esubst p\Theta} 
                   x 
                   {\ptrans{\sigma_i}}
                   {\esubst{\cpair{c_i}{p_i}}{\Theta-x}}}
    \end{align*}
    which we use to complete the next set of calculations.
    \begin{description}
        \item[Cost] $n\leq \esubst c \Theta$ and 
            $n_i\leq \esubst{c_i}{\Theta'}$, so
            \begin{align*}
            n+n_i 
              &\leq \esubst c \Theta + \esubst{c_i}{\Theta'} \\
              &\leq \ccost{\cpair{\esubst c \Theta+\esubst{c_i}{\Theta'}}{\esubst{p_i}{\Theta'}}} \\
              &\leq \ccost{(\costpluscpy
                            {\esubst c \Theta}
                            {\ccaset* 
                              {\esubst p\Theta} 
                              x 
                              {\ptrans{\sigma_i}}
                              {\esubst{\cpair{c_i}{p_i}}{\Theta-x}}})}.
            \end{align*}
        \item[Value]\hfill
            \begin{align*}
            v_i 
              &\vboundedby \esubst{p_i}{\Theta'} \\
                &\szleq\cpot{\cpair{\esubst c \Theta+\esubst{c_i}{\Theta'}}{\esubst{p_i}{\Theta'}}} \\
                &\szleq\cpot{(\costpluscpy
                            {\esubst c \Theta}
                            {\ccaset* 
                              {\esubst p\Theta} 
                              x 
                              {\ptrans{\sigma_i}}
                              {\esubst{\cpair{c_i}{p_i}}{\Theta-x}}})}.
            \end{align*}
    \end{description}

\item[$\typejudge\sctx{\slam x e}{\sarr{\sigma'}{\sigma}}$]
Fix $\vbounded[\sctx]\theta\Theta$; we must
show that
${\cl*{\slam x e}\theta}
\boundedby
{\esubst{\cpair{0}{\clam* x {\ptrans{\sigma'}} {\ctrans e}}}\Theta} = 
\cpair{0}{\clam* x {\ptrans{\sigma'}} {\esubst{\ctrans e}{\Theta-x}}}$.  
The evaluation of $\cl*{\slam x e}{\theta}$ has the form
\begin{prooftree}
\ndAXC{$\evalin {\cl*{\slam x e}{\theta}} {\cl*{\slam x e}{\theta}} 0$}
\end{prooftree}
so the cost claim is immediate.
\begin{description}
\item[Value]
Fix any $\vbounded{v'}{E'}$.  We must show that
$\bounded%
{\cl e{\subst{\theta}{v'}{x}}}
{\capp*{\clam* x {\ptrans{\sigma'}} {\esubst{\ctrans e}{\Theta-x}}}{E'}}$; 
by definition, \infruleref{beta-to}, and Weakening, it suffices to show
that 
$\bounded{\subst{e\theta} {v'} x}{\esubst{\ctrans e}{\extend\Theta{E'}{x}}}$.
Since $\vbounded\theta\Theta$ and $\vbounded v{E'}$, this follows from
the induction hypothesis.
\end{description}

\item[$\typejudge\sctx {\sapp {e_0} {e_1}}{\sigma}$]
Fix $\vbounded\theta\Theta$.  We must show that
$\cl*{\sapp{e_0}{e_1}}\theta%
\boundedby%
\esubst*{\costpluscpy{(c_0+c_1)}{\capp {p_0}{p_1}}}{\Theta}$,
where
${\ctrans{e_i}} = \cpair{c_i}{p_i}$.
The evaluation of $\cl*{\sapp{e_0}{e_1}}\theta$ has the form
\begin{prooftree}
  \AXC{$\evalin {\cl{e_0}\theta} {\cl*{\slam x {e_0'}}{\theta'}} {n_0}$}
  \AXC{$\evalin {\cl{e_1}\theta} {v_1} {n_1}$}
  \AXC{$\evalin {\cl{e_0'}{\subst{\theta'}{v_1}{x}}} {v} {n}$}
\ndTIC{$\evalin {\cl*{\sapp{e_0}{e_1}}{\theta}} {v} {n_0+n_1+n}$}
\end{prooftree}
By the IH, $n_0\leq \esubst{c_0}{\Theta}$, 
$\vbounded{\cl*{\slam x {e_0'}}{\theta'}}{\esubst{p_0}{\Theta}}$,
$n_1\leq \esubst{c_1}{\Theta}$, and 
$\vbounded{v_1}{\esubst{p_1}{\Theta}}$.  By definition of~$\vboundedby$,
${\cl{e_0'}{\subst{\theta'}{v_1}{x}}}
\boundedby%
{\capp*{\esubst{p_0}\Theta}{(\esubst{p_1}\Theta)}} =
\esubst*{\capp{p_0}{p_1}}\Theta$,
so $n\leq \ccost{(\esubst*{\capp{p_0}{p_1}}\Theta)}$ and 
$\vbounded v {\cpot{(\esubst*{\capp{p_0}{p_1}}\Theta)}}$.
\begin{description}
\item[Cost]
  $n_0 + n_1 + n \leq 
  \esubst{c_0}\Theta + \esubst{c_1}\Theta + 
      \ccost{(\esubst*{\capp{p_0}{p_1}}\Theta)} \szleq
  \ccost{(\esubst*{\costpluscpy{(c_0+c_1)}{\capp{p_0}{p_1}}}\Theta)}$.
\item[Value]
  $v\vboundedby
  {\cpot{(\esubst*{\capp{p_0}{p_1}}\Theta)}} \szleq 
  \cpot{(\esubst*{\costpluscpy{(c_0+c_1)}{\capp{p_0}{p_1}}}{\Theta})}$.
\end{description}

\item[$\typejudge\sctx{\sdelay e}{\ssusp\sigma}$]
Fix $\vbounded\theta\Theta$.  We must show that
$\cl*{\sdelay e}\theta \boundedby
\esubst{\cpair 0 {\ctrans e}}{\Theta} = \cpair 0 {\esubst{\ctrans e}\Theta}$.
The evaluation of $\cl*{\sdelay e}\theta$ has the form
\begin{prooftree}
\ndAXC{$\evalin{\cl*{\sdelay e}\theta} {\cl*{\sdelay e}\theta} 0$}
\end{prooftree}
so (cost) $0\leq 0$ and (value) since 
$\bounded{\cl e\theta}{\esubst{\ctrans e}{\Theta}}$ by the IH,
$\vbounded{\cl*{\sdelay e}{\theta}}{\esubst{\ctrans e}\Theta}$ 
by the definition of $\vboundedby[\ssusp\sigma]$.

\item[$\typejudge\sctx{\sforce e}{\sigma}$]
Fix $\vbounded\theta\Theta$.  We must show that
$\cl*{\sforce e}\theta \boundedby
\esubst*{\costpluscpy c p}{\Theta}$,
where ${\ctrans e} = \cpair c p$.  The
evaluation of $\cl*{\sforce e}\theta$ has the form
\begin{prooftree}
  \AXC{$\evalin {\cl e\theta} {\cl*{\sdelay {e'}}{\theta'}} n$}
  \AXC{$\evalin {\cl{e'}{\theta'}} {v} {n'}$}
\ndBIC{$\evalin {\cl*{\sforce e}\theta} v {n+n'}$}
\end{prooftree}
By the IH, $n\leq \esubst c \Theta$ and 
$\vbounded{\cl*{\sdelay{e'}}{\theta'}} {\esubst p \Theta}$, so by definition
of $\vboundedby$, $\bounded{\cl{e'}{\theta'}}{\esubst p\Theta}$ and hence
$n' \leq \ccost{(\esubst p\Theta)}$ and 
$\vbounded v {\cpot{(\esubst p\Theta)}}$.
So (cost) 
$n+n' \leq 
\esubst c\Theta + \ccost {\esubst p\Theta} \szleq 
\ccost{(\esubst*{\costpluscpy c p}{\Theta})}$ and
(value) 
$v\vboundedby 
{\cpot {\esubst p \Theta}} \szleq 
\cpot{(\esubst*{\costpluscpy c p}{\Theta})}$.

\item[$\typejudge\sctx {\scons\delta e} {\delta}$]
Fix $\vbounded\theta\Theta$.  We must show that
$\cl*{\scons\delta e}{\theta}\boundedby 
\esubst{\cpair{c}{\ccons{\ptrans\delta} p}}{\Theta}$, 
where
${\ctrans e} = \cpair c p$.  The evaluation of
$\cl*{\scons\delta e}{\theta}$ has the form
\begin{prooftree}
  \AXC{$\evalin {\cl e\theta} v n$}
\ndUIC{$\evalin {\cl*{\scons\delta e}\theta} {\scons\delta v} n$}
\end{prooftree}
\begin{description}
\item[Cost]
$n\leq \esubst c\Theta$ by the IH.
\item[Value]
By the IH we have that 
$\vbounded[\ptrans{\sfsubst{F}{\delta}}] 
          v {\esubst p\Theta}$, and
so by Lemma~\ref{lem:bounding-datatype-inversion},
$\vbounded[F,\delta] v {\esubst p \Theta}$.
Since
$\ccons{\ptrans\delta} p \leq
\ccons{\ptrans\delta} p$, the value bound 
follows by definition of $\vboundedby[\delta]$.
\end{description}

\item[$\typejudge\sctx {\sdest\delta e} {\sfsubst{F}{\delta}}$]
Fix $\vbounded\theta\Theta$.  We must show that
$\cl*{\sdest\delta e}\theta \boundedby
\esubst{\cpair{c}{\cdest{\ptrans\delta} p}}{\Theta} =
\cpair{\esubst c\Theta}
      {\cdest{\ptrans\delta}{(\esubst p\Theta)}}$, 
where
${\ctrans e} = \cpair c p$.  The evaluation of
$\cl*{\sdest\delta e}{\theta}$ has the form
\begin{prooftree}
  \AXC{$\evalin {\cl e\theta} {\scons\delta v} {n}$}
\ndUIC{$\evalin {\cl*{\sdest\delta e}{\theta}} {v} n$}
\end{prooftree}
\begin{description}
\item[Cost]
$n\leq \esubst c \Theta$ by the IH.
\item[Value]
By the IH, $\vbounded{\scons\delta v}{\esubst p \Theta}$, and so by
definition of $\vboundedby[\delta]$,
there is $E$ such that
$\vbounded[F,\delta] v E$ and 
$\ccons{\ptrans{\delta}} E 
 \szleq \esubst p\Theta$.
This latter fact along with the axioms for~$\szleq$ tell us that
$E \szleq \cdest{\ptrans{\delta}}(%
                \ccons{\ptrans{\delta}} E)
   \szleq \cdest{\ptrans{\delta}}
                (\esubst p\Theta)$.
\end{description}

\item[$\typejudge\sctx{\sfold \delta {e'} x e}{\sigma}$]
The type derivation has the form
\begin{prooftree}
  \AXC{$\typejudge\sctx{e'}{\delta}$}
  \AXC{$\typejudge{\sctx,x\oftype\sfsubst{F}{\ssusp\sigma}}
                  e
                  \sigma$}
\ndBIC{$\typejudge\sctx{\sfold \delta {e'} x e}\sigma$}
\end{prooftree}
Fix
$\vbounded[\sctx]\theta\Theta$ and without loss of generality assume
that $x\notin\dom\sctx\union\dom\theta\union\dom\Theta$; we must show that
$\cl*{\sfold\delta{e'} x e}{\theta}
 \boundedby
 \costpluscpy{c'}
       {\cfold{\ptrans\delta}
                {p'}
                {(x : {\sfsubst {\ptrans F}{\ctrans\sigma}})}
                {\costpluscpy{1}{\ctrans{e}}}}$
where $\ctrans{e'} = (c', p')$.
The evaluation of $\cl*{\sfold\delta{e'} x e}{\theta}$ has the form
\begin{prooftree}
  \AXC{$\evalin{\cl{e'}\theta}{\scons\delta{v'}}{n'}$}
  \AXC{$\eval{\smapv F y {\cl*{\sdelay*{\sfold\delta y x e}}\theta} {v'}}
             {v''}$}
  \AXC{$\evalin{\cl e {\bindin\theta x {v''}}}{v}{n}$}
\ndTIC{$\evalin{\cl*{\sfold\delta{e'} x e}{\theta}}{v}{n'+n+1}$}
\end{prooftree}
and so the following is also an evaluation, where we write
$\theta'$ for $\bindin\theta z {\scons\delta{v'}}$:
\begin{prooftree}
  \AXC{$\evalin{\cl z {\theta'}}{\scons\delta{v'}}{0}$}
  \AXC{$\eval{\smapv F y {\cl*{\sdelay*{\sfold\delta y x e}}{\theta'}} {v'}}
             {v''}$}
  \AXC{$\evalin{\cl e {\bindin{\theta'} x {v''}}}{v}{n}$}
\ndTIC{$\evalin{\cl*{\sfold\delta z x e}{\theta'}}{v}{n+1}$}
\end{prooftree}

The IH for~$e'$ tells us that
$n'\leq \esubst{c'}{\Theta}$ and
$\vbounded{\scons\delta{v'}}{\esubst{p'}\Theta}$; combined with the
IH for~$e$, Lemma~\ref{lem:bounding-fold} tells us that
$\bounded{\cl*{\sfold\delta z x e}{\theta'}}
         {\cfold{\ptrans\delta}
         {\esubst {p'}\Theta}
         {(x : {\sfsubst{\ptrans{F}}{\ctrans\sigma}})}
         {\costpluscpy 1 {\esubst{\ctrans e}{\Theta}}}}$
as required.

\item[$\typejudge\sctx{\slet x {e'} e}{\sigma}$]
The type derivation has the form
\begin{prooftree}
  \AXC{$\typejudge\sctx {e'} {\sigma'}$}
  \AXC{$\typejudge{\sctx,x\oftype{\cforall{\vec\alpha}{\sigma'}}}{e}{\sigma}$}
  \AXC{$\vec\alpha$ not free in any~$\Gamma(y)$}
\ndTIC{$\typejudge\sctx{\slet x {e'} {e}}{\sigma}$}
\end{prooftree}
Fix $\vbounded[\sctx]\theta\Theta$ and without loss of generality
assume $x$ is fresh for $\sctx$, $\theta$, and~$\Theta$ and that
no $\alpha_i$ is free in any~$\Theta(y)$.  We must show that
$\cl*{\slet x {e'} e}{\theta}
 \boundedby \esubst{(\costpluscpy
                     {c'}
                     {\substin{\ctrans e}{\ctylam{\vec\alpha}{p'}}{x}})}
                   \Theta
 = \costpluscpy
   {\esubst{c'}{\Theta}}
   {\substin{\esubst{\ctrans e}{\Theta}}
            {\ctylam{\vec\alpha}{\esubst{p'}{\Theta}}}
            {x}}$
where $\ctrans{e'} = (c', p')$.
The evaluation has the form
\begin{prooftree}
  \AXC{$\evalin {\cl{e'}\theta} {v'} {n'}$}
  \AXC{$\evalin {\cl e {\bindin\theta x{v'}}} v n$}
\ndBIC{$\evalin {\cl*{\slet x {e'} e}\theta} v {n'+n}$}
\end{prooftree}
The IH for~$e'$ tells us that $n'\leq \esubst{c'}{\Theta}$ and
$\vbounded[\sigma']{v'}{\esubst{p'}{\Theta}}$.  If we can show that
$\vbounded[\sforall{\vec\alpha}{\sigma'}]
          {v'}
          {\ctylam{\vec\alpha}{\esubst{p'}{\Theta}}}$,
then the induction hypothesis applied to~$e$ provides the remaining
pieces of the argument.
For this we need to show that for any closed~$\vec\rho$,
$\vbounded[\substin{\sigma'}{\vec\rho}{\vec\alpha}]
          {v'}
          {\ctyapp{(\ctylam{\vec\alpha}{\esubst{p'}\Theta})}
                  {\ptrans{\vec\rho}}}$,
and by~\infruleref{beta-all} and weakening, it suffices to show
$\vbounded[\substin{\sigma'}{\vec\rho}{\vec\alpha}]
          {v'}
          {\substin{\esubst{p'}{\Theta}}{\ptrans{\vec\rho}}{\vec\alpha}}$.
This in turn requires us to show that if
$\ftv(\sigma) = \set{\vec\alpha,\vec\beta}$, then for any
closed~$\vec{\rho'}$,
$\vbounded[\substin{\sigma'}{\vec\rho,\vec{\rho'}}{\vec\alpha,\vec\beta}]
          {v'}
          {\substin{\esubst{p'}{\Theta}}{\ptrans{\vec\rho,\vec{\rho'}}}
                                        {\vec\alpha,\vec\beta}}$,
which follows from the fact that
$\vbounded[\sigma']{v'}{\esubst{p'}{\Theta}}$.
\qedhere
\end{proofcases}
\end{proof}


\end{document}